\documentclass[10pt,letterpaper,oneside]{article}
\usepackage[T1]{fontenc}
\usepackage[utf8]{inputenc}
\usepackage[english]{babel}
\usepackage[textwidth=7.1in,textheight=9in,top=1in,headheight=14pt,headsep=20pt,footskip=30pt]{geometry}
\usepackage{amsmath,amssymb}
\usepackage{newtxtext,newtxmath,microtype}
\usepackage{graphicx,booktabs,colortbl,float,placeins,flafter,etoolbox,authblk,adjustbox}
\usepackage[font=small,labelfont=bf,labelsep=period]{caption}
\usepackage[numbers,sort&compress,square]{natbib}
\usepackage[hidelinks]{hyperref}

\newcommand{\tablenote}[1]{{\footnotesize\raggedright#1\par}}
\AtBeginEnvironment{tabular}{\sffamily\fontsize{7.5}{10}\selectfont}
\AtBeginEnvironment{tabular*}{\sffamily\fontsize{7.5}{10}\selectfont}
\usepackage{amsmath,amssymb,bm}
\usepackage{pifont}

\usepackage{amsthm}
\graphicspath{{./}}
\newcommand{\vr}{\mathbf r}
\newcommand{\vn}{\mathbf n}
\newcommand{\vz}{\mathbf z}
\newcommand{\vf}{\mathbf f}
\newcommand{\T}{\mathsf T}
\newcommand{\rmd}{\mathrm d}
\newcommand{\Fid}{F_{\rm id}}
\newcommand{\FC}{F_{\rm Coul}}
\newcommand{\Fex}{F_{\rm ex}}
\newcommand{\nb}{\bar n}
\newcommand{\methodsref}{\textit{Methods}}

\newcommand{\siref}{\textit{Supplementary Information}}

\newcommand{\PaperTitle}{A neural density functional for polymer electrolytes with exact structure and transferable Coulomb coupling}
\newcommand{\PaperAuthor}{Liyao Lyu}

\newcommand{\PaperAffiliationA}{School of Artificial Intelligence and Data Science, University of Science and Technology of China, Hefei, Anhui, China}
\newcommand{\PaperAffiliationB}{Suzhou Institute for Advanced Research, University of Science and Technology of China, Suzhou, Jiangsu, China}
\newcommand{\PaperAffiliationC}{Suzhou Big Data \& AI Research and Engineering Center, Suzhou, Jiangsu, China}
\newcommand{\PaperEmail}{lyuliyao@ustc.edu.cn}
\newcommand{\PaperKeywords}{classical density functional theory $|$ polymer electrolytes $|$ electrostatic correlations $|$ neural free energy functionals $|$ transferable coarse-grained models}
\newcommand{\PaperAcknowledgments}{The author thanks Huan Lei for helpful discussions. The author acknowledges the Supercomputing Center of University of Science and Technology of China for providing the computing resources.}
\newcommand{\PaperDataAvailability}{The analysis and training code, model configurations, trained parameters, the processed MD data used in this work (density and internal force-density profiles, applied potentials, bulk structure factors, Coulomb energies and the training, validation and test splits at all five permittivities) and the scripts that reproduce the figures and tables are publicly available in the GitHub repository \url{https://github.com/Lyuliyao/NDFT-polymer-electrolyte} (release v1.0.0) and archived at Zenodo (DOI: 10.5281/zenodo.23558960). Raw MD trajectories are not deposited because of their size and are available from the author upon request.}

\title{\PaperTitle}
\author[a,b,c]{\PaperAuthor\thanks{Correspondence: \PaperEmail}}
\affil[a]{\PaperAffiliationA}
\affil[b]{\PaperAffiliationB}
\affil[c]{\PaperAffiliationC}
\date{}
\hypersetup{pdfauthor={\PaperAuthor},pdftitle={\PaperTitle}}
\begin{document}
\maketitle
\begin{abstract}
Ion correlations set the structure of polymer electrolytes and depend on the host permittivity, which is low and varies between polymers. A functional that carries these correlations from molecular simulations to larger inhomogeneous systems must transfer across domain sizes, geometries and coupling strengths. We introduce a neural density functional for the ions of a salt-doped polymer melt with built-in integrability, continuous spatial symmetry, the Noether identities, perfect screening in stable bulk states, and parameters that refer to no grid or box. The permittivity enters only through the Bjerrum length, an explicit argument constrained by an exact relation: the free-energy derivative with respect to the Bjerrum length is the Coulomb correlation energy. Trained by force matching on planar profiles at five Bjerrum lengths spanning 8--30 monomer diameters, the functional predicts, without retraining, density profiles in fourfold longer boxes and under two-dimensional fields, bulk structure factors and the long-wavelength number response, also at a concentration and a permittivity left out of training. It reproduces this energy identity on Coulomb energies that never enter the training. Its nonlinear density dependence captures correlations that a quadratic pair closure misses: the number response grows several-fold with concentration, and at low concentration the coupling lowers it through ion association, below the ideal value at the strongest coupling, while the host sets it in the concentrated melt. At fixed coupling, it is more accurate than three neural architectures and loses little accuracy with a quarter of the training data.
\end{abstract}
\BeforeBeginEnvironment{tabular}{\begin{adjustbox}{max width=\linewidth}}
\AfterEndEnvironment{tabular}{\end{adjustbox}}
\section{Introduction}
Salt-doped polymers are attractive solid electrolytes for lithium batteries, offering improved safety and mechanical stability over liquid electrolytes \cite{hallinan2013polymer,bocharova2020perspectives}. Their performance is nonetheless hard to predict, since adding salt raises the number of charge carriers while also altering ion correlations \cite{son2020ion,fong2021ion,shen2020ion,tsamopoulos2024ion}: as concentration increases, ions form pairs and clusters \cite{borodin2006mechanism,molinari2018effect,france2019correlations}, and measured cation transference numbers may even turn negative \cite{pesko2017negative}. These correlations arise on two distinct length scales. In low-permittivity polymers the Bjerrum length extends over several to tens of monomer diameters, whereas packing constraints and solvation by the chains generate additional correlations at shorter range. A theory of these correlations must go beyond the Coulomb mean field and remain valid as concentration and external fields vary. Molecular simulations capture them accurately, and data-driven coarse-graining extends their time scales \cite{lyu2023construction,wang2025many,hernandez2026data}, but not the size or geometry of the domain. This calls for a continuum theory that retains their molecular fidelity while extending to the larger scales and more complex geometries of inhomogeneous electrolytes.

Classical density functional theory provides a framework that connects these two scales \cite{evans1979nature,hansen2013theory}. A free-energy functional determines the equilibrium density profiles through its first functional derivative and the bulk correlations through its second. The excess part of this functional is unknown, however, and has to be approximated.

For electrolytes, analytical approximations combine fundamental measure theory for ion packing \cite{rosenfeld1989free,roth2010fundamental} with electrostatic correlation terms based on the mean spherical approximation or a reference fluid \cite{rosenfeld1993free,gillespie2002coupling,hartel2017structure,jiang2018density}. These functionals are built on charged hard-sphere reference systems and do not include the solvation of ions by polymer chains. Theories of ion-containing polymers include solvation through Born-type and self-energy contributions \cite{wang2010fluctuation,nakamura2011thermodynamics,nakamura2012salt} and account for ionic correlations through liquid-state closures \cite{sing2014electrostatic}. In both approaches, each term is an analytical approximation derived for a different reference system, and neither the individual terms nor their sum is obtained from the polymer electrolyte itself.

Neural approaches instead learn the excess part from simulation data, either through the one-body direct correlation functions or as the excess free energy itself~\cite{sammuller2023neural,bui2025learning,dijkman2025learning}. Local learning of the one-body correlations describes inhomogeneous fluids accurately, even in planar domains far larger than those used for training \cite{sammuller2023neural}. With a separate treatment of the long-range electrostatics it extends to ionic mixtures \cite{bui2025learning}, and with machine-learned interatomic potentials to molecular liquids from first principles \cite{bui2026unified}. The learned correlations, however, are not guaranteed to derive from a free-energy functional, and they respect only the discrete symmetries of the grid.
Both properties matter: they make the predictions satisfy exact relations of statistical mechanics, and they restrict the functional to a narrower set of admissible forms, which can be learned from less data \cite{batzner20223} and, as we will show, generalizes more reliably to unseen conditions.

Recent work has begun to build some of these properties in. Learning the free energy directly ensures integrability \cite{dijkman2025learning}, and an equivariant formulation developed concurrently with the present work extends this approach by incorporating the spatial symmetries of a cubic grid \cite{cheng2026equivariant}. The parameters of both are defined on a fixed grid, and both were developed for simple fluids without long-range electrostatics.
Closest in form to the present work are learned functionals of weighted densities \cite{lin2019classical,kelley2024bridging,glitsch2025neural}, in particular the rotationally invariant functional of Kelley et al.\ \cite{kelley2024bridging}, which is trained on free energies that require thermodynamic integration, has been demonstrated for planar inhomogeneities, and does not treat long-range electrostatics.
Here these requirements are met simultaneously and by construction: integrability, continuous spatial symmetry and explicit long-range electrostatics are built into a functional that is learned from simulations alone and transfers across geometries without reference to a grid (Table~\ref{tab:architectures}).

\begin{table*}[tb]
\centering
\caption{Structural guarantees and transferability of the learned functionals compared in this work, in their original formulations. The first five rows are structural properties: \ding{51}, guaranteed by construction; N, verified numerically but not built in; G, holds only for the symmetry operations of the grid; \ding{55}, neither built in nor reported. The last two rows are transfer tests: \ding{51}, demonstrated without retraining; \ding{55}, not reported.}
\label{tab:architectures}
\begingroup
\renewcommand{\arraystretch}{1.2}
\small
\begin{tabular*}{\textwidth}{@{\extracolsep{\fill}}lccccc@{}}
\toprule
 & \multicolumn{2}{c}{Learned $c^{(1)}$} & \multicolumn{2}{c}{Learned $F_{\rm ex}$} & \\
\cmidrule(lr){2-3}\cmidrule(lr){4-5}
Property & Samm\"uller et al.\ \cite{sammuller2023neural} & Bui--Cox\ \cite{bui2025learning} & Dijkman et al.\ \cite{dijkman2025learning} & Cheng\ \cite{cheng2026equivariant} & This work\\
\midrule
Integrability & N & \ding{55} & \ding{51} & \ding{51} & \ding{51}\\
Spatial symmetries & G & G & \ding{55} & G & \ding{51}\\
Noether force/torque identities & N & \ding{55} & \ding{55} & \ding{55} & \ding{51}\\
Long-range electrostatics & \ding{55} & \ding{51} & \ding{55} & \ding{55} & \ding{51}\\
Grid-independent parameters & \ding{55} & \ding{55} & \ding{55} & \ding{55} & \ding{51}\\
\midrule
Larger domains & \ding{51} & \ding{51} & \ding{55} & \ding{51} & \ding{51}\\
Nonplanar geometries & \ding{55} & \ding{55} & \ding{55} & \ding{51} & \ding{51}\\
\bottomrule
\end{tabular*}
\endgroup
\end{table*}

\begin{figure}[!htb]
\centering
\includegraphics[width=0.95\textwidth,height=0.45\textheight,keepaspectratio]{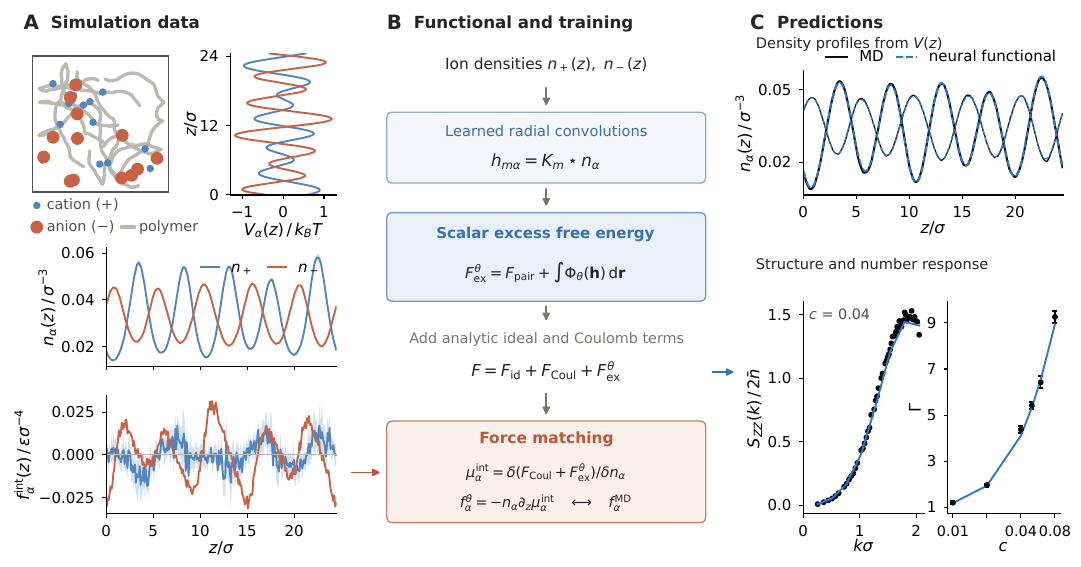}
\caption{A learned density functional for the ions of a salt-doped polymer melt. (\textit{A}) Molecular dynamics under a planar external potential provides ion densities $n_\pm(z)$ and internal force densities $f^{\rm int}_\pm(z)$. (\textit{B}) The excess free energy is parameterized by radial kernels and a pointwise neural network and trained by matching internal force densities to MD. The ideal-gas and Coulomb contributions are included analytically. (\textit{C}) Predicted density profiles for a held-out run, bulk charge structure, and number response at the smallest nonzero wavenumber, compared with MD.}
\label{fig:model}
\end{figure}

The polymer host imposes a further requirement. The ion correlations depend strongly on its permittivity, which varies from one polymer to another. In a coarse-grained description, the permittivity enters only through the Bjerrum length $l_B$, the separation at which the Coulomb energy of two ions equals $k_BT$. As $l_B$ increases, correlations shift from screening at weak coupling to ion pairing at strong coupling \cite{bjerrum1926,fisher1993criticality}. A functional trained at a single permittivity cannot capture this change, and its predictions at other permittivities are unsupported. A functional for polymer electrolytes must therefore take $l_B$ as an explicit argument and capture both how the coupling affects correlations and the length scales over which those effects persist.

In this paper, we propose a neural density functional for the ions in a salt-doped polymer melt \cite{tsamopoulos2024ion} (Fig.~\ref{fig:model}). The Coulomb interaction is kept analytically, and the remaining correlations are represented by a scalar functional of the cation and anion densities alone. It is built from radial convolution kernels and a pointwise neural network, both defined in continuous space rather than on a grid. The chemical potentials of both species are functional derivatives of this one scalar functional, so free-energy differences between density profiles do not depend on the integration path. The functional is invariant under all translations, rotations and reflections, so the total internal force and torque vanish, as the Noether identities require \cite{hermann2021noether}. Because the Coulomb term is explicit and the learned kernels are short ranged, the small-wavenumber charge correlations satisfy the perfect-screening sum rules \cite{stillinger1968general} in stable, noncritical bulk states. No parameter refers to a grid or box size, so the trained functional can be evaluated in domains of other sizes and geometries.

The permittivity enters the functional only through the Bjerrum length $l_B$, which we treat as an explicit argument rather than a fixed parameter. Because the Coulomb interaction is linear in $l_B$, the coupling-constant integration of statistical mechanics fixes the derivative of the free-energy functional with respect to $l_B$ as the Coulomb correlation energy, the classical analogue of the adiabatic connection in electronic density functional theory \cite{langreth1975exchange,gunnarsson1976exchange}. We build this structure in through a coupling law that writes the excess functional as the sum of an $l_B$-independent term, a term linear in $l_B$ and a term that saturates at strong coupling, instead of feeding $l_B$ to the network as an input. Through this relation, the Coulomb energies of the bulk simulations fix how the long-wavelength number response changes with $l_B$. These energies never enter the loss, and our functional reproduces the change.

We train this functional by matching species-resolved internal force densities measured in canonical molecular dynamics under static planar external potentials. The first Yvon--Born--Green equation connects these force densities to gradients of the interaction chemical potentials generated by the functional. The spatially constant chemical potentials drop out on taking the gradient, so the loss constrains the functional across inhomogeneous ionic environments without chemical-potential labels. This force-based supervision differs from fitting bulk pair correlations \cite{dijkman2025learning}, fitting free energies obtained by thermodynamic integration \cite{kelley2024bridging}, enforcing local chemical-potential balance \cite{sammuller2026determining,cheng2026equivariant}, matching the time evolution of density fields \cite{chen2026learning}, or fitting measure-dependent forces to particle trajectories \cite{lyu2026mvnn}. Neither bulk structure factors nor the associated number susceptibility enters the force-matching loss, so they test the learned free-energy curvature beyond the fitted forces.

We study five coupling strengths, with Bjerrum lengths spanning 8--30 monomer diameters, and train a single functional across all five. It predicts ionic structure in larger boxes, under external potentials that vary in two directions, at an untrained salt concentration, and at a permittivity excluded from training. In these tests, it recovers density profiles, bulk structure factors, and the long-wavelength number response. Its nonlinear density dependence captures concentration-dependent effective ion correlations that a fixed quadratic pair closure misses, including a long-wavelength number response that grows several-fold with concentration and, at strong coupling, falls below its ideal-solution value at the lowest concentration, where number fluctuations are enhanced. The functional also locates where the coupling acts: it changes the short-range charge correlations and promotes ion association at low concentration, but hardly alters the number response of the concentrated melt. When trained on the same ion data at a fixed coupling, the functional is more accurate than three other neural architectures. It loses little accuracy when trained on only a quarter of the runs, whereas the error of the best alternative grows by about two thirds, and it satisfies the integrability and Noether identities on the simulated profiles to numerical precision, which the alternatives do not.

\section{Theory}

\subsection{Free-energy formulation}
We consider a salt dissolved in a polymer melt. Treating the polymer implicitly, the ions are described by their number-density profiles $n_\alpha(\vr)$, with $\alpha\in\{+,-\}$, collected in $\vn=(n_+,n_-)$. Within classical density functional theory, these profiles are governed by the intrinsic free-energy functional
\begin{equation}
F[\vn]=\Fid[\vn]+\FC[\vn]+\Fex[\vn],
\label{eq:F}
\end{equation}
where
\(
\Fid[\vn]=k_BT\sum_\alpha\int \rmd\vr\,n_\alpha(\vr)\big[\ln\!\big(\Lambda^3n_\alpha(\vr)\big)-1\big]
\)
is the ideal-gas part, with thermal energy $k_BT$ and thermal wavelength $\Lambda$, and
\(
\FC[\vn]=\frac12\iint \rmd\vr\,\rmd\vr'\,\frac{\rho_Z(\vr)\rho_Z(\vr')}{4\pi\varepsilon_0\varepsilon_r|\vr-\vr'|}
\)
is the Coulomb mean-field energy of the charge density $\rho_Z(\vr)=\sum_\alpha e_\alpha n_\alpha(\vr)$. The ionic charges are $e_\alpha=z_\alpha e$, with valences $z_\pm=\pm1$ and elementary charge $e$. Here $\varepsilon_0$ is the vacuum permittivity and $\varepsilon_r$ the uniform relative permittivity of the medium. The excess part $\Fex[\vn]$ contains all correlations beyond the ideal gas and the Coulomb mean field and is the unknown component of the theory. With the polymer integrated out, $F$ is the intrinsic free energy of the ions in a melt that stays in equilibrium with their densities, so the response of the chains to the ions is part of $\Fex$. The functional therefore refers to one host state, here the melt at fixed temperature and pressure.

In external potentials $V_\alpha(\vr)$, the equilibrium profiles minimize $F+\sum_\alpha\int\rmd\vr\,V_\alpha n_\alpha$ at fixed particle numbers, which gives the Euler--Lagrange equations
\begin{equation}
k_BT\ln\!\big[\Lambda^3 n_\alpha(\vr)\big]+e_\alpha\phi(\vr)+\mu^{\rm ex}_\alpha(\vr;[\vn])+V_\alpha(\vr)=\mu_\alpha .
\label{eq:EL}
\end{equation}
The first three terms are the functional derivatives of $\Fid$, $\FC$ and $\Fex$, respectively. The mean electrostatic potential $\phi$ satisfies Poisson's equation, $-\nabla^2\phi=\rho_Z/(\varepsilon_0\varepsilon_r)$, the excess chemical potential is $\mu^{\rm ex}_\alpha=\delta\Fex/\delta n_\alpha(\vr)$, and the chemical potentials $\mu_\alpha$ are the Lagrange multipliers that fix the particle numbers.
 
Taking the gradient of Eq.~\ref{eq:EL} and combining it with the equilibrium force balance, the first equation of the Yvon--Born--Green hierarchy (derivation in \siref, section S2.2), gives
\begin{equation}
\vf_\alpha(\vr)=-n_\alpha(\vr)\,\nabla\mu^{\rm int}_\alpha(\vr;[\vn]),
\qquad
\mu^{\rm int}_\alpha=e_\alpha\phi+\mu^{\rm ex}_\alpha,
\label{eq:YBG}
\end{equation}
where $\vf_\alpha(\vr)$ is the equilibrium internal force density exerted on ions of species $\alpha$ by the other ions and the polymer monomers, excluding the external force. Equation~\ref{eq:YBG} therefore relates a quantity sampled directly from the interaction forces in molecular dynamics to the functional, without chemical-potential labels. Since only the gradient of $\mu^{\rm int}_\alpha$ enters, force matching determines $\Fex$ only up to terms linear in the particle numbers $N_\alpha=\int\rmd\vr\,n_\alpha$, which shift $\mu_\alpha$ by constants (\siref, section S2.2).

\subsection{The neural functional}
We parameterize the excess free energy as
\begin{align}
F^\theta_{\rm ex}[\vn]={}&
\frac12\sum_{\alpha\beta}\iint \rmd\vr\,\rmd\vr'\,
n_\alpha(\vr)\,u_{\alpha\beta}(|\vr-\vr'|)\,n_\beta(\vr')
\nonumber\\
&+\int \rmd\vr\,\Phi_\theta\big(\mathbf h(\vr)\big),
\label{eq:Fex}
\end{align}
where $\alpha,\beta$ denote the species, $u_{\alpha\beta}=u_{\beta\alpha}$ are radial pair kernels, and the neural network $\Phi_\theta$ is evaluated pointwise on the vector $\mathbf h(\vr)$ of weighted densities of both ionic species,
\[
h_{m\alpha}(\vr)=(K_m\star n_\alpha)(\vr).
\]
Here $\star$ denotes convolution. The learned radial kernels $K_m$ average each density over several length scales. The first term is quadratic in the densities and, on its own, is the pair closure used as a reference below. The second term makes the effective correlations depend nonlinearly on the surrounding densities. Because the kernels are continuous functions of distance and $\Phi_\theta$ acts pointwise, the functional is defined in continuous space: its parameters refer to no grid or box, and a discretization enters only when it is evaluated numerically. The trainable parameters $\theta$ comprise the weights and biases of $\Phi_\theta$, the parameters of the radial kernels $K_m$, including their Gaussian widths, and the pair coefficients $a_{\alpha\beta m}$ (\methodsref).

\subsection{Properties of the neural functional}
With the excess part of Eq.~\ref{eq:Fex}, the functional $F$ of Eq.~\ref{eq:F} has three properties by construction (proofs in \siref, section S1). First, because the network represents the scalar $\Fex$ rather than the excess chemical potentials, $\mu^{\rm ex}_\alpha=\delta\Fex/\delta n_\alpha$ holds exactly: the Hessian of $\Fex$, which gives the pair direct correlation functions, is symmetric, and excess free-energy differences between density profiles do not depend on the integration path. Second, $F$ is invariant under translations, rotations and reflections, because all kernels are radial and $\Phi_\theta$ acts pointwise, so by Noether's theorem the total internal force and torque vanish for any admissible density profile \cite{hermann2021noether}. Third, because the Coulomb term is explicit and the learned contribution is short ranged, with a finite limit as $k\to0$, the charge structure factor of an electroneutral monovalent electrolyte in a stable, noncritical bulk state satisfies the perfect-screening sum rules \cite{stillinger1968general},
\begin{equation}
\frac{S_{ZZ}(k)}{2\nb}
=\frac{k^2}{\kappa_D^2}+O(k^4),
\qquad
\kappa_D^2=8\pi l_B\nb,
\label{eq:SL}
\end{equation}
where $S_{ZZ}(k)$ is the charge structure factor per unit volume, $\nb=\nb_+=\nb_-$ the salt number density, $\kappa_D$ the Debye wavenumber and $l_B=e^2/(4\pi\varepsilon_0\varepsilon_r k_BT)$ the Bjerrum length. The vanishing constant term and the $k^2$ coefficient are the electroneutrality and Stillinger--Lovett conditions.

\paragraph{Dependence on the Coulomb coupling.}
At fixed polymer model, the permittivity enters the interaction energy only through the Bjerrum length $l_B$: in units of $k_BT$, this energy is $U_0+l_B U_C$, where $U_0$ collects the bonded, excluded-volume and solvation interactions and
\(
U_C=\sum_{i<j}z_iz_j/|\vr_i-\vr_j|
\)
is the Coulomb energy of the ions per unit Bjerrum length. The functionals at different permittivities therefore form a one-parameter family $F[\vn;l_B]$, and the coupling-constant integration of statistical mechanics fixes how its members are related: at fixed density profile,
\begin{equation}
\frac{\partial F[\vn;l_B]}{\partial l_B}=k_BT\,\langle U_C\rangle_{\vn,l_B},
\label{eq:HF}
\end{equation}
where the average is taken in the equilibrium ensemble with profile $\vn$ at coupling $l_B$ (\siref, section S1.5). The mean-field term $\FC$ is linear in $l_B$ and satisfies Eq.~\ref{eq:HF} with the mean-field Coulomb energy. The derivative of $\Fex$ with respect to $l_B$ must therefore equal the Coulomb correlation energy per unit Bjerrum length, the part of $k_BT\langle U_C\rangle$ beyond the mean field. The dependence of the excess functional on the coupling is thus not arbitrary: it is an energy.

We represent this dependence by a coupling law,
\begin{equation}
\Fex[\vn;l_B]=F_0[\vn]+\lambda\,F_1[\vn]+s(\lambda)\,F_2[\vn],
\label{eq:lb}
\end{equation}
with $\lambda=l_B/l_B^{(0)}$ and $s(\lambda)=1-e^{-\lambda/\lambda_s}$, where $F_0$, $F_1$ and $F_2$ have the form of Eq.~\ref{eq:Fex} and share the kernels $K_m$ and the hidden layers of $\Phi_\theta$, differing only in their output layers and pair coefficients. Here $l_B^{(0)}=7.96\sigma$ is a reference Bjerrum length and $\lambda_s$ is learned. At fixed $l_B$, Eq.~\ref{eq:lb} is a functional of the form of Eq.~\ref{eq:Fex}, so the properties above hold at every coupling. 
Equation~\ref{eq:lb} separates the correlation energy into three contributions: $F_0[\vn]$, independent of $l_B$; $\lambda F_1[\vn]$, linear in $l_B$, as in the mean spherical approximation at strong coupling \cite{waisman1972mean}; and $s(\lambda)F_2[\vn]$, which approaches a plateau as $l_B$ increases, analogous to saturation associated with ion pairing.

Equation~\ref{eq:HF} yields a test based on an observable absent from training. In a uniform bulk at salt density $\nb$, let $u_C(\nb;l_B)$ be the Coulomb energy per unit volume, which is entirely correlational because the mean-field energy of a neutral bulk vanishes. The long-wavelength number response $\Gamma(0)=1+(\nb/2k_BT)\,\mathbf q^{\T}W(0)\mathbf q$, defined below, is the second derivative of the excess free energy per unit volume with respect to $\nb$ (\siref, section S1.5), so that
\begin{equation}
\frac{\partial\Gamma(0)}{\partial l_B}
=\frac{\nb}{2k_BT}\,\frac{\partial^2}{\partial\nb^2}\,\frac{u_C(\nb;l_B)}{l_B}.
\label{eq:HFbulk}
\end{equation}
Force matching leaves $\Fex$ undetermined by terms linear in the particle numbers, which do not contribute to this second derivative. The right-hand side is measured from the Coulomb energies of zero-field simulations, and the left-hand side follows from the trained functional by differentiation with respect to $l_B$.
\subsection{Training by force matching}
We train $F^\theta_{\rm ex}$ by matching Eq.~\ref{eq:YBG} to the ion-density and internal force-density profiles of canonical molecular dynamics under static planar external potentials. For each run, the right-hand side of Eq.~\ref{eq:YBG} is evaluated with $F^\theta_{\rm ex}$ on the sampled density profiles. The parameters minimize the squared residuals against the MD force densities, weighted by the inverse variance of the MD estimate, together with the squared Fourier amplitudes of the residuals weighted by $1/k^2$, which compensates for the factor $k$ by which the gradient in Eq.~\ref{eq:YBG} suppresses long-wavelength errors of $\mu^{\rm ex}_\alpha$ (\methodsref).

\FloatBarrier
\section{Methods}

\subsection{Molecular dynamics}
We simulate the coarse-grained model of Tsamopoulos and Wang \cite{tsamopoulos2024ion}: 400 chains of 30 beads joined by FENE bonds \cite{KremerGrest1990}, with 120 to 960 cation--anion pairs for $c=0.01$ to $0.08$, in the Lennard-Jones units $\sigma$, $\epsilon$ and $\tau$ of the monomers. Both ions interact with the monomers through the solvation potential of Hall and coworkers \cite{shen2020ion}, of strength $4.33\,k_BT$ and range $5\sigma$. Electrostatics uses PPPM \cite{hockney2021computer} with dielectric constants 7.5, 5, 4, 3 and 2. All runs use LAMMPS \cite{thompson2022lammps} with a Nos\'e--Hoover thermostat \cite{nose1984unified,hoover1985canonical} at $k_BT=\epsilon$ and a time step of $0.005\tau$. Each state point is equilibrated at zero pressure and then simulated at the mean volume. 

\subsection{External potentials and profiles}
The potentials $V_\pm(z)=V_N(z)\pm\psi(z)$ combine a part $V_N$ that acts equally on both species with a part $\psi$ that acts with opposite signs, drawn from four families that span wavelengths from the box length down to $1.2\,\sigma$ (\siref, section S2.1). At $\varepsilon_r=7.5$ and 2, 22 to 30 potentials are applied at each concentration, plus 22 at the untrained concentration $c=0.05$. At $\varepsilon_r=5$, 4 and 3, the same 15 potentials are applied at every permittivity, with eight further potentials at $c=0.05$ at $\varepsilon_r=4$. Density and internal force density profiles are recorded on $0.1\sigma$ bins, with standard errors from eight consecutive blocks, after $6\times10^3$ to $4\times10^4\,\tau$ of equilibration and $3.7\times10^4$ to $2.0\times10^5\,\tau$ of sampling. Zero-field runs provide the structure factors and the Coulomb energies, computed by Ewald summation. In total, 276 runs train the functional, and 86 held-out runs and the 52 runs at $c=0.05$ test it (\siref, section S2.1).

\subsection{Neural functional and training}
The kernels $K_m(r)=B_m(r)[1+g_m(r)]$, $m=1$ to 6, combine normalized Gaussians $B_m$ of learned widths, bounded between $0.1\sigma$ and $8\sigma$, with the outputs $g_m$ of a perceptron of $r/\sigma$ with two hidden layers of 32 units. The pair kernels are $u_{\alpha\beta}=\sum_m a_{\alpha\beta m}K_m$ with learned symmetric coefficients, and $\Phi_\theta$ is a perceptron with two hidden layers of 128 units that acts on the 12 weighted densities. Both networks use softplus activations. The three functionals of Eq.~\ref{eq:lb} share $K_m$ and the hidden layers of $\Phi_\theta$, which has three outputs, and every pair coefficient has the form $a^{(0)}+\lambda a^{(1)}+s(\lambda)a^{(2)}$. The functional has 19{,}942 parameters, 19{,}647 at fixed coupling, and the pair closure 1{,}342. For the test across the coupling, three further functionals are trained with the runs at $\varepsilon_r=5$, 4 or 3 removed. In the conditioning ablation, $\ln\lambda$ enters $\Phi_\theta$ as an additional input or by feature-wise modulation of its hidden layers \cite{perez2018film} (\siref, section S5.2).

For the planar profiles, with $\Delta_{r\alpha i}=f^{\rm MD}_{r\alpha}(z_i)-f^\theta_{r\alpha}(z_i)$ the force-density residual of run $r$ in bin $i$, the loss is
\begin{equation}
\begin{aligned}
\mathcal L(\theta)={}&\sum_{r\alpha}\sum_{i=1}^{N_r}w_{r\alpha i}\,\Delta_{r\alpha i}^2\\
&+\sum_{r\alpha}\bar w_{r\alpha}\sum_{m=1}^{\lfloor N_r/2\rfloor}
\frac{2}{N_r}\,\frac{|\widehat\Delta_{r\alpha}(k_m)|^2}{(k_m\sigma)^2},
\end{aligned}
\label{eq:loss}
\end{equation}
where $N_r$ is the number of bins, $\widehat\Delta_{r\alpha}(k_m)$ the unnormalized discrete Fourier transform of the residual at $k_m=2\pi m/L_r$, $w_{r\alpha i}=1/(s_{r\alpha i}^2+\bar s_{r\alpha}^2)$ the inverse variance of the MD estimate with standard error $s_{r\alpha i}$ and its median $\bar s_{r\alpha}$ over the run, and $\bar w_{r\alpha}$ the mean bin weight of the run. We minimize the loss with AdamW \cite{LoshchilovHutter2019} (learning rate $10^{-3}$ with cosine decay over 12{,}000 steps, weight decay $10^{-5}$, and a 30-fold larger learning rate for the kernel widths) and keep the parameters with the lowest validation residual, evaluated every 20 steps (\siref, section S2.2).

\paragraph{Energy identity and channel decomposition.}
The Coulomb energy per unit volume of each zero-field run is computed by Ewald summation and differentiated twice with respect to $\nb$ by second differences on the concentration grid, with errors propagated from the time-block errors. The functionals are differentiated by automatic differentiation of $\mathbf q^{\T}W(0)\mathbf q$ with respect to $l_B$ and, for the comparison with MD, by the same second differences. The charge channel beyond the mean field is obtained from the zero-field $S_{ZZ}(k)$ by subtracting the ideal and Debye terms. The paired fraction of Bjerrum theory uses a contact distance of $\sigma$ with Debye--H\"uckel activity coefficients of the free ions (\siref, section S5.3).
\subsection{Other learned functionals}
We adapt three published forms to the two ion species and the analytic Coulomb term and train them with the same data, splits and loss (\siref, section S4.2). The one-body network \cite{sammuller2023neural,bui2025learning} is a perceptron with two hidden layers of 128 units that maps the densities of both species within $\pm3\sigma$ of a point to $\mu^{\rm ex}_\pm$ at that point. The lattice free energy \cite{cheng2026equivariant} uses $O_h$-invariant products of Cartesian density moments on $0.5\sigma$ voxels with a stencil radius of $1.5\sigma$. The convolutional free energy \cite{dijkman2025learning} passes the two densities through six periodic dilated convolutions and a linear read-out. The Lennard-Jones control is the truncated and shifted fluid of ref.~\cite{cheng2026equivariant} ($r_c=2.5\sigma$, $T=1.5$) in a cubic box of side $24\sigma$, trained at $\rho\sigma^3=0.1$, 0.2, 0.4, 0.6 and 0.7 with 22 planar potentials each and tested at $\rho\sigma^3=0.5$.

\section{Results}
\begin{figure}[!htb]
\centering
\includegraphics[width=0.95\textwidth,height=0.45\textheight,keepaspectratio]{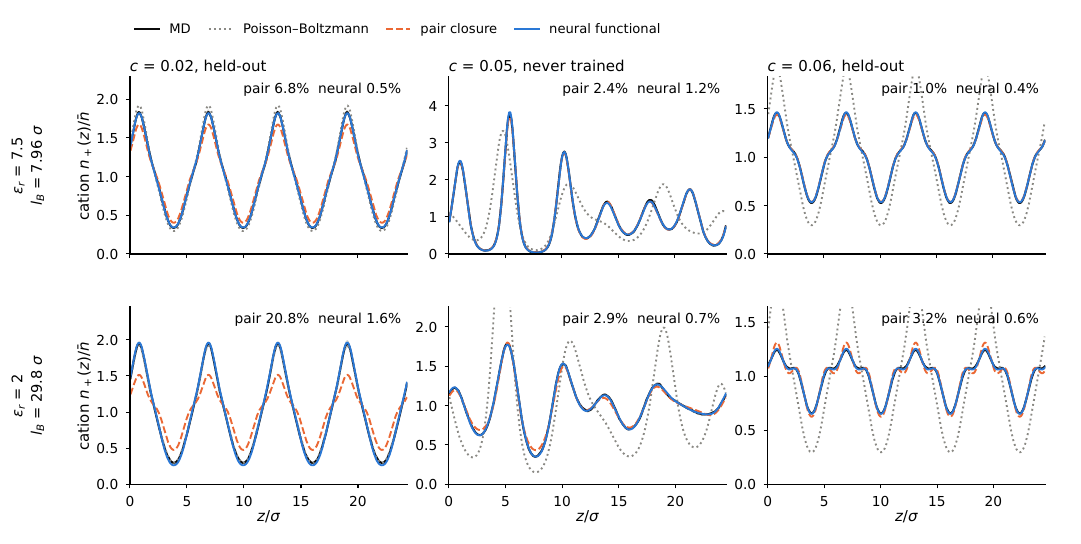}
\caption{Density profiles predicted for runs excluded from training at the two ends of the coupling range: $\varepsilon_r=7.5$ ($l_B=7.96\sigma$, top) and $\varepsilon_r=2$ ($l_B=29.8\sigma$, bottom). Columns: a periodic array of neutral Gaussian wells at $c=0.02$, the strongest Fourier potential ($2.8\,k_BT$ peak to peak in both parts) at the untrained concentration $c=0.05$, and the same array of wells at $c=0.06$. Cation densities relative to the mean. Black curves and bands, MD averages with $\pm2$ standard errors; dotted, Poisson--Boltzmann theory; dashed, pair closure; solid, neural functional. Numbers are the relative $L_2$ errors of the pair closure and the neural functional, averaged over both species.}
\label{fig:response}
\end{figure}

\subsection{Density profiles and concentration transfer}
We apply the method to a coarse-grained model of a salt-doped polymer melt \cite{tsamopoulos2024ion}, with cation and anion diameters of $0.4\sigma$ and $1.6\sigma$ in a melt of monomers of diameter $\sigma$ that solvate both species (\methodsref). We collect data at five dielectric constants, $\varepsilon_r=7.5$, 5, 4, 3 and 2, corresponding to Bjerrum lengths $l_B=7.96\sigma$ to $29.8\sigma$, and at five salt concentrations, $c=0.01$ to $0.08$ cations per monomer, under static planar potentials $V_\alpha(z)$ with amplitudes of up to a few $k_BT$. A single functional of the form of Eq.~\ref{eq:lb} is trained on 276 runs spanning the five permittivities and evaluated on held-out runs and on runs at the untrained concentration $c=0.05$ (\siref, section S2.1).

We test the learned functional by solving Eq.~\ref{eq:EL} to predict equilibrium density profiles for the held-out runs (Fig.~\ref{fig:response}). We compare with two reference functionals. Poisson--Boltzmann theory is the mean-field limit $\Fex=0$. The pair closure serves as a quadratic reference: it keeps only the first term of Eq.~\ref{eq:Fex}, with density-independent kernels fitted to the same training data. Across all test runs, the mean relative $L_2$ errors of the neural functional trained across the five permittivities are $1.1\%$ at $\varepsilon_r=7.5$ and $1.6\%$ at $\varepsilon_r=2$, compared with $2.8\%$ and $5.4\%$ for the pair closure and 25\% and 23\% for Poisson--Boltzmann theory. At the untrained concentration $c=0.05$, the neural functional's errors are $0.80\%$ and $0.75\%$ at the two permittivities.
This gap arises because one density-independent kernel has to describe all concentrations. At $\varepsilon_r=7.5$, a pair closure fitted at $c=0.04$ alone predicts the held-out profiles at that concentration with an error of $1.1\%$, as accurately as the neural functional ($1.0\%$), but fitted jointly to all five concentrations its error there rises to $2.7\%$ (\siref, section S3.1). This comparison indicates that the effective correlations depend on concentration, a dependence we examine below.

\subsection{Transfer across domain sizes and geometries}
\begin{figure}[!htb]
\centering
\includegraphics[width=0.95\textwidth,height=0.45\textheight,keepaspectratio]{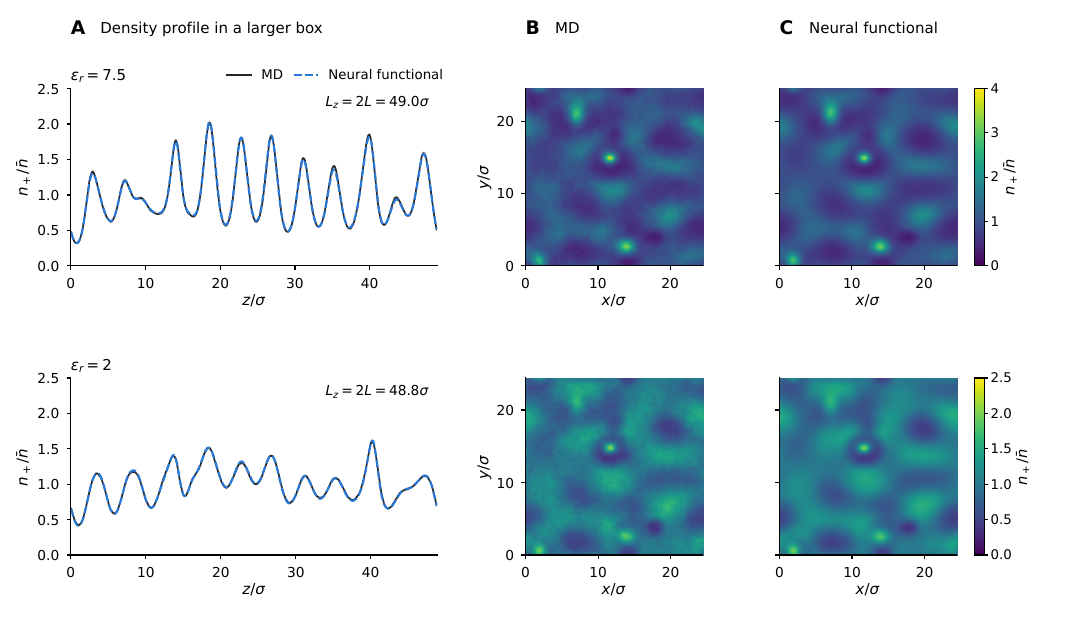}
\caption{Generalization to larger boxes and nonplanar density fields at $c=0.04$. Rows correspond to $\varepsilon_r=7.5$ and 2. (\textit{A}) Cation density profiles in a box twice the training length under an aperiodic potential. Black curves and gray bands show MD averages and $\pm2$ standard errors. Blue dashed curves show the neural-functional prediction. (\textit{B}, \textit{C}) MD and predicted cation densities under an aperiodic two-dimensional potential of oblique plane waves and irregularly placed Gaussian wells and barriers, using a shared color scale within each row. All predictions use parameters trained on planar profiles in the original box.}
\label{fig:generalization}
\end{figure}

Because its parameters refer to no grid or box, the trained functional can be evaluated in larger domains without changing its parameters. As a check of grid independence, we reevaluate the held-out profiles on grids with spacings $0.05\sigma$ and $0.2\sigma$. The excess chemical potentials change by less than $10^{-4}$ of their variation across the box (\siref, section S4.3). We also test transfer to boxes up to four times the training length under external potentials with longer wavelengths than those used for training (Fig.~\ref{fig:generalization}\textit{A}). At twice and four times the training length, the profile errors are 0.9--1.7\% and 0.3--0.5\%, against 3.0--5.5\% and 1.7--2.0\% for the pair closure.

The training profiles vary only along one direction, but the functional is defined for general three-dimensional densities. We therefore test it on six external potentials varying in two dimensions at $c=0.04$ and at $\varepsilon_r=7.5$ and 2, including both periodic and aperiodic fields (Fig.~\ref{fig:generalization} \textit{B} and \textit{C}; \siref, section S4.1). For the aperiodic potentials, the relative profile errors of the predicted density maps are 1.7--3.3\%, at the level of the MD noise of 1.7--3.0\%, against 3.7--5.0\% for the pair closure.

\begin{figure}[!htb]
\centering
\includegraphics[width=0.55\textwidth,height=0.45\textheight,keepaspectratio]{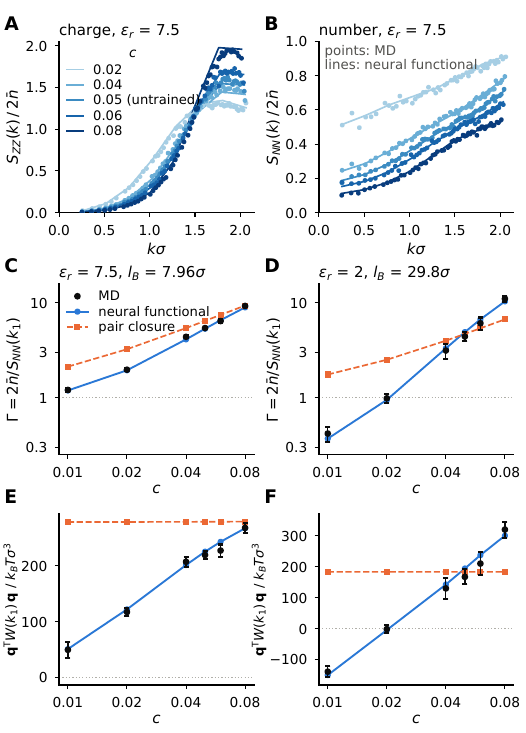}
\caption{Bulk structure and long-wavelength number response. (\textit{A}, \textit{B}) Normalized charge and number structure factors at $\varepsilon_r=7.5$ and $c=0.02$ to $0.08$. Points show zero-field MD results averaged over $k$-shells. Curves show predictions from Eq.~\ref{eq:OZ}. (\textit{C}, \textit{D}) Finite-wavenumber response factor $\Gamma(k_1)=2\nb/S_{NN}(k_1)$ at $\varepsilon_r=7.5$ and $2$. The dotted line marks the ideal value $\Gamma=1$. (\textit{E}, \textit{F}) Combined correlation kernel $\mathbf q^{\T}W(k_1)\mathbf q$ versus concentration at $\varepsilon_r=7.5$ and 2. Points show estimates from zero-field MD (\siref, section S3.2). In \textit{C} to \textit{F}, solid curves show the neural functional and dashed curves the pair closure. None of these quantities enters the force-matching loss.}
\label{fig:structure}
\end{figure}

\subsection{Bulk correlations and number response}
In a uniform bulk state, density fluctuations are determined by the Hessian of the free energy, a quantity not used in training. We define the Fourier-space excess kernel as
\[
W_{\alpha\beta}(k)
=\int\rmd\vr\,e^{-i\mathbf k\cdot\vr}\,
\frac{\delta^2\Fex}
{\delta n_\alpha(\vr)\,\delta n_\beta(\mathbf 0)}
\bigg|_{\vn=(\nb_+,\nb_-)}.
\]
Including the ideal-gas and Coulomb contributions, the Ornstein--Zernike relation gives
\begin{equation}
S(k)^{-1}
=\bar N^{-1}
+\frac{4\pi l_B}{k^2}\,\vz\vz^{\T}
+\frac{W(k)}{k_BT},
\label{eq:OZ}
\end{equation}
where $S_{\alpha\beta}(k)$ are the partial structure factors per unit volume. Here $\bar N={\rm diag}(\nb_+,\nb_-)$ contains the bulk densities, and $\vz=(1,-1)^{\T}$ is the valence vector.

Projecting $S(k)$ onto the charge and number-density channels gives
\[
S_{ZZ}(k)=\vz^{\T}S(k)\vz,
\qquad
S_{NN}(k)=\mathbf q^{\T}S(k)\mathbf q,
\]
where $\mathbf q=(1,1)^{\T}$.
Over $0.25\leq k\sigma\leq2.1$, the neural functional reproduces the zero-field MD structure factors in both channels with relative RMS errors of 2--7\% at most state points (Fig.~\ref{fig:structure}\textit{A} and \textit{B}). The errors are largest, 7--10\%, at $c=0.01$, where the MD structure factors are noisiest, and in the charge channel at $c=0.08$ and $\varepsilon_r=7.5$, where they reflect the height of the main peak of $S_{ZZ}(k)$ (\siref, section S3.1).

We quantify the long-wavelength number response by $\Gamma(k)=2\nb/S_{NN}(k)$, which equals 1 for an ideal solution. In the finite simulation box we report $\Gamma(k_1)$ at the smallest wavenumber $k_1=2\pi/L$. At $\varepsilon_r=2$ the MD estimate rises from $0.42\pm0.07$ at $c=0.01$ through the ideal value at $c=0.02$ to $11.1\pm0.8$ at $c=0.08$ (Fig.~\ref{fig:structure}\textit{D}): number fluctuations are enhanced at the lowest concentration and strongly suppressed at high concentration. The neural functional reproduces this concentration dependence, including the crossover, at both permittivities. The pair closure places $\Gamma(k_1)$ at $c=0.01$ on the wrong side of the ideal value and misses most of the concentration dependence. 

To identify which correlations contribute most of the concentration dependence of $\Gamma(k_1)$, including the crossover at $\varepsilon_r=2$, we return to Eq.~\ref{eq:OZ}. At small wavenumbers, its Coulomb term strongly suppresses charge fluctuations, so the number channel nearly decouples and
\begin{equation}
\Gamma(k_1)\simeq1+\frac{\nb}{2k_BT}\,\mathbf q^{\T}W(k_1)\mathbf q,
\label{eq:Gamma}
\end{equation}
where $\mathbf q^{\T}W\mathbf q=W_{++}+W_{--}+2W_{+-}$ is the excess kernel in the number channel. Number fluctuations are therefore enhanced when $\mathbf q^{\T}W\mathbf q<0$, and suppressed when $\mathbf q^{\T}W\mathbf q>0$. A fixed quadratic pair closure has a density-independent $W(k)$, so $\Gamma(k_1)-1$ keeps its sign at all concentrations. The neural term instead allows $W(k)$ to vary with the surrounding ion densities. The number-channel kernel inferred from zero-field MD grows with concentration at both permittivities and changes sign at $\varepsilon_r=2$, from about $-140\,k_BT\sigma^3$ at $c=0.01$ through zero at $c=0.02$ to about $320\,k_BT\sigma^3$ at $c=0.08$, whereas the pair closure keeps one value. The neural functional reproduces this concentration dependence (Fig.~\ref{fig:structure}\textit{E} and \textit{F}).

\begin{figure}[!htb]
\centering
\includegraphics[width=0.95\textwidth,height=0.45\textheight,keepaspectratio]{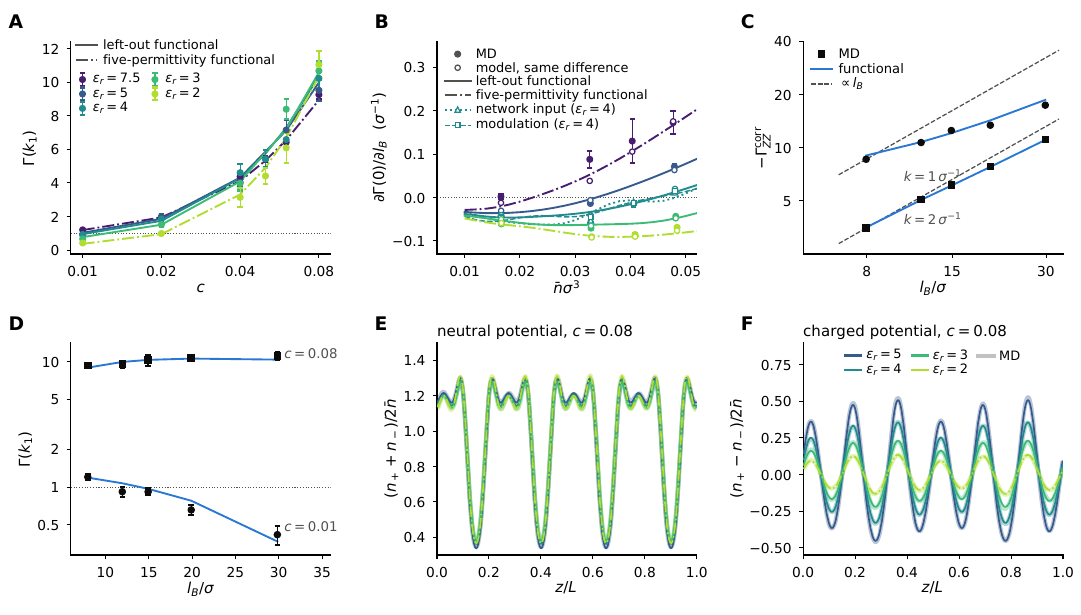}
\caption{Transfer across the Coulomb coupling. (\textit{A}) Response factor $\Gamma(k_1)$ versus concentration at the five permittivities. Points, zero-field MD; curves at $\varepsilon_r=5$, 4 and 3, functionals trained with that permittivity left out; curves at $\varepsilon_r=7.5$ and 2 (solid and dash-dotted), the functional trained on all five; dotted line, $\Gamma=1$. (\textit{B}) Energy identity, Eq.~\ref{eq:HFbulk}: $\partial\Gamma(0)/\partial l_B$ versus salt density. Filled points, MD (second differences of the Coulomb energy on the concentration grid); open points, the functionals differentiated alike; curves, the coupling law, line styles as in \textit{A}; dotted curves at $\varepsilon_r=4$, $l_B$ entering the network as an input (triangles) or by modulating the hidden layers (squares). (\textit{C}) Charge channel beyond the mean field, $-\Gamma_{ZZ}^{\rm corr}$, at $k=2\sigma^{-1}$ and $1\sigma^{-1}$ versus $l_B$ at $c=0.08$ (logarithmic axes); dashed lines are proportional to $l_B$. (\textit{D}) $\Gamma(k_1)$ at $c=0.08$ and $0.01$ versus $l_B$. In \textit{C} and \textit{D}: points, MD; curves, the five-permittivity functional. (\textit{E}, \textit{F}) Number response to a neutral potential and charge response to a charged potential at $c=0.08$ for $\varepsilon_r=5$, 4, 3 and 2: MD (wide light curves) and the functionals (thin curves, line styles as in \textit{A}).}
\label{fig:coupling}
\end{figure}

\paragraph{Transfer across the Coulomb coupling.}
The functional of Eq.~\ref{eq:lb} is trained on five permittivities. To test whether it carries the dependence on the coupling, we train functionals with one of the interior permittivities ($\varepsilon_r=5$, 4 or 3) left out and predict the runs at that permittivity (Fig.~\ref{fig:coupling}). The density profiles there are predicted with mean relative $L_2$ errors of 1.2--1.5\%, as accurate as held-out runs at the trained permittivities. Without the coupling law one functional cannot describe four permittivities at once: trained on the same runs with an $l_B$-independent $\Fex$, its profile errors rise to 7--23\% depending on the permittivity and a fifth of its Euler--Lagrange solutions do not converge (\siref, section S5.1). The response factor $\Gamma(k_1)$ at the 16 unseen state points agrees with MD within the MD uncertainty (Fig.~\ref{fig:coupling}\textit{A}). It rises with concentration at every permittivity and depends on the coupling most at the lowest concentration, where it falls from $1.20\pm0.06$ at $\varepsilon_r=7.5$ to $0.42\pm0.07$ at $\varepsilon_r=2$. At the lowest concentration the number fluctuations thus turn from suppressed to enhanced as the coupling grows, with the crossover between $\varepsilon_r=4$ and 3, and the functionals trained without a permittivity reproduce this trend.

The energy identity of Eq.~\ref{eq:HFbulk} tests the coupling dependence of the functional against the Coulomb energies of the zero-field simulations, which enter neither training nor the tests above (Fig.~\ref{fig:coupling}\textit{B}). In MD, $\partial\Gamma(0)/\partial l_B$ is negative at low density, where a stronger coupling promotes association and weakens the suppression of number fluctuations, and positive at high density, where Coulomb ordering suppresses them. The coupling law reproduces this behavior, including the sign change, at the trained permittivities and at the one left out, with deviations comparable to the MD uncertainty (\siref, section S5). Two alternative forms in which $l_B$ enters the network directly, as an additional input of $\Phi_\theta$ or by modulating its hidden layers, reproduce it as well: at the precision of the MD energies, the identity does not discriminate between the forms. We retain the coupling law for its structure. Its linear and saturating terms correspond to the two behaviors of the charge channel identified below, and its derivative with respect to $l_B$ follows from three terms rather than from a network's interpolation between the five trained values.

To locate where the coupling acts, we separate the bulk kernel into the number channel, $\Gamma(k)-1$, and the charge channel beyond the mean field, $\Gamma_{ZZ}^{\rm corr}(k)=2\nb/S_{ZZ}(k)-1-\kappa_D^2/k^2$ (Fig.~\ref{fig:coupling}\textit{C} and \textit{D}). The charge channel carries nearly the entire dependence on the coupling, with the two behaviors of the coupling law on two scales. At $k=2\sigma^{-1}$, the scale of ion contact, $\Gamma_{ZZ}^{\rm corr}$ grows in proportion to $l_B$, within 15\% over its 3.7-fold range: the correlation between neighboring opposite charges is set by their Coulomb energy, as the mean spherical approximation predicts \cite{waisman1972mean}. At $k=1\sigma^{-1}$, the scale of an ion pair, it only doubles: once the ions are associated, a stronger coupling adds little correlation, the saturation that $s(\lambda)F_2$ represents. The number channel, in contrast, hardly depends on the coupling in the concentrated melt. At $c=0.08$, $\Gamma(k_1)$ varies by less than 20\% across the five permittivities, so the neutral response is set by the packing and solvation of the host. At low concentration the coupling lowers $\Gamma(k_1)$ through association, since an ideal mixture of free ions and pairs has $\Gamma=1/(1+\alpha)$ with paired fraction $\alpha$. The pairing is weaker than for bare ions: Bjerrum theory with a contact distance of $\sigma$ \cite{bjerrum1926} predicts $\alpha=0.54$ at $\varepsilon_r=7.5$ and $c=0.01$, where MD gives a suppressed response, $\Gamma(k_1)=1.20\pm0.06$, because the solvation shells of the host raise the free energy of a contact pair. The functional learns this host-mediated interaction. The same separation holds for the nonlinear response to the applied potentials (Fig.~\ref{fig:coupling}\textit{E} and \textit{F}): the number response to a neutral potential is independent of the permittivity within the MD noise at $c\geq0.06$, whereas the charge response to a charged potential shrinks with the coupling, and the functionals trained without a permittivity predict both within the MD noise. At $c=0.01$ and 0.02 they underestimate the change of the number response with coupling by about 40\%, the least accurate state of the functional (\siref, section S5).
\subsection{Comparison of learned architectures}
We compare our functional with three published architectures: a one-body network that predicts excess chemical potentials directly from local density windows \cite{sammuller2023neural,bui2025learning}, a convolutional free energy that maps the whole planar profile to a free energy \cite{dijkman2025learning}, and a lattice free energy built from Cartesian moments of local density environments \cite{cheng2026equivariant}. These forms differ in the structural properties guaranteed by the published architectures (Table~\ref{tab:architectures}). To test whether the differences affect predictive performance, we adapt each architecture to the two ion species and train it with the force-matching loss, data and splits of our functional. The comparison therefore tests the architectures under one training protocol, not the published methods with their own data and training targets (\methodsref\ and \siref, section S4.2). It is made at fixed coupling, where Eq.~\ref{eq:lb} reduces to Eq.~\ref{eq:Fex}: each form is trained separately at $\varepsilon_r=7.5$ and at $\varepsilon_r=2$ on the runs of that permittivity. We test on the present ionic system, in which every form receives the same analytic Coulomb term as our functional, and, as a control, on the truncated Lennard-Jones fluid of ref.~\cite{cheng2026equivariant}, a one-component uncharged fluid of the kind for which these forms were developed. The forms are trained on the full training set and on about a quarter of it (\siref, section S4.3).

On the ions, our functional has the lowest held-out errors with the full training set, followed by the one-body network (\siref, section S4.3). The difference grows when the training set is reduced to a quarter: our errors grow by about a quarter, those of the one-body network by about two thirds, and some of its Euler--Lagrange solutions fail (\siref, section S4.3). With a quarter of the runs, our functional is thus about as accurate as the one-body network trained on all of them. The lattice free energy is less accurate, and its Euler--Lagrange iteration fails for a tenth of the test runs at $\varepsilon_r=7.5$ and a quarter at $\varepsilon_r=2$, either by not converging or by collapsing onto a narrow density spike. The Lennard-Jones control gives the same ordering, with our error less than half that of the best alternative with the full training set and with a quarter of it. The convolutional free energy, trained here by force matching on inhomogeneous profiles rather than by the pair-correlation matching of the original work, rarely yields a converged Euler--Lagrange solution in either system (\siref, sections S4.3 and S4.4), so we do not rank it. This is what one expects if built-in integrability and symmetry reduce what must be learned: the forms that lack them need more data to reach the same accuracy.

The structural checks on the MD profiles show the differences of Table~\ref{tab:architectures} in the adapted implementations (\siref, section S4.3). The one-body network has net-force errors of up to 14\% and Jacobian asymmetries of up to 55\%. The two free-energy baselines have symmetric Hessians, but their net-force errors reach 6\%, since the Noether identity is not guaranteed without invariance under continuous translations. Our functional satisfies all three identities to numerical precision (\siref, section S4.3).

\section{Discussion}
We construct a neural density functional for the ions of a salt-doped polymer melt with built-in integrability, continuous spatial symmetry, the Noether identities, explicit long-range electrostatics, and parameters that refer to no grid. It carries the Coulomb coupling as an explicit argument under an exact energy identity. Previous neural functionals each provide only some of these properties (Table~\ref{tab:architectures}). Among the architectures compared, ours is the most accurate and the least sensitive to the size of the training set. Trained only on planar force densities, the same functional predicts ionic structure in larger boxes, in fields that vary in two directions, at an untrained salt concentration, and at a permittivity left out of training. It also recovers the bulk structure factors and the long-wavelength number response, neither of which enters the training. Its nonlinear density dependence captures the concentration dependence of the effective ion correlations, including the sign change of the number-channel kernel at the strongest coupling, from enhanced to suppressed number fluctuations, which a density-independent pair kernel cannot represent. Across the coupling, the functional locates this dependence: linear in $l_B$ in the short-range charge correlations, as in the mean spherical approximation, sublinear at the pair scale, and nearly absent from the number response of the concentrated melt, which the host sets. The transfer is across the coupling at fixed host. Extending it across polymers, whose solvation differs, will require relearning the coupling-independent part of the functional and is a direction for future work.

A natural next step is to apply the functional to transport. Although the functional describes equilibrium, it supplies the chemical potentials whose gradients drive ion fluxes in stochastic and dynamical density functional theories \cite{dean1996langevin,avni2022conductivity}. Combined with a validated mobility model, it could predict conductivity and salt diffusion that include the ion correlations learned here. Such mobilities can themselves be learned from molecular dynamics \cite{chen2026learning}, and the equilibrium functional remains applicable as long as the unresolved degrees of freedom stay close to their equilibrium distribution \cite{lyu2025generalization}. The ion-only functional describes a homogeneous polymer host. At electrode interfaces and under confinement, where the polymer density itself is strongly inhomogeneous, an explicit polymer density field is needed (\siref, section S6). With it, the three-dimensional, grid-independent form of the functional applies once the boundary conditions are specified.
\FloatBarrier
\section*{Acknowledgments}
\PaperAcknowledgments

\section*{Data Availability}
\PaperDataAvailability

\clearpage
\begingroup
\small
\setcounter{equation}{0}
\setcounter{figure}{0}
\setcounter{table}{0}
\renewcommand{\theequation}{S\arabic{equation}}
\renewcommand{\thefigure}{S\arabic{figure}}
\renewcommand{\thetable}{S\arabic{table}}
\renewcommand{\theHequation}{S.\arabic{equation}}
\renewcommand{\theHfigure}{S.\arabic{figure}}
\renewcommand{\theHtable}{S.\arabic{table}}
\section*{Supplementary Information}
\begin{center}\large\bfseries\PaperTitle\end{center}

\section*{S1. Properties that hold by construction}

\subsection*{S1.1 Functional and notation}

The ions are described by their number densities $n_\alpha(\vr)$, $\alpha\in\{+,-\}$, collected in $\vn=(n_+,n_-)$, with valences $z_\pm=\pm1$. Energies are in units of $k_BT$ and charges in units of $e$. The free energy of the main text is
\begin{equation}
F[\vn]=\Fid[\vn]+\FC[\vn]+F^\theta_{\rm ex}[\vn],\qquad \Fid=\sum_\alpha\int \rmd^3r\,n_\alpha\big[\ln(n_\alpha\Lambda^3)-1\big],\qquad \FC=\frac{l_B}{2}\iint \rmd^3r\,\rmd^3r'\,\frac{\rho_Z(\vr)\rho_Z(\vr')}{|\vr-\vr'|},
\end{equation}
with $\rho_Z=\sum_\alpha z_\alpha n_\alpha$ the charge density in units of $e$, and the neural functional is
\begin{equation}
F^\theta_{\rm ex}[\vn]=\frac12\sum_{\alpha\beta}\iint \rmd^3r\,\rmd^3r'\,n_\alpha(\vr)\,u_{\alpha\beta}(|\vr-\vr'|)\,n_\beta(\vr')+\int \rmd^3r\,\Phi_\theta\big(\mathbf h(\vr)\big),\qquad h_{m\alpha}=K_m\star n_\alpha,
\end{equation}
with symmetric pair kernels $u_{\alpha\beta}=u_{\beta\alpha}=\sum_m a_{\alpha\beta m}K_m$. The radial kernels have the form $K_m(r)=B_m(r)[1+g_m(r)]$, where $B_m(r)=(2\pi s_m^2)^{-3/2}\exp[-r^2/(2s_m^2)]$ are normalized Gaussians with six learned widths $s_m$ (section S2.2), and $g_m$ is a smooth learned function of $r/\sigma$. For bounded $g_m$, the Gaussian envelope ensures that $K_m(r)\to0$ as $r\to\infty$. The same decay holds for the at-most-linear growth of the softplus networks used here. The kernels $K_m$ and $u_{\alpha\beta}$ are therefore integrable with finite radial moments. Their Fourier transforms are real, even, smooth functions of $k$, with $\tilde K_m(k)=\tilde K_m(0)+O(k^2)$ and $\tilde u_{\alpha\beta}(k)=\tilde u_{\alpha\beta}(0)+O(k^2)$.

For numerical evaluation, the radial Fourier integral is truncated at $5\sigma$ and computed by midpoint quadrature with spacing $0.01\sigma$:
\[
\tilde K_m(k)=4\pi\int_0^\infty r^2K_m(r)\frac{\sin(kr)}{kr}\,\rmd r
\simeq4\pi\int_0^{5\sigma}r^2K_m(r)\frac{\sin(kr)}{kr}\,\rmd r.
\]
In planar geometry, the weighted densities are evaluated as products in Fourier space, $\tilde h_{m\alpha}(k)=\tilde K_m(k)\tilde n_\alpha(k)$. The pointwise map $\Phi_\theta$ is a multilayer perceptron with smooth softplus activations and a linear output, so it is twice continuously differentiable and has no explicit dependence on position or direction. We write $\mu^{\rm int}_\alpha=\delta(\FC+F^\theta_{\rm ex})/\delta n_\alpha$ for the interaction chemical potential and $\vf^{\rm int}_\alpha=-n_\alpha\nabla\mu^{\rm int}_\alpha$ for the internal force density. 

With $\Phi_{m\alpha}=\partial\Phi_\theta/\partial h_{m\alpha}$ and $\Phi_{m\alpha,m'\beta}=\partial^2\Phi_\theta/\partial h_{m\alpha}\partial h_{m'\beta}$ evaluated at $\mathbf h(\vr)$, and since a radial kernel is its own adjoint under convolution, the first and second derivatives of the many-body term are
\begin{equation}
\frac{\delta}{\delta n_\alpha(\vr)}\int\Phi_\theta(\mathbf h)=\sum_m\big(K_m\star\Phi_{m\alpha}\big)(\vr),\qquad \frac{\delta^2}{\delta n_\alpha(\vr)\,\delta n_\beta(\vr')}\int\Phi_\theta(\mathbf h)=\sum_{mm'}\int \rmd^3r''\,K_m(|\vr-\vr''|)\,\Phi_{m\alpha,m'\beta}(\vr'')\,K_{m'}(|\vr''-\vr'|).
\label{eq:S:derivs}
\end{equation}
At uniform density the Hessian $H_{m\alpha,m'\beta}=\Phi_{m\alpha,m'\beta}$ is a constant symmetric matrix, and the bulk kernel of the excess functional is
\begin{equation}
W_{\alpha\beta}(k)=\tilde u_{\alpha\beta}(k)+\sum_{mm'}\tilde K_m(k)\,H_{m\alpha,m'\beta}\,\tilde K_{m'}(k).
\label{eq:S:W}
\end{equation}

\subsection*{S1.2 Covariance and integrability}
The functional derivatives inherit the symmetries of the scalar functional and satisfy the integrability condition of a gradient.

\textbf{Proposition 1 (translation and rotation covariance).} Let $\tau=(R,\mathbf a)$ with $R\in O(3)$ act on points by $\tau\vr=R\vr+\mathbf a$ and on densities by $(T_\tau\vn)_\alpha(\vr)=n_\alpha(\tau^{-1}\vr)$. For every such transformation compatible with the domain and every density field specified above:
\begin{enumerate}
    \item[i] $\FC[T_\tau\vn]+F^\theta_{\rm ex}[T_\tau\vn]=\FC[\vn]+F^\theta_{\rm ex}[\vn]$,
    \item[ii] $\mu^{\rm int}_\alpha[T_\tau\vn](\tau\vr)=\mu^{\rm int}_\alpha[\vn](\vr)$,
    \item[iii]  $\vf^{\rm int}_\alpha[T_\tau\vn](\tau\vr)=R\,\vf^{\rm int}_\alpha[\vn](\vr)$.
\end{enumerate}

\begin{proof}
The weighted densities are scalar fields: substituting $\vr = \tau{\mathbf x}, \vr'=\tau\mathbf s$ in 
\[
\begin{aligned}
    h_{m\alpha}[T_\tau\vn](\vr)= &
\int K_m(|\vr-\vr'|)\,(T_\tau\vn)_\alpha(\vr')\,\rmd^3r' = 
\int K_m(|\vr-\vr'|)\,n_\alpha(\tau^{-1}\vr')\,\rmd^3r' \\
= & \int K_m(|\tau\mathbf x-\tau\mathbf s|)\,n_\alpha(\mathbf s)\,\rmd^3\mathbf s  =  \int K_m(|\mathbf x-\mathbf s|)\,n_\alpha(\mathbf s)\,\rmd^3\mathbf s = h_{m\alpha}[\vn](\mathbf x) = h_{m\alpha}[\vn](\tau^{-1}\vr),
\end{aligned}
\]  
where $\rmd^3r'=|\det R|\,\rmd^3s=\rmd^3s$.
Since $\Phi_\theta$ has no explicit dependence on position,
the covariance of the weighted densities and the substitution
$\vr=\tau\mathbf s$ give
\[
\begin{aligned}
\int \rmd^3r\,
\Phi_\theta\big(\mathbf h[T_\tau\vn](\vr)\big)
&=
\int \rmd^3r\,
\Phi_\theta\big(\mathbf h[\vn](\tau^{-1}\vr)\big)\\
&=
\int \rmd^3s\,
\Phi_\theta\big(\mathbf h[\vn](\mathbf s)\big).
\end{aligned}
\]
Thus the many-body term is invariant.
The pair term and $\FC$ are double integrals of radial kernels and are invariant by the same two substitutions. This is (i).

Applying (i) to $\vn+\epsilon\boldsymbol\eta$ and using the
linearity of $T_\tau$ gives
\[
\begin{aligned}
&\FC[T_\tau\vn+\epsilon T_\tau\boldsymbol\eta]
+F^\theta_{\rm ex}[T_\tau\vn+\epsilon T_\tau\boldsymbol\eta]\\
&\qquad=
\FC[\vn+\epsilon\boldsymbol\eta]
+F^\theta_{\rm ex}[\vn+\epsilon\boldsymbol\eta].
\end{aligned}
\]
Differentiating with respect to $\epsilon$ at $\epsilon=0$ yields
\[
\sum_\alpha\int \rmd^3r\,
\mu^{\rm int}_\alpha[T_\tau\vn](\vr)\,
\eta_\alpha(\tau^{-1}\vr)
=
\sum_\alpha\int \rmd^3s\,
\mu^{\rm int}_\alpha[\vn](\mathbf s)\,
\eta_\alpha(\mathbf s).
\]
Substituting $\vr=\tau\mathbf s$ on the left-hand side and
rearranging, we obtain
\[
\sum_\alpha\int \rmd^3s\,
\Big[
\mu^{\rm int}_\alpha[T_\tau\vn](\tau\mathbf s)
-\mu^{\rm int}_\alpha[\vn](\mathbf s)
\Big]\eta_\alpha(\mathbf s)=0.
\]
Since this holds for every perturbation $\boldsymbol\eta$,
\[
\mu^{\rm int}_\alpha[T_\tau\vn](\tau\vr)
=
\mu^{\rm int}_\alpha[\vn](\vr),
\]
which proves (ii).

By (ii) and the chain rule, $\nabla\mu^{\rm int}_\alpha[T_\tau\vn](\vr)=R\,(\nabla\mu^{\rm int}_\alpha[\vn])(\tau^{-1}\vr)$. Multiplying by $-(T_\tau\vn)_\alpha(\vr)$ gives (iii).
\end{proof}
The proof includes reflections. No equivariant architecture is needed, because the covariance of the force follows from the invariance of one scalar.

\textbf{Proposition 2 (integrability).} For every density field specified above,
\begin{equation}
\frac{\delta\mu^{\rm int}_\alpha(\vr)}{\delta n_\beta(\vr')}=\frac{\delta\mu^{\rm int}_\beta(\vr')}{\delta n_\alpha(\vr)},
\end{equation}
so the direct correlation functions satisfy $c_{\alpha\beta}(\vr,\vr')=c_{\beta\alpha}(\vr',\vr)$, and in the bulk $W_{\alpha\beta}(k)=W_{\beta\alpha}(k)$ is real and depends on $|\mathbf k|$ only.

\begin{proof}
    Since $\Phi_\theta$ is twice continuously differentiable,
\[
\Phi_{m\alpha,m'\beta}(\vr'')
=
\Phi_{m'\beta,m\alpha}(\vr'').
\]
Using Eq.~\ref{eq:S:derivs}, the symmetry of the many-body
Hessian follows explicitly:
\[
\begin{aligned}
&\frac{\delta^2}{\delta n_\alpha(\vr)\,\delta n_\beta(\vr')}
\int \rmd^3s\,\Phi_\theta\big(\mathbf h(\mathbf s)\big)\\
&\quad=
\sum_{mm'}\int \rmd^3r''\,
K_m(|\vr-\vr''|)\,
\Phi_{m\alpha,m'\beta}(\vr'')\,
K_{m'}(|\vr''-\vr'|)\\
&\quad=
\sum_{mm'}\int \rmd^3r''\,
K_{m'}(|\vr'-\vr''|)\,
\Phi_{m'\beta,m\alpha}(\vr'')\,
K_m(|\vr''-\vr|)\\
&\quad=
\frac{\delta^2}{\delta n_\beta(\vr')\,\delta n_\alpha(\vr)}
\int \rmd^3s\,\Phi_\theta\big(\mathbf h(\mathbf s)\big),
\end{aligned}
\]
where the last equality follows by relabelling $m\leftrightarrow m'$.

The pair and Coulomb contributions to the interaction Hessian
are also symmetric:
\[
u_{\alpha\beta}(|\vr-\vr'|)
=
u_{\beta\alpha}(|\vr'-\vr|),
\qquad
\frac{l_Bz_\alpha z_\beta}{|\vr-\vr'|}
=
\frac{l_Bz_\beta z_\alpha}{|\vr'-\vr|}.
\]
Combining these contributions gives
\[
\frac{\delta\mu^{\rm int}_\alpha(\vr)}
     {\delta n_\beta(\vr')}
=
\frac{\delta\mu^{\rm int}_\beta(\vr')}
     {\delta n_\alpha(\vr)}.
\]

In a uniform state, $H_{m\alpha,m'\beta}=H_{m'\beta,m\alpha}$,
so Eq.~\ref{eq:S:W} gives
\[
\begin{aligned}
W_{\beta\alpha}(k)
&=
\tilde u_{\beta\alpha}(k)
+\sum_{mm'}\tilde K_m(k)\,
H_{m\beta,m'\alpha}\,\tilde K_{m'}(k)\\
&=
\tilde u_{\alpha\beta}(k)
+\sum_{mm'}\tilde K_{m'}(k)\,
H_{m'\beta,m\alpha}\,\tilde K_m(k)\\
&=
\tilde u_{\alpha\beta}(k)
+\sum_{mm'}\tilde K_m(k)\,
H_{m\alpha,m'\beta}\,\tilde K_{m'}(k)\\
&=W_{\alpha\beta}(k).
\end{aligned}
\]
The Fourier transforms of the real radial kernels are real
and depend only on $k=|\mathbf k|$. The same therefore holds for
$W_{\alpha\beta}(k)$. 
\end{proof}

\subsection*{S1.3 Noether sum rules}
We first record the Helmholtz condition, which the interaction chemical potential satisfies by Proposition 2, and then derive the sum rules from the invariance in Proposition 1.

\textbf{Lemma (Helmholtz condition).} A family $\psi_\alpha[\vn](\vr)$ on a convex set of densities is a functional gradient, $\psi_\alpha=\delta F/\delta n_\alpha$, if and only if $\delta\psi_\alpha(\vr)/\delta n_\beta(\vr')=\delta\psi_\beta(\vr')/\delta n_\alpha(\vr)$. In that case, with $\vn_\lambda=\vn_0+\lambda(\vn-\vn_0)$,
\begin{equation}
F[\vn]=F[\vn_0]+\int_0^1\rmd\lambda\sum_\alpha\int\rmd^3r\,\psi_\alpha[\vn_\lambda](\vr)\,\big(n_\alpha-n_{0,\alpha}\big)(\vr),
\label{eq:S:lineint}
\end{equation}
independent of the path.
\begin{proof}
Necessity follows from the symmetry of second derivatives:
if $\psi_\alpha=\delta F/\delta n_\alpha$, then
\[
\frac{\delta\psi_\alpha(\vr)}{\delta n_\beta(\vr')}
=
\frac{\delta^2F}{\delta n_\beta(\vr')\,\delta n_\alpha(\vr)}
=
\frac{\delta^2F}{\delta n_\alpha(\vr)\,\delta n_\beta(\vr')}
=
\frac{\delta\psi_\beta(\vr')}{\delta n_\alpha(\vr)}.
\]

For sufficiency, keep $\vn_0$ fixed and define $F$ by
Eq.~\ref{eq:S:lineint}, with
$\vn_\lambda=\vn_0+\lambda(\vn-\vn_0)$.
Since
\[
\frac{\delta n_{\lambda,\alpha}(\vr)}
     {\delta n_\beta(\vr')}
=
\lambda\,\delta_{\alpha\beta}\delta(\vr-\vr'),
\]
functional differentiation gives
\[
\begin{aligned}
\frac{\delta F}{\delta n_\beta(\vr')}
=
\int_0^1\rmd\lambda\,\Bigg[
&\psi_\beta[\vn_\lambda](\vr')\\
&+\lambda\sum_\alpha\int\rmd^3r\,
\left.
\frac{\delta\psi_\alpha[\vn](\vr)}
     {\delta n_\beta(\vr')}
\right|_{\vn=\vn_\lambda}
\big(n_\alpha-n_{0,\alpha}\big)(\vr)
\Bigg].
\end{aligned}
\]
By the assumed symmetry,
\[
\left.
\frac{\delta\psi_\alpha[\vn](\vr)}
     {\delta n_\beta(\vr')}
\right|_{\vn=\vn_\lambda}
=
\left.
\frac{\delta\psi_\beta[\vn](\vr')}
     {\delta n_\alpha(\vr)}
\right|_{\vn=\vn_\lambda}.
\]
Moreover, the chain rule along $\vn_\lambda$ gives
\[
\frac{\partial}{\partial\lambda}
\psi_\beta[\vn_\lambda](\vr')
=
\sum_\alpha\int\rmd^3r\,
\left.
\frac{\delta\psi_\beta[\vn](\vr')}
     {\delta n_\alpha(\vr)}
\right|_{\vn=\vn_\lambda}
\big(n_\alpha-n_{0,\alpha}\big)(\vr).
\]
Combining these identities, we obtain
\[
\begin{aligned}
\frac{\delta F}{\delta n_\beta(\vr')}
&=
\int_0^1\rmd\lambda\,
\left[
\psi_\beta[\vn_\lambda](\vr')
+\lambda\frac{\partial}{\partial\lambda}
\psi_\beta[\vn_\lambda](\vr')
\right]\\
&=
\int_0^1\rmd\lambda\,
\frac{\partial}{\partial\lambda}
\left[\lambda\,\psi_\beta[\vn_\lambda](\vr')\right]\\
&=
\left[\lambda\,\psi_\beta[\vn_\lambda](\vr')\right]_0^1
=
\psi_\beta[\vn](\vr').
\end{aligned}
\]
Thus $\psi_\alpha=\delta F/\delta n_\alpha$, and its line
integral along any smooth path between two densities equals
the difference in $F$ at the endpoints, establishing path
independence. 
\end{proof}

Eq.~\ref{eq:S:lineint} is the functional line integration used by neural functionals that learn the one-body direct correlation function $c^{(1)}=-\delta\Fex/\delta n$ directly. Such a network has no reason to satisfy the Helmholtz condition exactly, and the value of Eq.~\ref{eq:S:lineint} then depends on the path. The scalar form of Eq.~\ref{eq:Fex} satisfies it identically.

\textbf{Proposition 3 (Noether sum rules).} For every density field specified above, in or out of equilibrium: 
\begin{enumerate}
    \item [i]$\sum_\alpha\int\vf^{\rm int}_\alpha\,\rmd^3r=0$,
    \item [ii]on $\mathbb R^3$, $\sum_\alpha\int\vr\times\vf^{\rm int}_\alpha\,\rmd^3r=0$,
\end{enumerate}

\begin{proof}
We use the Einstein summation convention for Cartesian indices.

(i) For a translation by $\mathbf a$,
\[
(T_{\mathbf a}\vn)_\alpha(\vr)=n_\alpha(\vr-\mathbf a),
\qquad
\left.
\frac{\partial(T_{\mathbf a}\vn)_\alpha(\vr)}
     {\partial a_p}
\right|_{\mathbf a=0}
=-\partial_p n_\alpha(\vr).
\]
Differentiating Proposition 1(i) with respect to $a_p$ gives
\[
\begin{aligned}
0
&=
\left.
\frac{\partial}{\partial a_p}
\big(\FC[T_{\mathbf a}\vn]
+F^\theta_{\rm ex}[T_{\mathbf a}\vn]\big)
\right|_{\mathbf a=0}\\
&=
-\sum_\alpha\int\rmd^3r\,
\mu^{\rm int}_\alpha(\vr)\,\partial_p n_\alpha(\vr)\\
&=
\sum_\alpha\int\rmd^3r\,
n_\alpha(\vr)\,\partial_p\mu^{\rm int}_\alpha(\vr)\\
&=
-\sum_\alpha\int\rmd^3r\,f^{\rm int}_{\alpha,p}(\vr).
\end{aligned}
\]
Since this holds for each Cartesian component,
\[
\sum_\alpha\int\rmd^3r\,\vf^{\rm int}_\alpha(\vr)=0.
\]

(ii) On $\mathbb R^3$, let $\tau_\vartheta$ be a rotation by
angle $\vartheta$ about the unit vector
$\hat{\boldsymbol\omega}$. Then
\[
\left.
\frac{\partial(T_{\tau_\vartheta}\vn)_\alpha(\vr)}
     {\partial\vartheta}
\right|_{\vartheta=0}
=
-(\hat{\boldsymbol\omega}\times\vr)\cdot
\nabla n_\alpha(\vr).
\]
Using rotational invariance and
$\nabla\cdot(\hat{\boldsymbol\omega}\times\vr)=0$,
integration by parts gives
\[
\begin{aligned}
0
&=
\left.
\frac{\partial}{\partial\vartheta}
\big(\FC[T_{\tau_\vartheta}\vn]
+F^\theta_{\rm ex}[T_{\tau_\vartheta}\vn]\big)
\right|_{\vartheta=0}\\
&=
-\sum_\alpha\int\rmd^3r\,
\mu^{\rm int}_\alpha
(\hat{\boldsymbol\omega}\times\vr)\cdot\nabla n_\alpha\\
&=
\sum_\alpha\int\rmd^3r\,
n_\alpha(\hat{\boldsymbol\omega}\times\vr)
\cdot\nabla\mu^{\rm int}_\alpha\\
&=
-\hat{\boldsymbol\omega}\cdot
\sum_\alpha\int\rmd^3r\,
\vr\times\vf^{\rm int}_\alpha.
\end{aligned}
\]
Since $\hat{\boldsymbol\omega}$ is arbitrary,
\[
\sum_\alpha\int\rmd^3r\,
\vr\times\vf^{\rm int}_\alpha=0.
\]
\end{proof}

\subsection*{S1.4 Screening sum rules}

The screening sum rules follow from the bulk linear response of the same functional. For a uniform stationary state with densities $\nb_\alpha$, write $\bar N={\rm diag}(\nb_+,\nb_-)$ and $\vz=(1,-1)^{\T}$. Linearizing the Euler--Lagrange equation $\ln n_\alpha+\mu^{\rm int}_\alpha+V_\alpha={\rm const}_\alpha$ and Fourier transforming gives $[\bar N^{-1}+(4\pi l_B/k^2)\vz\vz^{\T}+W(k)]\delta\tilde{\vn}(\mathbf k)=-\delta\tilde{\mathbf V}(\mathbf k)$ for $k\neq0$. Thus the response matrix defined by $\delta\tilde{\vn}=-S(k)\delta\tilde{\mathbf V}$ satisfies
\begin{equation}
S(k)^{-1}=\bar N^{-1}+\frac{4\pi l_B}{k^2}\,\vz\vz^{\T}+W(k).
\label{eq:S:OZ}
\end{equation}
This is the Ornstein--Zernike relation $S^{-1}=\bar N^{-1}-c$, with $c=-(4\pi l_B/k^2)\vz\vz^{\T}-W$. For a stable uniform state, the fluctuation-response relation identifies $S(k)$ as the structure factor, $S_{\alpha\beta}(k)=V^{-1}\langle\delta\tilde n_\alpha(\mathbf k)\,\delta\tilde n_\beta(-\mathbf k)\rangle$. The explicit Coulomb term and the regular small-$k$ behavior of $W(k)$ yield the following screening identities.

\textbf{Proposition 4 (perfect-screening sum rules).} Let $A(k)=\bar N^{-1}+W(k)$ be the short-range part of $S(k)^{-1}$ in Eq.~\ref{eq:S:OZ}. If $A(0)$ is invertible and $s_0=\vz^{\T}A(0)^{-1}\vz\neq0$, then for every learned $W$
\begin{equation}
\frac{S_{ZZ}(k)}{2\nb}=\frac{k^2}{\kappa_D^2}+O(k^4),\qquad \mathbf a^{\T}S(k)\,\vz=O(k^2)\ \text{for every fixed }\mathbf a,
\end{equation}
where $S_{ZZ}=\vz^{\T}S\,\vz$ is the charge structure factor, $\nb_+=\nb_-=\nb$ and $\kappa_D^2=4\pi l_B\sum_\alpha\nb_\alpha z_\alpha^2=8\pi l_B\nb$.

\begin{proof}
By the small-$k$ expansions of the radial kernels and
Eq.~\ref{eq:S:W},
\[
W(k)=W(0)+O(k^2),
\qquad
A(k)=A(0)+O(k^2).
\]
Since $A(0)$ is invertible, $A(k)$ remains invertible for
sufficiently small $k$, with
\[
A(k)^{-1}=A(0)^{-1}+O(k^2).
\]
Consequently,
\[
s(k)\equiv\vz^{\T}A(k)^{-1}\vz
=s_0+O(k^2),
\qquad
s_0=\vz^{\T}A(0)^{-1}\vz\neq0.
\]

Equation~\ref{eq:S:OZ} can be written as
\[
S(k)^{-1}
=
A(k)+\frac{4\pi l_B}{k^2}\vz\vz^{\T}.
\]
Applying the Sherman--Morrison formula gives
\[
\begin{aligned}
S(k)
&=
A(k)^{-1}
-
\frac{
(4\pi l_B/k^2)\,
A(k)^{-1}\vz\vz^{\T}A(k)^{-1}
}{
1+(4\pi l_B/k^2)\,s(k)
}\\
&=
A(k)^{-1}
-
\frac{
A(k)^{-1}\vz\vz^{\T}A(k)^{-1}
}{
s(k)+k^2/(4\pi l_B)
}.
\end{aligned}
\]
The denominator is nonzero for sufficiently small $k$
because it tends to $s_0\neq0$.

Contracting with $\vz$ on both sides yields
\[
\begin{aligned}
S_{ZZ}(k)
&=\vz^{\T}S(k)\vz\\
&=
s(k)-\frac{s(k)^2}{s(k)+k^2/(4\pi l_B)}\\
&=
\frac{k^2}{4\pi l_B}\,
\frac{s(k)}{s(k)+k^2/(4\pi l_B)}\\
&=
\frac{k^2}{4\pi l_B}
\left[1+\frac{k^2}{4\pi l_Bs(k)}\right]^{-1}.
\end{aligned}
\]
Since $s(k)=s_0+O(k^2)$ and $s_0\neq0$,
\[
\frac{k^2}{4\pi l_Bs(k)}
=
\frac{k^2}{4\pi l_Bs_0}+O(k^4).
\]
Expanding the reciprocal therefore gives
\[
\begin{aligned}
S_{ZZ}(k)
&=
\frac{k^2}{4\pi l_B}
\left[
1-\frac{k^2}{4\pi l_Bs_0}+O(k^4)
\right]\\
&=
\frac{k^2}{4\pi l_B}
-\frac{k^4}{(4\pi l_B)^2s_0}
+O(k^6).
\end{aligned}
\]
Using $\kappa_D^2=8\pi l_B\nb$, we obtain
\[
\frac{S_{ZZ}(k)}{2\nb}
=
\frac{k^2}{\kappa_D^2}+O(k^4).
\]

For any fixed vector $\mathbf a$, the same inverse formula
gives
\[
\begin{aligned}
\mathbf a^{\T}S(k)\vz
&=
\mathbf a^{\T}A(k)^{-1}\vz
-
\frac{
\big[\mathbf a^{\T}A(k)^{-1}\vz\big]s(k)
}{
s(k)+k^2/(4\pi l_B)
}\\
&=
\frac{k^2}{4\pi l_B}\,
\frac{
\mathbf a^{\T}A(k)^{-1}\vz
}{
s(k)+k^2/(4\pi l_B)
}.
\end{aligned}
\]
Finally,
\[
\mathbf a^{\T}A(k)^{-1}\vz
=
\mathbf a^{\T}A(0)^{-1}\vz+O(k^2),
\qquad
\frac{1}{s(k)+k^2/(4\pi l_B)}
=
\frac{1}{s_0}+O(k^2),
\]
so
\[
\mathbf a^{\T}S(k)\vz
=
\frac{k^2}{4\pi l_Bs_0}\,
\mathbf a^{\T}A(0)^{-1}\vz
+O(k^4)
=
O(k^2).
\]
This proves both screening identities. 
\end{proof}

The leading $k^2$ coefficient is the Stillinger--Lovett second-moment condition \cite{SI-stillinger1968general}. It is fixed by the explicit Coulomb term, while short-range correlations enter at order $k^4$. 

\subsection*{S1.5 Coupling-constant identity}
Let $\mathbf R$ denote the coordinates of all ions and monomers and write the interaction energy, in units of $k_BT$, as $U(\mathbf R;l_B)=U_0(\mathbf R)+l_B\,U_C(\mathbf R)$, where $U_0$ contains the bonded, excluded-volume and solvation interactions and $U_C=\sum_{i<j}z_iz_j/|\vr_i-\vr_j|$ runs over ion pairs, with periodic images included by Ewald summation. For a configuration $\mathbf R$, define the microscopic number density of species $\alpha$ by $\hat n_\alpha(\vr;\mathbf R)=\sum_{i\in\alpha}\delta(\vr-\vr_i)$. A normalized distribution $P(\mathbf R)$ generates the density profile
\begin{equation}
n_\alpha(\vr)=\int\rmd\mathbf R\,P(\mathbf R)\,\hat n_\alpha(\vr;\mathbf R),\qquad \int\rmd\mathbf R\,P(\mathbf R)=1.
\label{eq:S:densityconstraint}
\end{equation}
The intrinsic free energy of the ions at coupling $l_B$ is the constrained minimum \cite{SI-evans1979nature}
\begin{equation}
F[\vn;l_B]=\min_{P\to\vn}\;k_BT\int\rmd\mathbf R\,P(\mathbf R)\big[U(\mathbf R;l_B)+\ln P(\mathbf R)\big]
\label{eq:S:cs}
\end{equation}
over normalized distributions $P$ of all coordinates satisfying Eq.~\ref{eq:S:densityconstraint}. The minimizer $P_{\vn,l_B}$ is the equilibrium distribution that produces the profile $\vn$. Because $U$ is linear in $l_B$ and the constraint does not involve $l_B$, the derivative of the minimum with respect to $l_B$ equals the partial derivative of the objective at the minimizer, the variation of $P_{\vn,l_B}$ contributing nothing to first order:
\begin{equation}
\frac{\partial F[\vn;l_B]}{\partial l_B}=k_BT\int\rmd\mathbf R\,P_{\vn,l_B}(\mathbf R)\,U_C(\mathbf R)=k_BT\,\langle U_C\rangle_{\vn,l_B}.
\label{eq:S:HF}
\end{equation}
This is Eq.~\ref{eq:HF} of the main text, the classical counterpart of the adiabatic connection of electronic density functional theory \cite{SI-langreth1975exchange,SI-gunnarsson1976exchange}. Writing $F=\Fid+\FC+\Fex$ with $\FC=\tfrac12 l_Bk_BT\iint\rho_Z(\vr)\rho_Z(\vr')/|\vr-\vr'|$, the mean-field term gives $\partial\FC/\partial l_B=k_BT\,U_C^{\rm mf}[\vn]$ with $U_C^{\rm mf}=\tfrac12\iint\rho_Z\rho_Z/|\vr-\vr'|$, and $\Fid$ does not depend on $l_B$. Hence
\begin{equation}
\frac{\partial\Fex[\vn;l_B]}{\partial l_B}=k_BT\big(\langle U_C\rangle_{\vn,l_B}-U_C^{\rm mf}[\vn]\big)=\frac{E_C^{\rm corr}[\vn;l_B]}{l_B},
\label{eq:S:HFex}
\end{equation}
where $E_C^{\rm corr}=k_BT\,l_B(\langle U_C\rangle-U_C^{\rm mf})$ is the Coulomb correlation energy, the part of the electrostatic energy beyond the mean field.

In a uniform electroneutral bulk at salt density $\nb$ the mean-field energy vanishes, because the $k=0$ term of the Ewald sum is excluded. With $f_{\rm ex}(\nb;l_B)=\Fex/V$ and $u_C(\nb;l_B)=E_C^{\rm corr}/V$,
\[
\frac{\partial f_{\rm ex}}{\partial l_B}=\frac{u_C(\nb;l_B)}{l_B}.
\]
At $k\to0$ the number--charge coupling term of Eq.~\ref{eq:S:gamma} vanishes and $\sum_{\alpha\beta}W_{\alpha\beta}(0)=\partial^2f_{\rm ex}/\partial\nb^2$ along $\nb_+=\nb_-=\nb$, so that $\Gamma(0)=1+(\nb/2k_BT)\,\partial^2f_{\rm ex}/\partial\nb^2$. Differentiating with respect to $l_B$ gives Eq.~\ref{eq:HFbulk} of the main text,
\[
\frac{\partial\Gamma(0)}{\partial l_B}=\frac{\nb}{2k_BT}\,\frac{\partial^2}{\partial\nb^2}\,\frac{u_C(\nb;l_B)}{l_B}.
\]
Force matching determines $\Fex$ up to terms $a(l_B)+\sum_\alpha b_\alpha(l_B)N_\alpha$, which are linear in $\nb$ and drop out of the second derivative. The identity therefore tests exactly the part of the coupling dependence that the force data do not fix. We test the identity by comparing the functional's derivative with respect to $l_B$ with the Coulomb energies measured in zero-field simulations (section S5).

\section*{S2. Data, force matching and prediction}

\subsection*{S2.1 Model and data}
\textit{Molecular dynamics.} We simulate the coarse-grained salt-doped polymer melt of Tsamopoulos and Wang \cite{SI-tsamopoulos2024ion}: 400 chains of 30 monomers of diameter $\sigma$, joined by FENE bonds \cite{SI-KremerGrest1990} with $K=30\epsilon/\sigma^2$ and $R_0=1.5\sigma$, and 120 to 960 cation--anion pairs for $c=0.01$ to $0.08$ cations per monomer. Cations and anions have diameters $0.4\sigma$ and $1.6\sigma$ and valences $z_\pm=\pm1$. Monomer pairs interact through the Lennard-Jones potential with a $2\sigma$ cutoff, and unlike pairs use the mixing rule $\sigma_{ij}=(\sigma_i+\sigma_j)/2$. Interactions involving ions are purely repulsive apart from electrostatics and the ion--monomer solvation potential of Hall and coworkers \cite{SI-tsamopoulos2024ion,SI-shen2020ion},
\[
U_{\rm solv}(r)=-S\left[\left(\frac{\sigma}{r}\right)^4-\left(\frac{\sigma}{r_c}\right)^4\right],\qquad r\le r_c,
\]
with $S=4.33\epsilon$, $r_c=5\sigma$ and $\epsilon=k_BT$. Electrostatics is computed by PPPM \cite{SI-hockney2021computer} at the dielectric constants $\varepsilon_r=7.5$, 5, 4, 3 and 2, corresponding to Bjerrum lengths $l_B=7.96\sigma$ to $29.8\sigma$. The five dielectric constants use the same interaction parameters apart from electrostatics. All runs use LAMMPS \cite{SI-thompson2022lammps} with a Nos\'e--Hoover thermostat \cite{SI-nose1984unified,SI-hoover1985canonical} at $k_BT=\epsilon$ and a time step of $0.005\tau$, in the Lennard-Jones units $\sigma$, $\epsilon$ and $\tau$ of the monomers. Each concentration is equilibrated at zero pressure and then simulated in the canonical ensemble at its mean volume, with box lengths $L=24.3$ to $25.0\,\sigma$ along $z$. Runs equilibrate for $6\times10^3$ to $4\times10^4\,\tau$ and sample for $3.7\times10^4$ to $2.0\times10^5\,\tau$. Zero-field runs at every state point provide the bulk structure factors and the Coulomb energies, the latter by Ewald summation.

\textit{Applied potentials.} The planar potentials $V_\pm(z)=V_N(z)\pm\psi(z)$ are drawn from four families of superposed cosines with wavenumbers $k_m=2\pi m/L$: box modes ($m=1$, 2), Fourier potentials (two or three of $m=3$ to 6), Fourier potentials with one or two added short modes ($m=8$, 12, 16 or 20, wavelengths $3.0$ to $1.2\,\sigma$) and periodic trains of one Gaussian well and one barrier per period $L/3$ or $L/4$. Phases and relative mode weights are random, each part present is rescaled to a peak-to-peak variation between $0.5$ and $3\,k_BT$, and a realization carries both parts, $V_N$ alone or $\psi$ alone. The potentials applied in the runs shown below are drawn above the corresponding density profiles.

At $\varepsilon_r=7.5$ and 2, four runs at each training concentration are held out, covering the different potential families, and three further runs are used for validation. At $\varepsilon_r=5$, 4 and 3, the held-out and validation runs are the potentials of the same tags, which leaves one to four held-out runs per concentration (16, 16 and 14 in total). The functional is trained on 276 runs and tested on 86 held-out runs and on 52 runs at the untrained concentration $c=0.05$ (22 at $\varepsilon_r=7.5$, 8 at $\varepsilon_r=4$, 22 at $\varepsilon_r=2$).
Densities and internal force densities are recorded on $0.1\sigma$ bins. Profile uncertainties use eight time blocks. Internal-force blocks are formed as $f_b^{\rm int}=f_b^{\rm tot}-f_b^{\rm ext}$ before computing their standard error.

\subsection*{S2.2 Force matching}
Let $\mathbf R$ denote all particle coordinates and
$\rmd\mathbf R$ the corresponding configurational measure.
The normalized canonical distribution is
\[
P(\mathbf R)
=
Z^{-1}\exp\left[
-\frac{1}{k_BT}
\left(
U_{\rm int}(\mathbf R)
+\sum_\alpha\sum_{j\in\alpha}V_\alpha(\vr_j)
\right)
\right],
\]
and the one-body density is
\[
n_\alpha(\vr)
=
\left\langle
\sum_{j\in\alpha}\delta(\vr-\vr_j)
\right\rangle
=
\sum_{j\in\alpha}\int\rmd\mathbf R\,
P(\mathbf R)\,\delta(\vr-\vr_j).
\]
Using
$\nabla_{\vr}\delta(\vr-\vr_j)
=-\nabla_j\delta(\vr-\vr_j)$
and integrating by parts over $\vr_j$, we obtain
\[
\begin{aligned}
\nabla n_\alpha(\vr)
&=
\sum_{j\in\alpha}\int\rmd\mathbf R\,
P(\mathbf R)\,\nabla_{\vr}\delta(\vr-\vr_j)\\
&=
-\sum_{j\in\alpha}\int\rmd\mathbf R\,
P(\mathbf R)\,\nabla_j\delta(\vr-\vr_j)\\
&=
\sum_{j\in\alpha}\int\rmd\mathbf R\,
\delta(\vr-\vr_j)\,\nabla_jP(\mathbf R).
\end{aligned}
\]
The boundary term vanishes for periodic boundaries
(with the periodic delta function), or for sufficiently
decaying distributions. For $j\in\alpha$,
\[
\nabla_jP(\mathbf R)
=
-\frac{P(\mathbf R)}{k_BT}
\left[
\nabla_jU_{\rm int}(\mathbf R)
+\nabla_jV_\alpha(\vr_j)
\right].
\]
Substitution therefore gives
\[
\begin{aligned}
k_BT\nabla n_\alpha(\vr)
={}&
-\left\langle
\sum_{j\in\alpha}\delta(\vr-\vr_j)
\nabla_jU_{\rm int}
\right\rangle\\
&-
\left\langle
\sum_{j\in\alpha}\delta(\vr-\vr_j)
\nabla_jV_\alpha(\vr_j)
\right\rangle.
\end{aligned}
\]
Since the delta function fixes $\vr_j=\vr$,
\[
\left\langle
\sum_{j\in\alpha}\delta(\vr-\vr_j)
\nabla_jV_\alpha(\vr_j)
\right\rangle
=
n_\alpha(\vr)\nabla V_\alpha(\vr).
\]
Defining the internal force density by
\[
\vf^{\rm int}_\alpha(\vr)
=
-\left\langle
\sum_{j\in\alpha}\delta(\vr-\vr_j)
\nabla_jU_{\rm int}
\right\rangle,
\]
we recover the local force balance
\cite{SI-tschopp2022force,SI-hansen2013theory},
\begin{equation}
k_BT\nabla n_\alpha(\vr)
=
\vf^{\rm int}_\alpha(\vr)
-n_\alpha(\vr)\nabla V_\alpha(\vr).
\label{eq:S:YBG}
\end{equation}
Taking the gradient of the Euler--Lagrange equation and using Eq.~\ref{eq:S:YBG} yields $\vf^{\rm int}_\alpha=-n_\alpha\nabla(e_\alpha\phi+\mu^{\rm ex}_\alpha)$, the force-matching relation of the main text.

Force matching constrains only $\nabla\mu^{\rm ex}_\alpha$. Two excess functionals that give the same forces on every profile differ by a functional $G[\vn]$ whose derivative $\delta G/\delta n_\alpha(\vr)=c_\alpha[\vn]$ is constant in space. Symmetry of the second derivative, $\delta c_\alpha/\delta n_\beta(\vr')=\delta c_\beta/\delta n_\alpha(\vr)$, with the left side independent of $\vr$ and the right side independent of $\vr'$, makes both independent of position, so $c_\alpha$ depends on $\vn$ only through the particle numbers $N_\beta=\int\rmd\vr\,n_\beta$, and $G=g(N_+,N_-)$ for some function $g$. In general, force matching therefore leaves $\Fex$ undetermined by an arbitrary function of the particle numbers. For the short-ranged form of Eq.~\ref{eq:Fex} of the main text, $\delta\Fex/\delta n_\alpha(\vr)$ depends on the density only within a few kernel ranges of $\vr$, whereas $\partial g/\partial N_\alpha$ for a nonlinear $g$ depends on the density arbitrarily far away. The only representable $g$ are linear, so $\Fex$ is determined up to $\sum_\alpha c_\alpha N_\alpha$, which shifts each $\mu_\alpha$ by a constant that is the same at every state point. Density profiles, differences of $\mu_\alpha$ between states and the bulk Hessian are unaffected. An absolute chemical potential would require one reference value per species, for instance from Widom insertion at a single state point.

The six Gaussian envelope widths are learned. They start at $0.25$, $0.43$, $0.76$, $1.32$, $2.30$ and $4.00\sigma$ and converge to $0.21$, $0.48$, $0.60$, $0.65$, $0.84$ and $1.18\sigma$ in the functional trained on all five permittivities. The saturation scale of Eq.~\ref{eq:lb} of the main text converges to $\lambda_s=0.98$. The decomposition of Eq.~\ref{eq:lb} into $F_0$, $F_1$ and $F_2$ is poorly determined over the sampled range, because $s(\lambda)$ varies only from 0.64 to 0.98 between $l_B=8\sigma$ and $30\sigma$ and no data lie below $8\sigma$. We therefore report only the total functional and its derivative with respect to $l_B$. The pointwise network receives $h_{m\alpha}/n_{\rm ref}$, with $n_{\rm ref}=0.02\sigma^{-3}$, and its output contributes $n_{\rm ref}\int\Phi_\theta\,\rmd\vr$ to the free energy. Other architecture parameters are given in the main text. The kernel range and activation were selected using only
the validation profiles at $\varepsilon_r=2$.

Training uses full batches and the AdamW schedule specified in the main text, with the learning rate decaying to $10^{-5}$ over 12{,}000 steps and the updates of the six widths scaled by 30. Validation is evaluated every 20 steps, and the parameters with the lowest validation residual are retained. For most models they lie within the last 2{,}000 steps.
Table~\ref{tab:S:spectrum} compares training with and without the Fourier term, using the same architecture and data splits. At $\varepsilon_r=7.5$ the Fourier term brings $\Gamma(k_1)$ closer to MD (4.47 against 4.24 without it) and reduces the amplitude deficits of the long-wavelength modes by about two percentage points. At $\varepsilon_r=2$ its effect on $\Gamma(k_1)$ is within the MD uncertainty, and the profile errors are slightly higher with it (2.85\% against 2.48\%).

\begin{table}[H]
\centering
\caption{Loss ablation with and without the Fourier term weighted by $(k\sigma)^{-2}$, for the functional at fixed coupling. Amplitude ratio: predicted over MD amplitude of the $k_1$ mode, summed over the six long-wavelength training runs at $c=0.04$ (MD: $\Gamma(k_1)=4.38\pm0.15$ at $\varepsilon_r=7.5$ and $3.14\pm0.56$ at $\varepsilon_r=2$). Last two columns: amplitude deficit $100\times(1-A_{\rm model}/A_{\rm MD})$ of the driven cation mode in boxes two and four times the training length, where it has wavenumber $k_1/2$ and $k_1/4$ (section S4.1); values before and after the slash are for neutral ($\psi=0$) and mixed ($V_N\neq0$, $\psi\neq0$) potentials.}
\label{tab:S:spectrum}
\begin{tabular}{c l c c c c c}
$\varepsilon_r$ & Fourier term & Ratio at $k_1$ & $\Gamma(k_1)$  & Profile (\%) & $k_1/2$ (\%) & $k_1/4$ (\%)\\
\midrule
7.5 & without & 1.02 & 4.24 & 1.39 & 11 / 1 & 6 / 1\\
7.5 & with & 0.96 & 4.47 & 1.43 & 9 / -2 & 4 / -1\\
2 & without & 0.98 & 3.46 & 2.48 & -1 / -1 & 4 / 6\\
2 & with & 0.93 & 3.63 & 2.85 & 1 / 0 & 5 / 8\\
\bottomrule
\end{tabular}
\end{table}

\subsection*{S2.3 Predictions and uncertainty}
At fixed particle numbers, the predicted planar profiles satisfy
\begin{equation}
n_\alpha(z)=N_\alpha\frac{e^{-u_\alpha(z)}}{A\int_0^L e^{-u_\alpha(z')}\,\rmd z'},\qquad
u_\alpha=V_\alpha+e_\alpha\phi[\vn]+\mu^{\rm ex}_\alpha[\vn],
\label{eq:S:canon}
\end{equation}
where $A$ is the cross-sectional area. The chemical potentials enforcing particle numbers drop out. We solve for $u_\alpha$ using Anderson acceleration with history eight and mixing parameter 0.1 \cite{SI-anderson1965iterative,SI-walker2011anderson}. The charge residual is preconditioned by $k^2/(k^2+\kappa_D^2)$ \cite{SI-kerker1981efficient}. Convergence requires the maximum density change divided by $\nb$ to be below $10^{-6}$. Profile errors are $\|n_\alpha^{\rm pred}-n_\alpha^{\rm MD}\|_2/\|n_\alpha^{\rm MD}\|_2$, averaged over runs.

The bulk kernel $W_{\alpha\beta}(k)$ follows by automatic differentiation of $\mu^{\rm ex}_\alpha$ at uniform density along a one-bin perturbation of species $\beta$, followed by Fourier transformation. Equation~\ref{eq:S:OZ} gives $S(k)$. The neural functionals used for the main results have no negative eigenvalues of $S(k)^{-1}$ at the sampled concentrations and wavenumbers of a box eight times longer than the simulation box.

MD structure factors are obtained from zero-field runs.
For each frame, we compute the microscopic Fourier densities
\[
\tilde n_\alpha(\mathbf k,t)
=
\sum_{j\in\alpha}
e^{-i\mathbf k\cdot\vr_j(t)},
\qquad
\mathbf k=\frac{2\pi}{L}\mathbf m,
\quad \mathbf m\in\mathbb Z^3.
\]
For the homogeneous system at nonzero wavevectors,
$\langle\tilde n_\alpha(\mathbf k,t)\rangle=0$.
The number structure factor is therefore
\[
S_{NN}(k)
=
\frac{1}{V}
\left\langle
\left|
\tilde n_+(\mathbf k,t)+\tilde n_-(\mathbf k,t)
\right|^2
\right\rangle_{t,\,|\mathbf k|=k},
\]
where $V=L^3$ and the average is over time and wavevector
directions with the same magnitude. The charge structure factor $S_{ZZ}(k)$ is defined in the same way from $\tilde n_+(\mathbf k,t)-\tilde n_-(\mathbf k,t)$.
The uncertainty of $\Gamma(k_1)=2\nb/S_{NN}(k_1)$ is a standard error from time blocks, corrected for the autocorrelation time. At $\varepsilon_r=7.5$, $\Gamma(k_1)$ and its uncertainty are obtained from nine independent runs at each concentration, and at $\varepsilon_r=2$ from a single run.

\section*{S3. Predictive accuracy and many-body correlations}

\subsection*{S3.1 Predictions across concentrations}
Table~\ref{tab:S:profiles} collects the concentration-resolved force residuals and profile errors at both dielectric constants. The force residual $\chi^2$ is the first term of the force-matching loss of the main text, averaged over bins, species and runs. With these weights, a model that reproduces the MD forces exactly would give $\chi^2\approx0.5$. Table~\ref{tab:S:sk} gives the corresponding bulk structure-factor errors and finite-wavenumber number response. The lowest concentration at strong coupling is the least accurate state for the number structure factor. The representative profiles and structure factors are shown in the main text; the structure factors at $\varepsilon_r=2$ are shown in Fig.~\ref{fig:S:structure2} and every held-out profile in Figs.~\ref{fig:S:heldout75} and \ref{fig:S:heldout2}.

\begin{figure}[tbp]
\centering
\includegraphics[width=\linewidth]{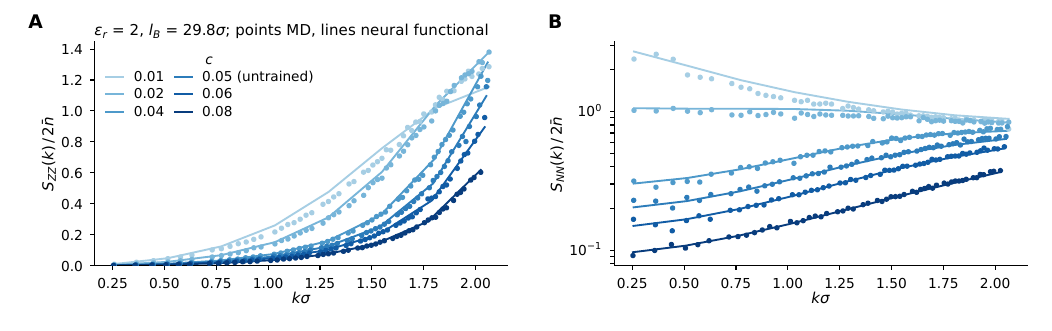}
\caption{Bulk structure at $\varepsilon_r=2$. (\textit{A}) Charge and (\textit{B}) number structure factors at $c=0.01$ to 0.08 (light to dark; $c=0.05$ untrained). Points, zero-field MD; lines, the neural functional from Eq.~\ref{eq:S:OZ}. Relative errors are listed in Table~\ref{tab:S:sk}; Fig.~4 of the main text shows $\varepsilon_r=7.5$.}
\label{fig:S:structure2}
\end{figure}

\begin{figure}[tbp]
\centering
\includegraphics[width=\linewidth,height=0.84\textheight,keepaspectratio]{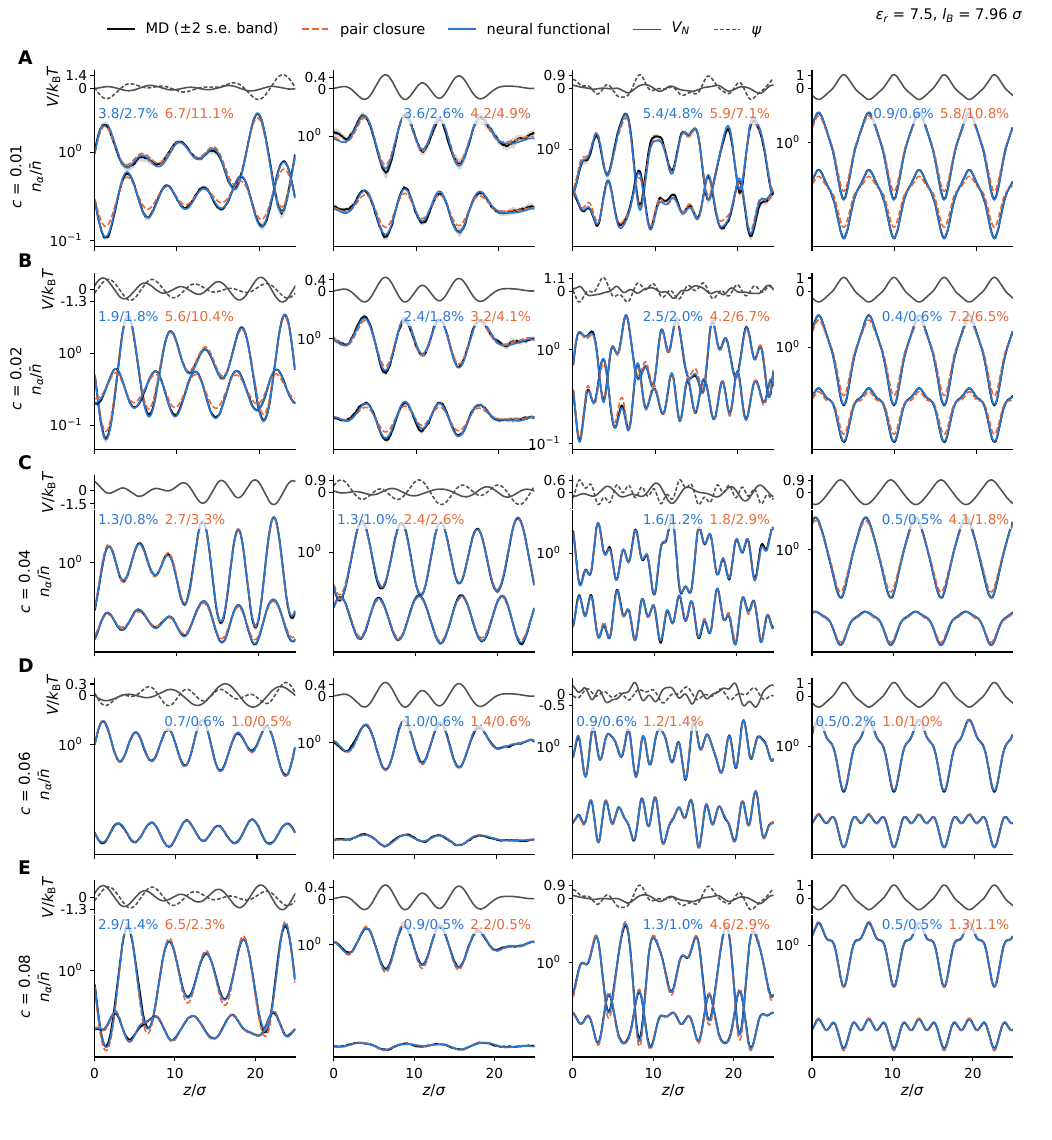}
\caption{All held-out profiles at $\varepsilon_r=7.5$. Rows \textit{A}--\textit{E}: $c=0.01$, 0.02, 0.04, 0.06 and 0.08; columns: the four held-out runs of each concentration, with the applied potential drawn above each panel ($V_N$ solid, $\psi$ dashed, in $k_BT$). Each panel shows $n_\alpha/\nb$ on a logarithmic axis, cations above, anions below (divided by three): MD with $\pm2$ standard errors (black, grey), the neural functional (blue) and the pair closure (orange dashed). Numbers are relative $L_2$ errors, cation / anion, of the functional (blue) and the pair closure (orange). Concentration averages are listed in Table~\ref{tab:S:profiles}.}
\label{fig:S:heldout75}
\end{figure}

\begin{figure}[tbp]
\centering
\includegraphics[width=\linewidth,height=0.84\textheight,keepaspectratio]{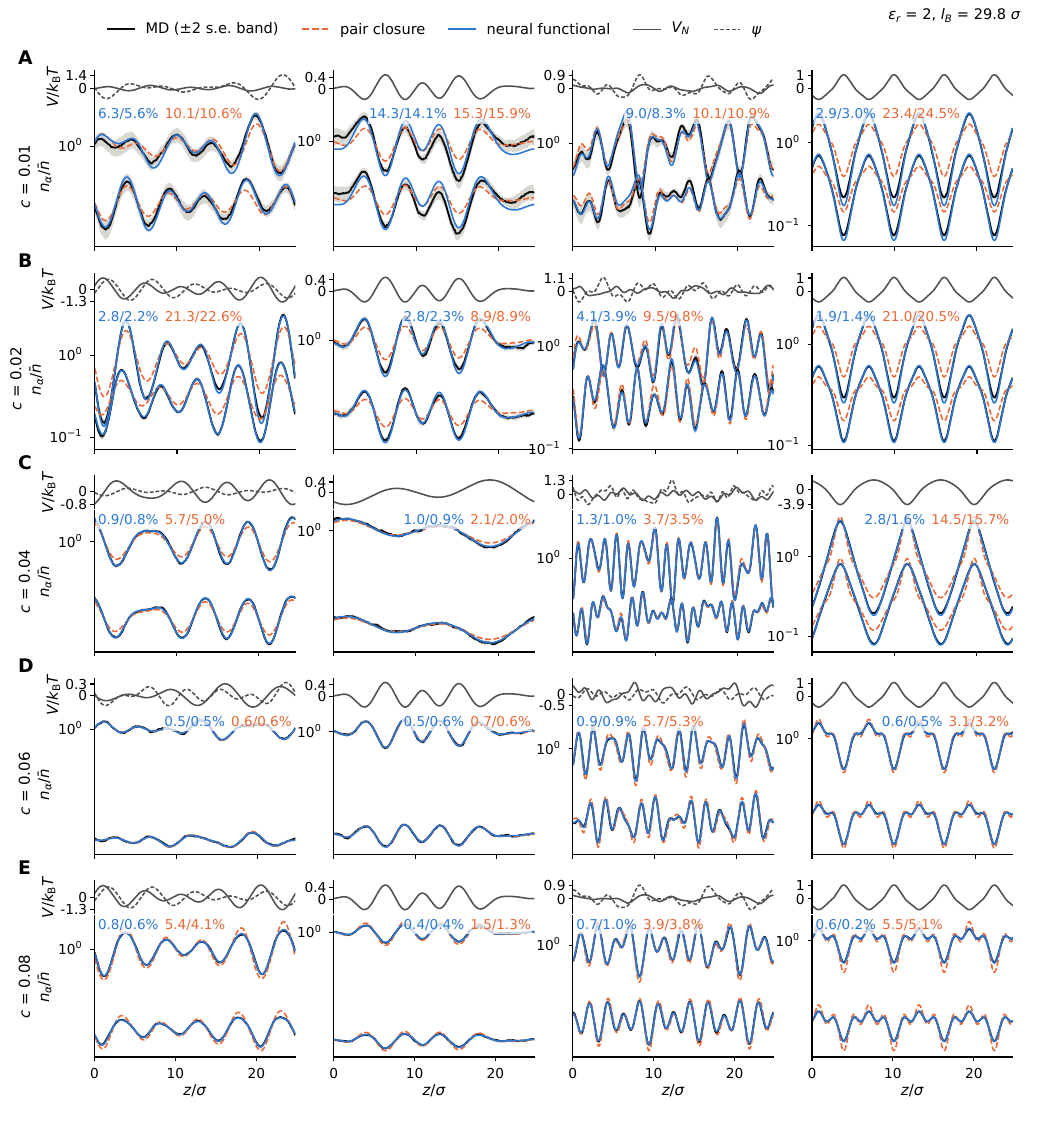}
\caption{All held-out profiles at $\varepsilon_r=2$, as in Fig.~\ref{fig:S:heldout75}.}
\label{fig:S:heldout2}
\end{figure}

\begin{table}[H]
\centering
\caption{Predictions by concentration. Force residual $\chi^2$ and relative profile error in percent (cation / anion), averaged over held-out runs. At $c=0.05$ all 22 runs are untrained.}
\label{tab:S:profiles}
\begin{tabular}{c l c c c c}
$\varepsilon_r$ & $c$ & NF $\chi^2$ & NF profile (\%) & Pair $\chi^2$ & Pair profile (\%)\\
\midrule
7.5 & 0.01 & 0.555 & 3.4 / 2.6 & 1.74 & 5.7 / 8.5\\
7.5 & 0.02 & 0.568 & 1.8 / 1.5 & 2.90 & 5.0 / 6.9\\
7.5 & 0.04 & 0.589 & 1.2 / 0.9 & 1.62 & 2.8 / 2.7\\
7.5 & 0.05 & 0.589 & 0.9 / 0.7 & 1.50 & 1.7 / 1.8\\
7.5 & 0.06 & 0.613 & 0.8 / 0.5 & 0.77 & 1.1 / 0.9\\
7.5 & 0.08 & 0.741 & 1.4 / 0.9 & 3.01 & 3.7 / 1.7\\
2 & 0.01 & 0.655 & 8.1 / 7.7 & 6.18 & 14.7 / 15.5\\
2 & 0.02 & 0.574 & 2.9 / 2.5 & 19.65 & 15.2 / 15.4\\
2 & 0.04 & 0.984 & 1.5 / 1.1 & 19.78 & 6.5 / 6.5\\
2 & 0.05 & 0.622 & 0.8 / 0.7 & 2.93 & 2.3 / 2.4\\
2 & 0.06 & 0.649 & 0.6 / 0.6 & 2.74 & 2.5 / 2.5\\
2 & 0.08 & 0.688 & 0.6 / 0.6 & 20.46 & 4.1 / 3.6\\
\bottomrule
\end{tabular}
\end{table}

\begin{table}[H]
\centering
\caption{Bulk predictions at both dielectric constants. Structure-factor errors are relative rms deviations in percent over the MD wavevectors up to $k=2.1\sigma^{-1}$, shown as neural functional / pair closure. $\Gamma=2\nb/S_{NN}(k_1)$ uses the MD uncertainty procedure of section S2.3.}
\label{tab:S:sk}
\begin{tabular}{c c c c c c c}
$\varepsilon_r$ & $c$ & $S_{ZZ}$ error & $S_{NN}$ error & $\Gamma$ MD & $\Gamma$ NF & $\Gamma$ pair\\
\midrule
7.5 & 0.01 & 3.3 / 12.3 & 2.9 / 18.5 & $1.20\pm0.06$ & 1.19 & 2.12\\
7.5 & 0.02 & 4.5 / 12.7 & 2.5 / 16.1 & $1.97\pm0.07$ & 1.94 & 3.24\\
7.5 & 0.04 & 3.0 / 7.5 & 2.0 / 8.0 & $4.38\pm0.15$ & 4.11 & 5.41\\
7.5 & 0.05 & 2.2 / 4.0 & 1.9 / 4.3 & $5.42\pm0.15$ & 5.33 & 6.46\\
7.5 & 0.06 & 2.8 / 2.6 & 2.4 / 4.7 & $6.42\pm0.26$ & 6.56 & 7.46\\
7.5 & 0.08 & 7.0 / 13.2 & 2.3 / 14.5 & $9.26\pm0.26$ & 8.94 & 9.36\\
2 & 0.01 & 7.6 / 19.5 & 10.5 / 51.6 & $0.42\pm0.07$ & 0.37 & 1.75\\
2 & 0.02 & 3.4 / 27.4 & 7.0 / 38.2 & $0.98\pm0.10$ & 0.95 & 2.51\\
2 & 0.04 & 3.4 / 21.8 & 2.8 / 21.1 & $3.14\pm0.56$ & 3.31 & 3.99\\
2 & 0.05 & 3.4 / 12.7 & 3.1 / 11.7 & $4.41\pm0.49$ & 4.91 & 4.70\\
2 & 0.06 & 4.2 / 2.5 & 2.3 / 6.0 & $6.09\pm0.91$ & 6.69 & 5.39\\
2 & 0.08 & 2.8 / 27.3 & 2.3 / 34.7 & $11.05\pm0.81$ & 10.35 & 6.69\\
\bottomrule
\end{tabular}
\end{table}

A pair closure can reproduce profiles at a single concentration but cannot adjust its bulk kernel with density. Fig.~\ref{fig:S:kernels} shows the bulk kernel $W(k)$ of Eq.~\ref{eq:S:W} in real space: the effective pair interactions of the neural functional shift with concentration, whereas the pair closure has a single kernel. At $\varepsilon_r=7.5$, fitting only $c=0.04$ gives a held-out residual of 0.65 and cation/anion profile errors of 1.2/1.1\%. Its held-out residual rises to 1.36 when $c=0.01$ and 0.02 are added, to 1.56 after adding $c=0.06$, and to 2.01 after adding $c=0.08$. This comparison motivates the density-dependent correlations of the neural functional.

\begin{figure}[tbp]
\centering
\includegraphics[width=\linewidth]{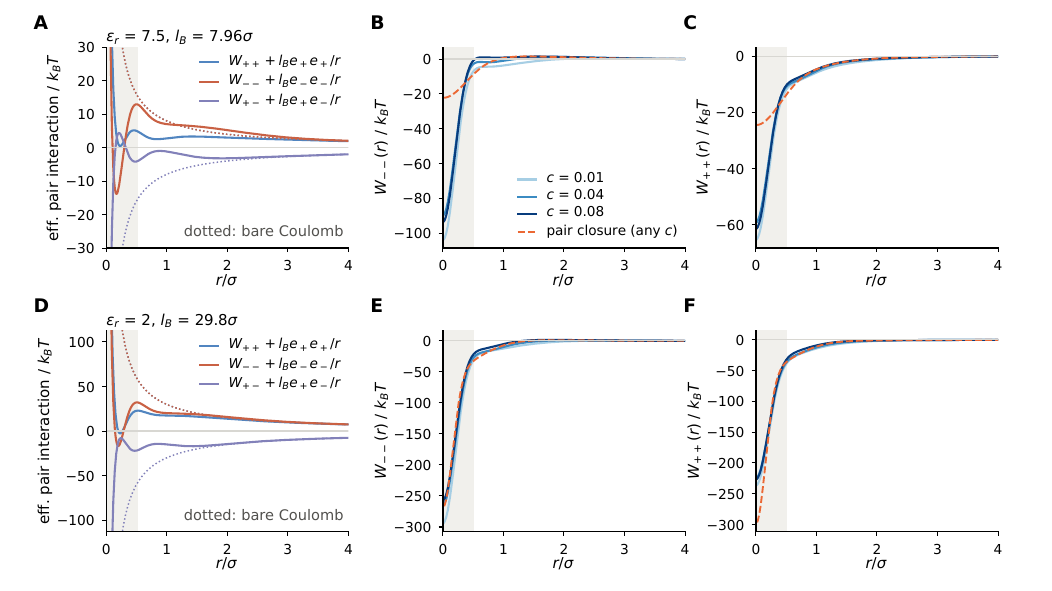}
\caption{Effective pair interactions of the learned functional in real space at $\varepsilon_r=7.5$ (\textit{A}--\textit{C}) and $\varepsilon_r=2$ (\textit{D}--\textit{F}). (\textit{A}, \textit{D}) Effective pair interactions $W_{\alpha\beta}(r)+l_Be_\alpha e_\beta/r$ at $c=0.04$ for the three species pairs (solid) and the bare Coulomb term alone (dotted); the shaded strip marks $r<0.5\sigma$. (\textit{B}, \textit{E}) Anion--anion kernel $W_{--}(r)$ of the neural functional at $c=0.01$, 0.04 and 0.08 (light to dark) and the density-independent kernel of the pair closure (dashed). (\textit{C}, \textit{F}) The same for $W_{++}(r)$. $W(r)$ is the inverse Fourier transform of the bulk kernel $W(k)$ of Eq.~\ref{eq:S:W}, the short-range part of $-c_{\alpha\beta}$.}
\label{fig:S:kernels}
\end{figure}

\subsection*{S3.2 Number response}
For $\nb_+=\nb_-=\nb$, write the regular part of $S^{-1}$ in the orthonormal number and charge basis, $\mathbf e_N=(1,1)^{\T}/\sqrt2$ and $\mathbf e_Z=(1,-1)^{\T}/\sqrt2$. In the energy units of section S1,
\[
A_{NN}=\nb^{-1}+\tfrac12(W_{++}+W_{--}+2W_{+-}),\qquad
A_{ZZ}=\nb^{-1}+\tfrac12(W_{++}+W_{--}-2W_{+-}),\qquad
A_{NZ}=\tfrac12(W_{++}-W_{--}).
\]
Inverting Eq.~\ref{eq:S:OZ} gives
\begin{equation}
\Gamma(k)\equiv\frac{2\nb}{S_{NN}(k)}
=1+\frac{\nb}{2}\sum_{\alpha\beta}W_{\alpha\beta}(k)
-\frac{\nb A_{NZ}^2}{A_{ZZ}+8\pi l_B/k^2}.
\label{eq:S:gamma}
\end{equation}
The last term accounts for coupling between number and charge fluctuations. It vanishes as $k\to0$ and contributes 1--4\% of $\Gamma$ at $k_1$ in the reported model. Neglecting that correction yields the MD estimate used in the main text,
\[
\mathbf q^{\T}W(k_1)\mathbf q\simeq\frac{2[\Gamma(k_1)-1]}{\nb},\qquad \mathbf q=(1,1)^{\T}.
\]
Multiplying by $k_BT$ restores energy units.

\section*{S4. Transfer and comparisons with other functionals}

\subsection*{S4.1 Larger boxes and nonplanar densities}
At $c=0.04$ and both dielectric constants, we test transfer
without retraining using the models trained on planar profiles.
The tests include four aperiodic potentials absent from training,
applied in a box twice the training length. They combine irregular Gaussian wells and barriers with Fourier modes at and between the training wavenumbers, with amplitudes within the trained range. Two contain no wavelength longer than the training box. The other two add a mode at $k_1/2$ with a peak-to-peak amplitude of $1\,k_BT$, in the neutral part alone or in the neutral and charged parts. The neural functional reproduces the density profiles with relative $L_2$ errors of 1.3--1.7\% without the long mode and 0.9--1.2\% with it, compared with 3.0--5.5\% for the pair closure. The profiles are shown in Fig.~\ref{fig:S:aperiodic}.

In a box four times the training length, a single mode at $k_1/4$ is applied in the neutral part alone or in the neutral and charged parts (Table~\ref{tab:S:box4}). The amplitude deficit is $1-A_{\rm model}/A_{\rm MD}$, where $A$ is the amplitude of the driven mode of the cation density and the MD value is the long-time limit of a relaxation fit to its time series. A fourth potential superposes the $k_1/4$ mode on aperiodic in-range structure. The neural functional predicts it with profile errors of 0.9\% and 0.5\% at $\varepsilon_r=7.5$ and 2. Zero-field simulations in the boxes two and four times the training length give the number response at $k_1/2$ and $k_1$ directly; the functional deviates from these MD values by $-0.1\%$ and $-2.1\%$ at $\varepsilon_r=7.5$ and by $+0.9\%$ and $-1.8\%$ at $\varepsilon_r=2$, with uncertainties of 2--3\%.

\begin{figure}[tbp]
\centering
\includegraphics[width=\linewidth]{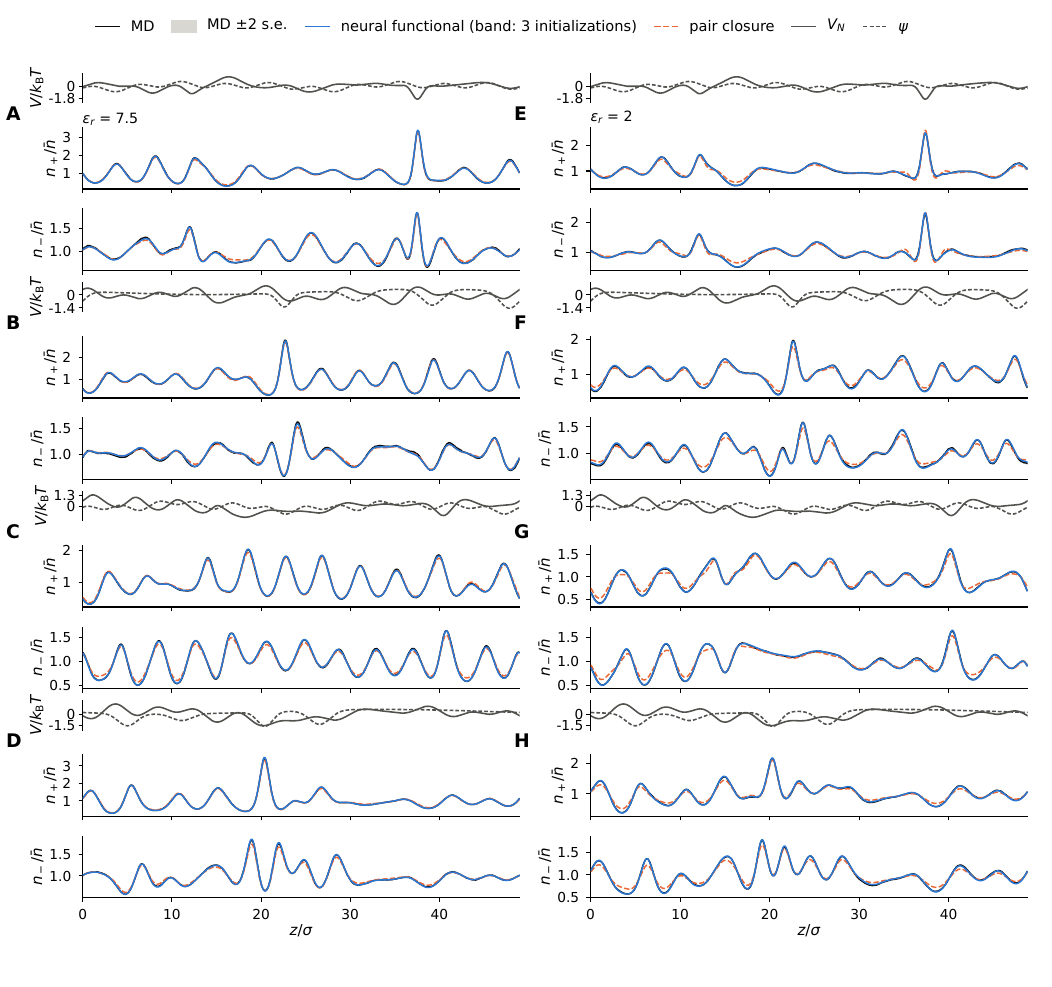}
\caption{The four aperiodic potentials in the box twice the training length, $c=0.04$. (\textit{A}--\textit{D}) $\varepsilon_r=7.5$ and (\textit{E}--\textit{H}) $\varepsilon_r=2$: cation (upper) and anion (lower) densities under four aperiodic potentials, drawn above each panel over the twofold box ($V_N$ solid, $\psi$ dashed, in $k_BT$). Two combine irregular Gaussian wells and barriers with modes at and between the training wavenumbers, the other two add the box mode $k_1/2$, the one wavelength longer than the training box. MD with $\pm2$ standard errors (black, grey), the neural functional (blue; band, spread over three initializations) and the pair closure (orange dashed).}
\label{fig:S:aperiodic}
\end{figure}

\begin{table}[H]
\centering
\caption{Box four times the training length: relative profile errors and amplitude deficits of the driven $k_1/4$ mode (both in \%), shown as neural functional / pair closure.}
\label{tab:S:box4}
\begin{tabular}{c l c c}
$\varepsilon_r$ & potential & Profile error & Amplitude deficit\\
\midrule
7.5 & neutral & 0.43 / 1.09 & 5 / 15\\
7.5 & neutral and charged & 0.41 / 0.81 & 0 / 11\\
2 & neutral & 0.35 / 2.52 & 0 / 24\\
2 & neutral and charged & 0.36 / 2.62 & 2 / 26\\
\bottomrule
\end{tabular}
\end{table}

Six potentials varying in $x$ and $y$ test transfer beyond planar geometry: an egg carton, neutral and charged Gaussian rods on a square lattice, a checkerboard with oblique wavevectors, and two aperiodic potentials. Aperiodic I combines six irregularly placed Gaussian wells and barriers and two oblique plane waves in the neutral part with three charged Gaussian wells. Aperiodic II combines three oblique plane waves and three Gaussian barriers with four irregularly placed charged wells. Single bins of the MD maps carry a sampling noise of 5--6\%, so Table~\ref{tab:S:field2d} and Fig.~\ref{fig:S:field2d} report the relative profile error on the Fourier modes that MD resolves, those with amplitudes above three standard errors, together with the MD noise level of the same quantity. On the aperiodic potentials, the neural functional reaches this noise level, with errors of 1.7--3.3\% against noise levels of 1.7--3.0\%, while the pair closure errs by 3.7--5.0\% and Poisson--Boltzmann theory by 31--62\%. On the periodic potentials, its errors are 0.3--1.5\%, against 2.3--8.5\% for the pair closure, with one exception: for the charged rods at $\varepsilon_r=7.5$ the error is 4.1\%, against 2.4\% for the pair closure. The rods concentrate the cations to 5.4 times the mean density, and the functional overestimates this peak by about 20\%. Across the three initializations of the functional trained on all five permittivities, the driven charge and number amplitudes under this potential exceed MD by 3--9\% and 7--18\%, whereas the functional at fixed coupling with the same form stays within 2--4\%. The weighted densities at the rod centres lie within the range of the planar training profiles, channel by channel, so the overestimate is not caused by local densities beyond the training data.

\begin{figure}[tbp]
\centering
\includegraphics[width=\linewidth]{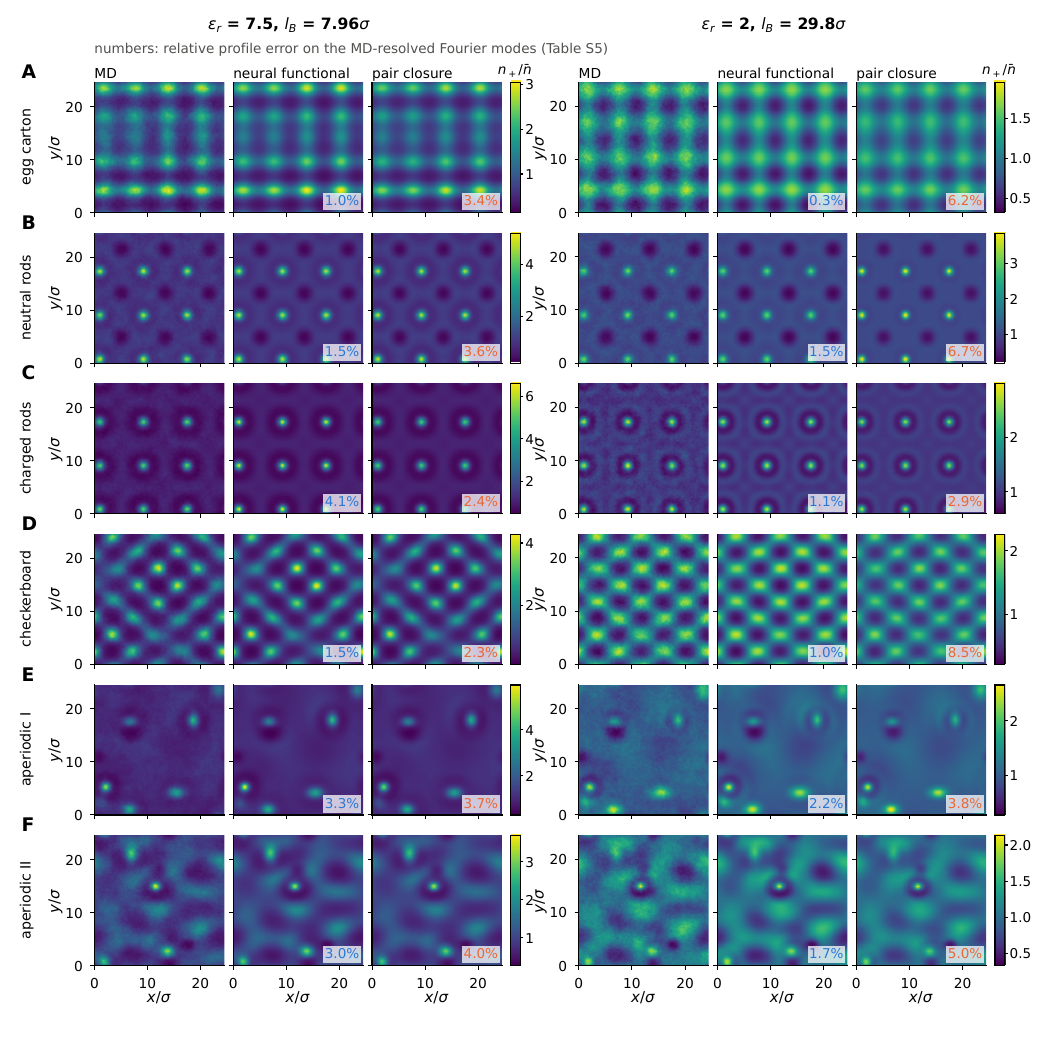}
\caption{The six two-dimensional potentials at $c=0.04$ (rows \textit{A}--\textit{F}) at $\varepsilon_r=7.5$ (left) and $\varepsilon_r=2$ (right): cation density $n_+(x,y)/\nb$ from MD, the neural functional and the pair closure, on a common colour scale per row and permittivity. Both models were trained on planar profiles only. Numbers are the relative profile errors of Table~\ref{tab:S:field2d}; aperiodic I and II also appear in Fig.~3 of the main text.}
\label{fig:S:field2d}
\end{figure}

\begin{table}[H]
\centering
\caption{Two-dimensional potentials at $c=0.04$: relative profile errors (\%) of the density maps predicted by the functionals trained on planar profiles, $\|n^{\rm pred}_\alpha-n^{\rm MD}_\alpha\|_2/\|n^{\rm MD}_\alpha\|_2$ averaged over both species. The error is evaluated on the Fourier modes that MD resolves (amplitudes above three standard errors; their number for both species is given), since single bins of the MD maps carry 5--6\% sampling noise. The last column is the MD noise level of the same quantity. The lowest error in each row is in bold.}
\label{tab:S:field2d}
\begin{tabular}{llccccc}
$\varepsilon_r$ & potential & resolved modes & neural functional & pair closure & PB & MD noise\\
\midrule
7.5 & egg carton & 32 & \textbf{1.0} & 3.4 & 40 & 1.0\\
 & neutral rods & 115 & \textbf{1.5} & 3.6 & 68 & 1.3\\
 & charged rods & 102 & 4.1 & \textbf{2.4} & 26 & 1.2\\
 & checkerboard & 54 & \textbf{1.5} & 2.3 & 42 & 1.1\\
 & aperiodic I & 327 & \textbf{3.3} & 3.7 & 61 & 3.0\\
 & aperiodic II & 218 & \textbf{3.0} & 4.0 & 33 & 2.6\\
2 & egg carton & 10 & \textbf{0.3} & 6.2 & 33 & 0.4\\
 & neutral rods & 114 & \textbf{1.5} & 6.7 & 70 & 1.1\\
 & charged rods & 96 & \textbf{1.1} & 2.9 & 15 & 1.0\\
 & checkerboard & 20 & \textbf{1.0} & 8.5 & 34 & 0.6\\
 & aperiodic I & 235 & \textbf{2.2} & 3.8 & 62 & 2.3\\
 & aperiodic II & 154 & \textbf{1.7} & 5.0 & 31 & 1.7\\
\bottomrule
\end{tabular}
\end{table}

\subsection*{S4.2 Comparison protocol}
Table~1 of the main text refers to the published architectures. The one-body networks of refs.~\cite{SI-sammuller2023neural,SI-bui2025learning} apply a single network to a sliding window of a planar density profile on a fixed grid. They are therefore equivariant under translations by the grid spacing and can be applied to planar domains larger than their training boxes. Ref.~\cite{SI-sammuller2023neural} includes mirror symmetry only through data augmentation and verifies the thermal Noether sum rules and the symmetry of the pair direct correlation function numerically. Ref.~\cite{SI-bui2025learning} trains separate networks for cations and anions, so their cross derivatives need not coincide, and verifies the contact theorem instead. The convolutional free energy of ref.~\cite{SI-dijkman2025learning} applies average pooling after each layer to an input of fixed length, which breaks invariance under lattice translations and ties the model to the training box and grid. Ref.~\cite{SI-cheng2026equivariant} builds invariance under the cubic point group $O_h$ into finite-range environments on a fixed voxel grid and demonstrates transfer to larger cells and three-dimensional geometries at unchanged grid spacing. For the present functional, the guarantees hold for the continuous formulation (section S1). Their finite-grid residuals are reported in section S4.3, and transfer to larger boxes and nonplanar densities in section S4.1.

\textit{Common protocol.} Each adapted form replaces only the excess free energy $\Fex$ of our functional or, for the one-body network, its excess chemical potentials $\mu^{\rm ex}_\alpha$. Its input is the density profile on the $0.1\sigma$ bins of the MD data, continued periodically and divided by $n_{\rm ref}=0.02\sigma^{-3}$ ($0.4\sigma^{-3}$ for the Lennard-Jones fluid). For the ions, the Coulomb mean field $\FC$ is kept analytic, as in our functional, so each network represents only the excess part. All forms are trained with the force-matching loss, optimizer, step budget and model selection of our functional, on the same splits (71 training runs at $\varepsilon_r=7.5$ and 82 at $\varepsilon_r=2$). The comparison therefore tests the architectures under one training protocol, not the published methods with their own data and training targets.

\textit{One-body network.} Refs.~\cite{SI-sammuller2023neural,SI-bui2025learning} fit the one-body direct correlation function $c_1=-\mu^{\rm ex}/k_BT$, obtained from grand-canonical simulations at known chemical potential, with a perceptron that acts on a window of the density profile. In ref.~\cite{SI-sammuller2023neural}, the window spans $\pm2.56\sigma$ at a resolution of $0.01\sigma$ (513 inputs) and the perceptron has three hidden layers of 512 units with smooth activations. Ref.~\cite{SI-bui2025learning} adds the long-range electrostatics through local molecular field theory.
Our adaptation maps the cation and anion densities in a window of $\pm3\sigma$ around $z$, $2\times61$ values, to both excess chemical potentials $\mu^{\rm ex}_{\theta,+}(z)$ and $\mu^{\rm ex}_{\theta,-}(z)$ with a single perceptron instead of one network per species. The perceptron has two hidden layers of 128 units with SiLU activations (32{,}514 parameters). The two outputs share hidden features, but $\partial\mu^{\rm ex}_{\theta,+}/\partial n_-$ and $\partial\mu^{\rm ex}_{\theta,-}/\partial n_+$ are not constrained to agree. The output enters force matching only through the force density $-n_\alpha\partial_z(e_\alpha\phi+\mu^{\rm ex}_{\theta,\alpha})$, so canonical data without chemical-potential labels suffice. A window of $\pm6\sigma$ (47{,}874 parameters) gives a validation residual 32\% and 48\% larger at $\varepsilon_r=7.5$ and $2$, and about ten times larger for the Lennard-Jones fluid.

\textit{Lattice free energy.} Ref.~\cite{SI-cheng2026equivariant} writes the excess free energy of a one-component fluid on a cubic grid of spacing $\Delta L=0.5\sigma$ and voxel volume $\Delta V=\Delta L^3$ as $\Fex=\Delta V\sum_g\rho_g\,[a_{\rm loc}(\rho_g)+a_{\rm CACE}(\rho_g,\mathbf B_g)]$. The features $\mathbf B_g$ are built from the Cartesian moments $A_{\mathbf l}(g)=\sum_{\mathbf q}\rho_{g+\mathbf q}\,q_x^{l_x}q_y^{l_y}q_z^{l_z}$ with $|\mathbf l|\le3$, summed with unit weight over the 122 nonzero integer offsets with $|\mathbf q|\le3$. Products of up to two moments, averaged over the 48 elements of $O_h$, give 15 invariants. Both read-outs have hidden layers of 32 and 16 SiLU units and also take the temperature (1{,}762 parameters in total).  We form the moments separately for cations and anions. The first-order invariants of each species and the second-order invariants within and across species, averaged over $O_h$ and with vanishing and duplicate forms removed, give 47 invariants (15 for one species, as in the original). Both read-outs have one output per species, and $\Fex=A\,\Delta z\sum_i\sum_\alpha\bar n_\alpha(z_i)\,a_\alpha(z_i)$, where $A$ is the cross-sectional area and $\Delta z$ the bin width. The voxel density $\bar n_\alpha(z_i)$ is the average of the profile over the five bins ($0.5\sigma$) centred on bin $i$. For a planar density, the lateral sums in the moments reduce to fixed weights, $A^\alpha_{\mathbf l}(z_i)=\sum_{q_z}C_{\mathbf l}(q_z)\,\bar n_\alpha(z_i+q_z\Delta L)$ with $C_{\mathbf l}(q_z)=\sum_{q_x,q_y}q_x^{l_x}q_y^{l_y}q_z^{l_z}$ over the stencil, which are evaluated exactly. Six of the 47 invariants (two of the 15) then vanish identically. Every bin is a voxel centre, so $\Fex$ averages the lattice free energy over the registrations of the lattice on the bin grid and is invariant under translations by one bin. Densities are divided by $n_{\rm ref}$, each invariant by its root-mean-square value over the training profiles, and temperature is not an input. The read-out $a_{\rm CACE}$ is widened to two hidden layers of 128 units, the width of our read-out, while $a_{\rm loc}$ keeps 32 and 16 (23{,}828 parameters). The excess chemical potentials $\mu^{\rm ex}_\alpha=\delta\Fex/\delta n_\alpha$ follow by automatic differentiation and are trained by force matching. We keep the stencil radius of the original, $1.5\sigma$. A radius of $3\sigma$, with the same invariants and number of parameters, raises the validation residual eightfold at $\varepsilon_r=7.5$ and 27-fold for the Lennard-Jones fluid. At $\varepsilon_r=2$, it lowers the validation residual by about a fifth and the mean held-out profile error from 5.0\% to 3.9\%, but the Euler--Lagrange solution fails for 12 instead of 9 of the 38 test runs.

\textit{Convolutional free energy.} Ref.~\cite{SI-dijkman2025learning} represents the excess free energy of a one-component fluid on 320 grid points ($\Delta z=\sigma/32$) by six periodic dilated convolutions (kernel size 3, dilation 2, channels 16, 16, 32, 32, 64, 64), each followed by average pooling over two points, and a scalar output. Its stated size of about 24{,}400 parameters implies a linear read-out of the flattened final feature map, 64 channels at five positions. The activation is not specified. The network is trained by pair-correlation matching: rows of its Hessian are fitted to the laterally integrated direct correlation functions of bulk grand-canonical simulations, together with an offset term for the first derivative. We keep the convolution and pooling stack. The input has two channels, $n_+/n_{\rm ref}$ and $n_-/n_{\rm ref}$ (one for the Lennard-Jones fluid), padding is periodic and the activations are softplus. Pooling drops an odd last point, so six poolings leave three positions for any grid of 192 to 255 bins, which covers all boxes here (243 to 250 bins for the ions, 240 for the Lennard-Jones fluid). A linear layer maps the $3\times64$ final features to $\Fex$ (24{,}321 parameters). The network is trained by force matching on the inhomogeneous profiles instead of by pair-correlation matching on bulk data.

\textit{Lennard-Jones fluid.} The truncated and shifted fluid has $r_c=2.5\sigma$, $T=1.5$ and $L=24\sigma$. It has a single species and no Coulomb term, so each form takes one density. The forms then have 18{,}867 (ours), 24{,}577 (one-body network), 19{,}426 (lattice free energy) and 24{,}273 (convolutional free energy) parameters. The planar potentials use the same families as the ion data, 22 per density. One run each at 0.5 and $0.6\sigma^{-3}$ violates the force balance $k_BT\,\partial_zn=f^{\rm ext}+f^{\rm int}$ beyond its sampling error and is discarded. At the trained densities 0.1, 0.2, 0.4, 0.6 and $0.7\sigma^{-3}$, four runs per density are held out and three are used for validation, which leaves 74 training runs. The 21 runs at $0.5\sigma^{-3}$ are never trained. This comparison tests the adapted implementations on a simple fluid, but does not isolate the cause of their different performance on the ions.

\subsection*{S4.3 Data efficiency and structural consistency}
\begin{table}[H]
\centering
\caption{Profile errors and structural checks of the adapted baselines, our functional at fixed coupling and the pair closure. Profile errors are mean relative $L_2$ errors (\%) as in Fig.~2 of the main text (MD statistical error about 1\%); ``quarter'', training on a quarter of the runs; bold, lowest error in each column. Means exclude failed solutions (footnotes); entries in parentheses are final iterates of mostly unconverged Euler--Lagrange iterations (footnote $c$). Structural checks are maxima over the MD ion profiles of the relative net internal force, the Hessian asymmetry and the path dependence of the free energy.}
\label{tab:S:compare}
\small
\setlength{\tabcolsep}{3pt}\begin{tabular}{lcccccccccccc}
 & & \multicolumn{2}{c}{Lennard-Jones} & \multicolumn{6}{c}{Ions} & \multicolumn{3}{c}{Structural checks}\\
\cmidrule(lr){3-4}\cmidrule(lr){5-10}\cmidrule(lr){11-13}
 & & & & \multicolumn{2}{c}{held-out} & \multicolumn{2}{c}{$c=0.05$} & \multicolumn{2}{c}{quarter} & & & \\
\cmidrule(lr){5-6}\cmidrule(lr){7-8}\cmidrule(lr){9-10}
Form & Parameters & all & quarter & $\varepsilon_r=7.5$ & 2 & 7.5 & 2 & 7.5 & 2 & Force & Hessian & Path\\
\midrule
Neural functional (this work) & 19{,}647 & \textbf{0.23} & \textbf{0.27} & \textbf{1.43} & \textbf{2.85} & \textbf{0.79} & 0.88 & \textbf{1.81} & \textbf{3.56} & $4\times10^{-8}$ & $3\times10^{-16}$ & $2\times10^{-11}$\\
One-body network~\cite{SI-sammuller2023neural,SI-bui2025learning}$^{a}$ & 32{,}514 & 0.64 & 0.70 & 1.75 & 3.41 & 1.01 & \textbf{0.85} & 3.00 & 5.55 & 0.14 & 0.55 & 0.009\\
Lattice free energy~\cite{SI-cheng2026equivariant}$^{b}$ & 23{,}828 & 0.54 & 0.80 & 2.97 & 7.16 & 2.06 & 2.72 & 4.53 & 6.31 & $7\times10^{-4}$ & $2\times10^{-16}$ & $7\times10^{-8}$\\
Convolutional free energy~\cite{SI-dijkman2025learning}$^{c}$ & 24{,}321 & (2.03) & (3.20) & (13.8) & (15.1) & (16.1) & (16.2) & -- & -- & 0.06 & $10^{-16}$ & $10^{-15}$\\
Pair closure & 1{,}342 & 12.7 & -- & 3.90 & 8.66 & 1.75 & 2.37 & -- & -- & \multicolumn{3}{c}{exact}\\
\bottomrule
\end{tabular}\par\smallskip
{\footnotesize\raggedright $^{a}$With a quarter of the runs, it fails for 3 of the 20 held-out ion runs at $\varepsilon_r=7.5$ and for 1 of the 41 Lennard-Jones runs. $^{b}$Trained on all runs, it fails for 4 of the 42 ion test runs at $\varepsilon_r=7.5$ (3 not converged, 1 collapsed) and for 10 at $\varepsilon_r=2$ (6 not converged, 4 collapsed); trained on a quarter of the runs, it fails for 2 and 7 of the 20 held-out runs at $\varepsilon_r=7.5$ and 2, respectively. $^{c}$Trained on all runs, it converges for 5 of the 42 ion test runs at $\varepsilon_r=7.5$, for none at $\varepsilon_r=2$, and for 7 of the 41 Lennard-Jones runs; it is not trained on a quarter of the ion runs.}
\end{table}

Table~\ref{tab:S:archfrac} reports the errors for training on 25\%, 50\% and 100\% of the ion training runs. The subsets are nested and preserve the proportions of concentrations and potential families; validation and test runs are the same throughout. The architecture comparison of the main text quotes the fits on the full training set and on a quarter of it. Figure~\ref{fig:S:arch} plots these errors together with representative held-out profiles and the structural residuals of Table~\ref{tab:S:compare}; Fig.~\ref{fig:S:lj} shows the corresponding Lennard-Jones control.

\begin{figure}[tbp]
\centering
\includegraphics[width=\linewidth]{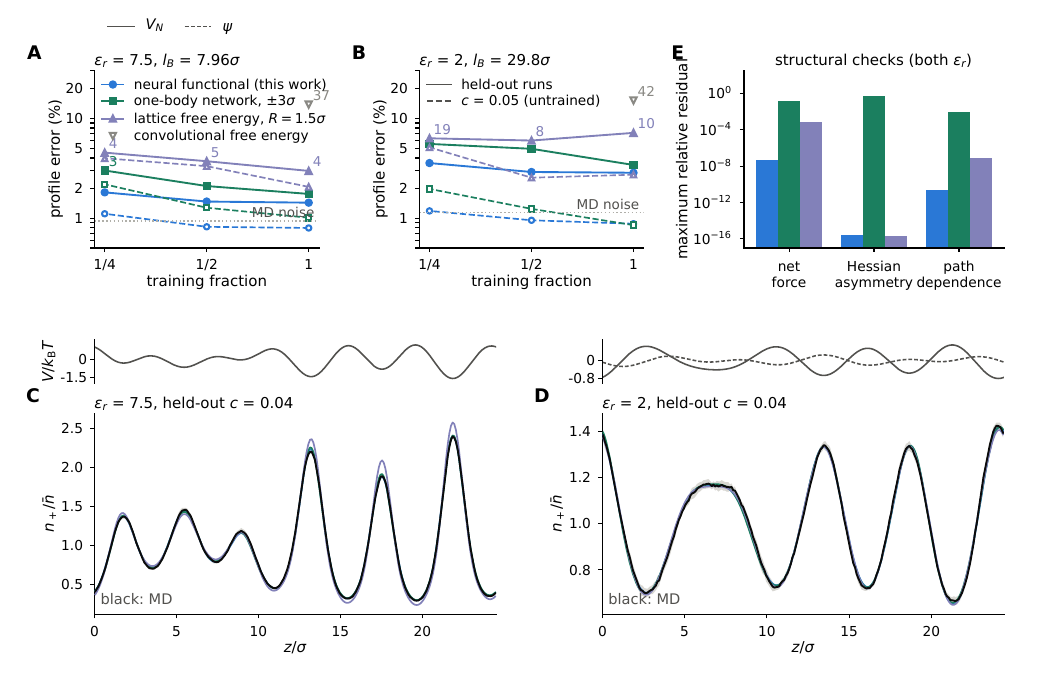}
\caption{Data efficiency and structural consistency on the ions. (\textit{A}, \textit{B}) Mean relative profile error of the 20 held-out runs (solid) and of the 22 runs at the untrained $c=0.05$ (dashed, open markers) versus the fraction of training runs, at $\varepsilon_r=7.5$ and 2, for our functional, the one-body network and the lattice free energy; the convolutional free energy (trained on all runs; final iterates) is a single open triangle. Numbers next to markers count failed Euler--Lagrange solutions among the 42 test runs; the dotted line is the MD noise. Values as in Table~\ref{tab:S:archfrac}. (\textit{C}, \textit{D}) One held-out run at $c=0.04$ at each permittivity: cation density from MD (black, $\pm2$ standard errors) and from each form, with the applied potential of that run drawn above the panel ($V_N$ solid, $\psi$ dashed, in $k_BT$). (\textit{E}) Largest relative residual of the net internal force, the Hessian asymmetry and the free-energy path dependence over all runs (Table~\ref{tab:S:compare}).}
\label{fig:S:arch}
\end{figure}

\begin{figure}[tbp]
\centering
\includegraphics[width=\linewidth]{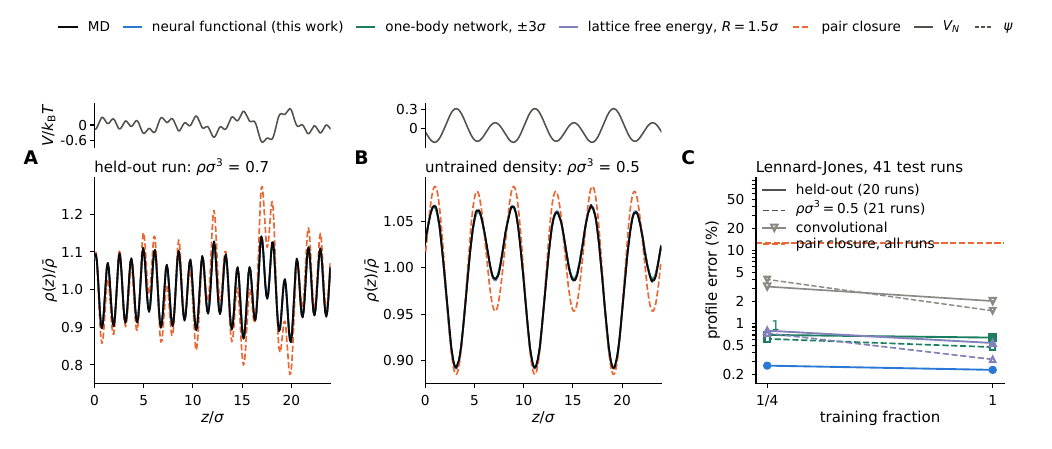}
\caption{Lennard-Jones control (truncated and shifted, $r_c=2.5\sigma$, $T=1.5$, $L=24\sigma$; trained at $\rho\sigma^3=0.1$, 0.2, 0.4, 0.6 and 0.7). (\textit{A}) A held-out run at $\rho\sigma^3=0.7$ and (\textit{B}) a run at the untrained density $\rho\sigma^3=0.5$: MD against the four forms, with the applied potential of that run drawn above the panel (in $k_BT$). Our functional, the one-body network and the lattice free energy lie within 0.1--0.4\% of MD and are hidden under the MD curve; only the pair closure (orange dashed) is visibly different. (\textit{C}) Mean relative profile error of the 20 held-out runs (solid) and the 21 runs at $\rho\sigma^3=0.5$ (dashed) for the full training set and a quarter of it; the pair closure trained on all runs is the horizontal line. Held-out values as in Table~\ref{tab:S:compare}.}
\label{fig:S:lj}
\end{figure}

\begin{table}[H]
\centering
\caption{Data efficiency on the ions. Mean relative profile errors (\%) of the held-out runs / of the runs at the untrained concentration $c=0.05$, for the training fractions shown. Means exclude Euler--Lagrange solutions that do not converge or collapse onto a density spike (error above 50\%). The one-body window is $\pm3\sigma$ and the lattice stencil radius is $1.5\sigma$. The lowest error in each row is in bold.}
\label{tab:S:archfrac}
\begin{tabular}{c c c c c}
$\varepsilon_r$ & Training fraction & Neural functional & One-body network & Lattice free energy\\
\midrule
7.5 & 0.25 & \textbf{1.81} / \textbf{1.10} & 3.00 / 2.18 & 4.53 / 3.98\\
7.5 & 0.5 & \textbf{1.46} / \textbf{0.82} & 2.09 / 1.27 & 3.72 / 3.32\\
7.5 & 1 & \textbf{1.43} / \textbf{0.79} & 1.75 / 1.01 & 2.97 / 2.06\\
2 & 0.25 & \textbf{3.56} / \textbf{1.18} & 5.55 / 1.96 & 6.31 / 5.12\\
2 & 0.5 & \textbf{2.90} / \textbf{0.95} & 4.95 / 1.24 & 6.00 / 2.54\\
2 & 1 & \textbf{2.85} / 0.88 & 3.41 / \textbf{0.85} & 7.16 / 2.72\\
\bottomrule
\end{tabular}
\end{table}

Structural errors in Table~\ref{tab:S:archstruct} are evaluated on the MD profiles. The relative net internal force is $|\sum_\alpha\int f^{\rm int}_\alpha\,\rmd z|/\sum_\alpha\int|f^{\rm int}_\alpha|\,\rmd z$, including Coulomb forces. Integrability is tested by $\|J-J^{\T}\|/\|J\|$ for $J=\partial\mu^{\rm ex}/\partial n$. Reflection error is $\|\mu[Rn]-R\mu[n]\|/\|\mu-\langle\mu\rangle\|$, with $R:z\mapsto-z$. Path dependence is $|\oint\mu^{\rm ex}\cdot\rmd n|/\sum|{\rm legs}|$ on closed paths $\nb\to n_1\to n_2\to\nb$ through held-out profiles. The continuum identities of section S1 and their finite-grid residuals are distinct from predictive accuracy.

Grid independence is tested by evaluating each form, with unchanged parameters, on the test profiles resampled by band-limited interpolation from the $0.1\sigma$ bins to grids of $0.05\sigma$ and $0.2\sigma$. The rms change of $\mu^{\rm ex}_\theta$ relative to its variation over the box, averaged over the 42 test runs of each $\varepsilon_r$, is below $2\times10^{-10}$ on the finer and $7\times10^{-6}$ on the coarser grid for the neural functional (maxima $4\times10^{-9}$ and $3\times10^{-5}$), and below $10^{-14}$ and $4\times10^{-5}$ for the pair closure. The one-body network and the lattice free energy have a native grid spacing, since their window and voxel are fixed numbers of bins, and the read-out of the convolutional free energy fixes the number of grid points. Profiles on another grid must first be interpolated to the training spacing before these forms can be evaluated. The test therefore applies to the neural functional and the pair closure, whose parameters refer to no grid.

\begin{table}[H]
\centering
\caption{Structural errors for the selected implementations. Values are maxima over runs and state points. Parentheses give arithmetic means over runs where available.}
\label{tab:S:archstruct}
\begin{tabular}{l c c c c c}
Form & $\varepsilon_r$ & Net force & Jacobian asymmetry & Reflection & Path dependence\\
\midrule
Neural functional & 7.5 & $4\times10^{-8}$ ($7\times10^{-10}$) & $3\times10^{-16}$ ($2\times10^{-16}$) & $2\times10^{-14}$ & $8\times10^{-16}$\\
Neural functional & 2 & $2\times10^{-8}$ ($2\times10^{-10}$) & $3\times10^{-16}$ ($3\times10^{-16}$) & $4\times10^{-14}$ & $2\times10^{-11}$\\
One-body network & 7.5 & 0.14 (0.008) & 0.55 (0.31) & 0.38 & 0.008\\
One-body network & 2 & 0.08 (0.005) & 0.46 (0.25) & 0.16 & 0.009\\
Lattice free energy & 7.5 & $7\times10^{-4}$ ($7\times10^{-6}$) & $2\times10^{-16}$ ($2\times10^{-16}$) & $6\times10^{-15}$ & $2\times10^{-10}$\\
Lattice free energy & 2 & $2\times10^{-5}$ ($2\times10^{-7}$) & $2\times10^{-16}$ ($10^{-16}$) & $10^{-14}$ & $7\times10^{-8}$\\
\bottomrule
\end{tabular}
\end{table}

\subsection*{S4.4 Convolutional free-energy baseline}
Table~\ref{tab:S:dijkman} compares the convolutional free energy of section S4.2 with our functional, both trained by force matching on the same data. Its scalar output preserves integrability, but pooling and the read-out break translation and reflection invariance, so its bulk Hessian is not diagonal in $k$ and a bulk structure factor is not uniquely defined. The convolutional form has larger force residuals, and few of its Euler--Lagrange solutions converge. Since the original work trains this architecture by pair-correlation matching on bulk data, these results characterize the architecture under force matching, not the published method.

\begin{table}[H]
\centering
\caption{Force-matching comparison of the neural and convolutional free energies. Structure-factor errors are the largest relative rms errors across $c=0.01$--0.08 ($S_{ZZ}$ / $S_{NN}$). Profile errors in parentheses are those of final iterates that mostly did not converge.}
\label{tab:S:dijkman}
\begin{tabular}{c l c c c c}
$\varepsilon_r$ & Form & Held-out $\chi^2$ & Profile (\%) & EL converged & $S(k)$ error (\%)\\
\midrule
7.5 & neural functional & 0.61 & 1.43 & 42/42 & 6.8 / 2.6\\
7.5 & convolutional & 83.4 & (13.8) & 5/42 & --\\
2 & neural functional & 0.85 & 2.85 & 42/42 & 8.0 / 6.5\\
2 & convolutional & 97.0 & (15.1) & 0/42 & --\\
\bottomrule
\end{tabular}
\par\smallskip\tablenote{Convolutional profile errors include Euler--Lagrange iterations that did not converge. EL counts cover all 42 test runs.}
\end{table}

\section*{S5. Transfer across the Coulomb coupling}

\subsection*{S5.1 Leave-one-out transfer}
Three functionals of the form of Eq.~\ref{eq:lb} of the main text are trained with the runs at $\varepsilon_r=5$, 4 or 3 removed, and each predicts the runs at the permittivity it never saw. Table~\ref{tab:S:loo} lists the profile errors by concentration next to the held-out errors of the functional trained on all five permittivities, and the control without the coupling law. The errors at the permittivity left out follow the concentration dependence of the in-sample errors, largest at $c=0.01$, the most dilute state, where the MD profiles are noisiest, and are nowhere larger than the in-sample errors by more than their run-to-run spread. The control fails: one $l_B$-independent functional cannot represent four couplings, and its Euler--Lagrange solutions diverge at $c\geq0.04$. Its held-out errors at the four trained permittivities are 7.3 to 9.5\% at $\varepsilon_r=5$, 3 and 2 and 23\% at $\varepsilon_r=7.5$ (12\% over all 70 held-out runs), and 14 of the 70 Euler--Lagrange solutions do not converge. Table~\ref{tab:S:loo} lists its errors at the permittivity left out. The response factor $\Gamma(k_1)$ at the 16 state points left out agrees with MD within its uncertainty, with a largest deviation of 2.0 standard errors at $\varepsilon_r=3$ and $c=0.01$.
\begin{table}[H]
\centering
\caption{Leave-one-out transfer across the Coulomb coupling. Relative profile errors (\%, cation / anion) of the runs at the permittivity left out of training, predicted by the functional trained on the other four permittivities, compared with the held-out error of the functional trained on all five at the same state points. The last column is the control without the coupling law, an $l_B$-independent functional of Eq.~\ref{eq:Fex} trained on the same four permittivities (its errors at $\varepsilon_r=4$, with the number of converged Euler--Lagrange solutions). Statistical errors of the MD profiles are about 1\%.}
\label{tab:S:loo}
\begin{tabular}{c c c c c c c}
$\varepsilon_r$ left out & $c$ & runs & trained without $\varepsilon_r$ & held-out runs & trained on all five & no coupling law\\
\midrule
5 & 0.01 & 14 & 3.9 / 3.2 & 3 & 3.5 / 2.7 & --\\
 & 0.02 & 15 & 1.8 / 1.5 & 4 & 1.7 / 1.3 & --\\
 & 0.04 & 14 & 1.1 / 0.9 & 1 & 0.9 / 0.8 & --\\
 & 0.06 & 14 & 0.8 / 0.6 & 4 & 0.7 / 0.6 & --\\
 & 0.08 & 14 & 0.7 / 0.5 & 4 & 0.7 / 0.5 & --\\
4 & 0.01 & 14 & 3.1 / 2.5 & 3 & 3.3 / 2.8 & 8.9 / 10.3 (14/14)\\
 & 0.02 & 15 & 1.5 / 1.4 & 4 & 1.6 / 1.5 & 7.1 / 9.0 (15/15)\\
 & 0.04 & 15 & 0.9 / 0.8 & 1 & 0.9 / 0.7 & 9.2 / 11.2 (13/15)\\
 & 0.06 & 15 & 0.7 / 0.6 & 4 & 0.6 / 0.5 & 11.1 / 16.0 (12/15)\\
 & 0.08 & 15 & 0.7 / 0.5 & 4 & 0.6 / 0.5 & 11.0 / 17.2 (8/15)\\
3 & 0.01 & 13 & 3.9 / 3.6 & 2 & 5.0 / 5.0 & --\\
 & 0.02 & 14 & 1.8 / 1.5 & 3 & 1.7 / 1.3 & --\\
 & 0.04 & 13 & 0.9 / 0.8 & 1 & 0.7 / 0.7 & --\\
 & 0.06 & 15 & 0.7 / 0.6 & 4 & 0.6 / 0.5 & --\\
 & 0.08 & 14 & 0.5 / 0.4 & 4 & 0.5 / 0.6 & --\\
\bottomrule
\end{tabular}
\end{table}

\subsection*{S5.2 Conditioning on the coupling}
Table~\ref{tab:S:conditioning} compares, with $\varepsilon_r=4$ left out, the coupling law of Eq.~\ref{eq:lb} with the same kernels and network conditioned on $\ln\lambda$ through an additional input or through feature-wise modulation of the hidden layers. All conditioned forms predict the profiles at the unseen permittivity equally well, and all reproduce $\Gamma(k_1)$ there within the MD uncertainty. For the energy identity of Eq.~\ref{eq:HFbulk}, the MD derivative is obtained by second differences of the Coulomb energy on the concentration grid. It changes sign at $\nb\sigma^3\approx0.035$ at $\varepsilon_r=5$ and $0.044$ at $\varepsilon_r=4$, is positive throughout the studied range at $\varepsilon_r=7.5$ and negative throughout at $\varepsilon_r=3$ and 2. Evaluated on the same basis, with each functional differentiated by the same second differences at the MD state points, the root-mean-square deviations from MD at the four points of $\varepsilon_r=4$ are $0.012\,\sigma^{-1}$ for the coupling law, $0.005\,\sigma^{-1}$ for the network input and $0.009\,\sigma^{-1}$ for the modulation, and over all 18 points $0.014$, $0.020$ and $0.027\,\sigma^{-1}$. These differences are below the root-mean-square MD standard error of the derivative, $0.015\,\sigma^{-1}$, and comparable to the difference between the analytic derivative of a functional and its second-difference estimate, about $0.014\,\sigma^{-1}$. The larger 18-point deviations of the conditioned forms arise at the trained permittivities, not at the one left out. At the precision of the MD energies, the energy identity therefore does not discriminate between the three forms.
\begin{table}[H]
\centering
\caption{Conditioning on the coupling, with $\varepsilon_r=4$ left out of training. Columns: mean relative profile error (\%) at $\varepsilon_r=4$ (converged Euler--Lagrange solutions / runs); held-out error at the four trained permittivities; root-mean-square deviation of $\partial\Gamma(0)/\partial l_B$ from MD over the 18 interior state points of Fig.~5\textit{B} and over the four points at $\varepsilon_r=4$, both differentiated by the same second differences (MD root-mean-square value 0.076$\,\sigma^{-1}$, standard error 0.015$\,\sigma^{-1}$); root-mean-square deviation of $\Gamma(k_1)$ at $\varepsilon_r=4$ in MD standard errors (six concentrations).}
\label{tab:S:conditioning}
\begin{tabular}{l c c c c}
Form & $\varepsilon_r=4$ (\%) & held-out (\%) & energy identity, 18 points / $\varepsilon_r=4$ ($\sigma^{-1}$) & $\Gamma(k_1)$ at $\varepsilon_r=4$ (MD standard errors)\\
\midrule
coupling law, Eq.~\ref{eq:lb} & 1.2 (74/74) & 1.8 (70/70) & 0.014 / 0.012 & 0.8\\
$\ln\lambda$ as a network input & 1.2 (74/74) & 1.7 (70/70) & 0.020 / 0.005 & 0.7\\
feature-wise modulation (16 units) & 1.3 (74/74) & 1.7 (70/70) & 0.027 / 0.009 & 0.9\\
no dependence on $l_B$ & 10.7 (62/74) & 12.7 (56/70) & -- & --\\
\bottomrule
\end{tabular}
\end{table}

\subsection*{S5.3 Channel decomposition and reference theories}
Table~\ref{tab:S:channels} separates the bulk correlations into the charge channel beyond the mean field and the number channel at three concentrations and the five permittivities, for MD, the functional trained on all five permittivities and the mean spherical approximation of the restricted primitive model with contact distance $\sigma$ \cite{SI-waisman1972mean}. At $k=2\sigma^{-1}$ the MD charge channel exceeds the mean spherical approximation by a nearly constant factor of 1.4--1.6 at every permittivity and concentration, so the approximation has the dependence on the coupling right and misses only the magnitude set by the host. At $k=1\sigma^{-1}$ the ratio falls from 3.1 at $\varepsilon_r=7.5$ to 2.2 at $\varepsilon_r=2$: the sublinear growth at the pair scale is beyond the approximation. Bjerrum's association constant with the same contact distance, $K_B=4\pi\int_\sigma^{l_B/2}r^2e^{l_B/r}\,\rmd r$, is $1.2\times10^4\sigma^3$ at $\varepsilon_r=7.5$ and $4.5\times10^{12}\sigma^3$ at $\varepsilon_r=2$. The paired fraction $\alpha$ solves the mass-action law $\alpha/(1-\alpha)^2=K_B\nb\,\gamma_\pm^2$ with the Debye--H\"uckel activity coefficient of the free ions, $\ln\gamma_\pm=-l_B\kappa_f/[2(1+\kappa_f\sigma)]$ and $\kappa_f^2=8\pi l_B\nb(1-\alpha)$ \cite{SI-bjerrum1926,SI-fisher1993criticality}. It predicts a paired fraction of 0.54 at $\varepsilon_r=7.5$ and $c=0.01$, where the MD number response $\Gamma(k_1)=1.20\pm0.06$ is suppressed. The suppression turns into enhancement between $\varepsilon_r=4$ and 3, with $\Gamma(k_1)=0.66\pm0.06$ at $\varepsilon_r=3$ and $0.42\pm0.07$ at $\varepsilon_r=2$. The functionals trained without a permittivity reproduce the charge channel there to 7\% and the number channel to 17\%.
\begin{table}[H]
\centering
\caption{Channel decomposition of the bulk correlations across the coupling. Charge channel beyond the mean field, $\Gamma_{ZZ}^{\rm corr}(k)=2\bar n/S_{ZZ}(k)-1-\kappa_D^2/k^2$, at $k=2\sigma^{-1}$ (divided by $l_B$) and $k=1\sigma^{-1}$, and the number channel $\Gamma(k_1)=2\bar n/S_{NN}(k_1)$: zero-field MD / the functional trained on all five permittivities / the mean spherical approximation of the restricted primitive model with contact distance $\sigma$. MD charge-channel values are from single zero-field runs without error estimates; $\Gamma(k_1)$ uncertainties are those of section S2.3.}
\label{tab:S:channels}
\begin{tabular}{c c c c c}
$c$ & $\varepsilon_r$ & $\Gamma_{ZZ}^{\rm corr}(2\sigma^{-1})/l_B$ ($\sigma^{-1}$) & $\Gamma_{ZZ}^{\rm corr}(1\sigma^{-1})$ & $\Gamma(k_1)$\\
\midrule
0.01 & 7.5 & -0.063 / -0.069 / -0.049 & -1.37 / -1.35 / -0.49 & $1.20\pm0.06$ / 1.19\\
 & 5 & -0.062 / -0.065 / -0.046 & -1.72 / -1.69 / -0.69 & $0.92\pm0.09$ / 1.07\\
 & 4 & -0.062 / -0.062 / -0.045 & -1.89 / -1.93 / -0.84 & $0.92\pm0.06$ / 0.97\\
 & 3 & -0.061 / -0.060 / -0.044 & -2.17 / -2.30 / -1.08 & $0.66\pm0.06$ / 0.78\\
 & 2 & -0.058 / -0.056 / -0.041 & -2.27 / -3.02 / -1.53 & $0.42\pm0.07$ / 0.37\\
0.04 & 7.5 & -0.243 / -0.246 / -0.162 & -4.90 / -4.73 / -1.59 & $4.38\pm0.15$ / 4.11\\
 & 5 & -0.230 / -0.235 / -0.153 & -6.05 / -5.89 / -2.25 & $4.03\pm0.23$ / 4.27\\
 & 4 & -0.226 / -0.229 / -0.149 & -6.76 / -6.77 / -2.73 & $4.61\pm0.51$ / 4.24\\
 & 3 & -0.219 / -0.221 / -0.144 & -8.78 / -8.24 / -3.49 & $3.97\pm0.43$ / 4.03\\
 & 2 & -0.210 / -0.213 / -0.136 & -10.96 / -11.25 / -4.95 & $3.14\pm0.56$ / 3.31\\
0.08 & 7.5 & -0.440 / -0.444 / -0.280 & -8.59 / -9.02 / -2.74 & $9.26\pm0.26$ / 8.94\\
 & 5 & -0.425 / -0.424 / -0.268 & -10.69 / -10.76 / -3.89 & $9.52\pm0.64$ / 9.92\\
 & 4 & -0.410 / -0.408 / -0.261 & -12.50 / -12.08 / -4.72 & $10.23\pm0.98$ / 10.30\\
 & 3 & -0.393 / -0.391 / -0.252 & -13.40 / -14.28 / -6.05 & $10.68\pm0.56$ / 10.55\\
 & 2 & -0.372 / -0.371 / -0.240 & -17.40 / -18.64 / -8.61 & $11.05\pm0.81$ / 10.35\\
\bottomrule
\end{tabular}
\end{table}

\subsection*{S5.4 Nonlinear response across the coupling}
The runs at $\varepsilon_r=5$, 4, 3 and 2 apply the same potentials, with the same seeds and amplitudes, at every concentration, so the driven profiles compare directly across the coupling. We separate each response into the number channel, $n_++n_--2\bar n$, and the charge channel, $n_+-n_-$. Table~\ref{tab:S:nonlinear} summarizes the change of the responses from $\varepsilon_r=5$ to 3 in MD and as predicted by the functionals trained without the respective permittivity. The charge response to a charged potential shrinks with the coupling at every concentration and is predicted to a few percent of its amplitude. The number response to a neutral potential is independent of the coupling within the MD noise at $c\geq0.06$, which the functionals reproduce, and changes with the coupling at $c=0.01$ and 0.02, where the functionals underestimate the change by about 40\%.
\begin{table}[H]
\centering
\caption{Nonlinear response at permittivities left out of training: the same potentials at $\varepsilon_r=5$, 4 and 3, each predicted by the functional trained without that permittivity. Neutral potentials: change of the number response from $\varepsilon_r=5$ to 3 (\% of amplitude; MD with the noise removed in quadrature / functional) and prediction errors at $\varepsilon_r=5$, 4 and 3 (\% of amplitude, MD noise in parentheses). Charged potentials: charge-response amplitude at $\varepsilon_r=4$ and 3 relative to 5 (MD / functional) and prediction errors at $\varepsilon_r=5$, 4 and 3 (\% of amplitude).}
\label{tab:S:nonlinear}
\begin{tabular}{c c c c c c c}
$c$ & potentials & $\Delta N$ MD / functional & number error 5 / 4 / 3 & $|Z|$ ratio MD & $|Z|$ ratio functional & charge error 5 / 4 / 3\\
\midrule
0.01 & 3 / 2 & 38 / 23 & 11.1 / 16.5 / 12.8 (0.0) & 0.92 / 0.79 & 0.90 / 0.78 & 4.1 / 3.2 / 4.2\\
0.02 & 4 / 2 & 25 / 15 & 9.0 / 6.4 / 6.5 (0.0) & 0.90 / 0.78 & 0.89 / 0.76 & 2.4 / 2.3 / 3.1\\
0.04 & 5 / 1 & 9 / 3 & 7.4 / 5.9 / 6.1 (0.0) & 0.83 / 0.63 & 0.82 / 0.62 & 2.2 / 1.9 / 2.0\\
0.06 & 5 / 2 & 4 / 4 & 6.7 / 5.7 / 6.4 (0.0) & 0.84 / 0.66 & 0.84 / 0.66 & 2.4 / 2.3 / 2.5\\
0.08 & 5 / 2 & 8 / 10 & 6.3 / 7.7 / 6.9 (0.0) & 0.78 / 0.56 & 0.80 / 0.57 & 2.1 / 2.3 / 2.9\\
\bottomrule
\end{tabular}
\end{table}

\section*{S6. Implicit treatment of the polymer host}
The functional describes the ions alone. The polymer enters through the solvation and packing it imposes on them, as a host in equilibrium with the ion densities, in the same way as the solvent enters implicit-solvent functionals of the primitive model \cite{SI-bui2025learning}. A polymer density field would add a third component to $F$ and to the Ornstein--Zernike relation. At the length scales of the ion correlations studied here, this component is nearly inert in a melt. Zero-field simulations at $c=0.04$ that record all particles give a normalized monomer structure factor $S_{pp}(k)/n_p$ of $0.060$ and $0.050$ at $k_1$ for $\varepsilon_r=7.5$ and 2 ($0.078$ and $0.058$ at $2k_1$), flat as in a neat melt, and a correlation coefficient between the monomer and ion number densities of $-0.15$ and $-0.16$ at $k_1$, the sign of volume exclusion. Inverting the three-component Ornstein--Zernike relation separates the ion number kernel into a direct ion--ion part and a polymer-mediated part, $\nb\,n_p\,c_{Np}^2/(1-n_pc_{pp})$ in terms of the direct correlation functions $c_{Np}$ and $c_{pp}$. The polymer-mediated share of the kernel is 7\% and 5\% at $k_1$, 11\% and 15\% at $2k_1$, 40\% and 34\% at $3k_1$, and dominant above about $4k_1$, the scale of the solvation shell. The polymer therefore mediates the short-range part of the ion correlations, which the learned kernels absorb, and contributes little at the wavelengths of the number response in Figs.~4 and 5 of the main text. Ions are not bound to chains: the chain that solvates an ion changes with a half-life of 50 and $100\,\tau$ at $\varepsilon_r=7.5$ and 2, short compared with the sampling times. The ion-only description corresponds to a homogeneous host. Where the polymer density itself is strongly inhomogeneous, at electrodes or under confinement, the host is no longer a background in equilibrium with the ions, and an explicit polymer density field, with its own chain-connectivity term, is required.

\FloatBarrier
\renewcommand{\refname}{Supplementary References}

\endgroup

\begin{thebibliography}{54}
\providecommand{\natexlab}[1]{#1}
\providecommand{\url}[1]{\texttt{#1}}
\expandafter\ifx\csname urlstyle\endcsname\relax
  \providecommand{\doi}[1]{doi: #1}\else
  \providecommand{\doi}{doi: \begingroup \urlstyle{rm}\Url}\fi

\bibitem[Hallinan and Balsara(2013)]{hallinan2013polymer}
Daniel~T. Hallinan, Jr. and Nitash~P. Balsara.
\newblock Polymer electrolytes.
\newblock \emph{Annual Review of Materials Research}, 43\penalty0 (1):\penalty0
  503--525, 2013.

\bibitem[Bocharova and Sokolov(2020)]{bocharova2020perspectives}
Vera Bocharova and Alexei~P Sokolov.
\newblock Perspectives for polymer electrolytes: {A} view from fundamentals of
  ionic conductivity.
\newblock \emph{Macromolecules}, 53\penalty0 (11):\penalty0 4141--4157, 2020.

\bibitem[Son and Wang(2020)]{son2020ion}
Chang~Yun Son and Zhen-Gang Wang.
\newblock Ion transport in small-molecule and polymer electrolytes.
\newblock \emph{J. Chem. Phys.}, 153\penalty0 (10):\penalty0 100903, 2020.

\bibitem[Fong et~al.(2021)Fong, Self, McCloskey, and Persson]{fong2021ion}
Kara~D Fong, Julian Self, Bryan~D McCloskey, and Kristin~A Persson.
\newblock Ion correlations and their impact on transport in polymer-based
  electrolytes.
\newblock \emph{Macromolecules}, 54\penalty0 (6):\penalty0 2575--2591, 2021.

\bibitem[Shen and Hall(2020)]{shen2020ion}
Kuan-Hsuan Shen and Lisa~M Hall.
\newblock Ion conductivity and correlations in model salt-doped polymers:
  {E}ffects of interaction strength and concentration.
\newblock \emph{Macromolecules}, 53\penalty0 (10):\penalty0 3655--3668, 2020.

\bibitem[Tsamopoulos and Wang(2024)]{tsamopoulos2024ion}
Alexandros~J Tsamopoulos and Zhen-Gang Wang.
\newblock Ion conductivity in salt-doped polymers: {C}ombined effects of
  temperature and salt concentration.
\newblock \emph{ACS Macro Letters}, 13\penalty0 (3):\penalty0 322--327, 2024.

\bibitem[Borodin and Smith(2006)]{borodin2006mechanism}
Oleg Borodin and Grant~D Smith.
\newblock Mechanism of ion transport in amorphous poly(ethylene oxide)/{LiTFSI}
  from molecular dynamics simulations.
\newblock \emph{Macromolecules}, 39\penalty0 (4):\penalty0 1620--1629, 2006.

\bibitem[Molinari et~al.(2018)Molinari, Mailoa, and
  Kozinsky]{molinari2018effect}
Nicola Molinari, Jonathan~P Mailoa, and Boris Kozinsky.
\newblock Effect of salt concentration on ion clustering and transport in
  polymer solid electrolytes: {A} molecular dynamics study of {PEO--LiTFSI}.
\newblock \emph{Chemistry of Materials}, 30\penalty0 (18):\penalty0 6298--6306,
  2018.

\bibitem[France-Lanord and Grossman(2019)]{france2019correlations}
Arthur France-Lanord and Jeffrey~C Grossman.
\newblock Correlations from ion pairing and the {Nernst--Einstein} equation.
\newblock \emph{Physical Review Letters}, 122\penalty0 (13):\penalty0 136001,
  2019.

\bibitem[Pesko et~al.(2017)Pesko, Timachova, Bhattacharya, Smith, Villaluenga,
  Newman, and Balsara]{pesko2017negative}
Danielle~M Pesko, Ksenia Timachova, Rajashree Bhattacharya, Mackensie~C Smith,
  Irune Villaluenga, John Newman, and Nitash~P Balsara.
\newblock Negative transference numbers in poly(ethylene oxide)-based
  electrolytes.
\newblock \emph{J. Electrochem. Soc.}, 164\penalty0 (11):\penalty0
  E3569--E3575, 2017.

\bibitem[Lyu and Lei(2023)]{lyu2023construction}
Liyao Lyu and Huan Lei.
\newblock Construction of coarse-grained molecular dynamics with many-body
  non-{M}arkovian memory.
\newblock \emph{Physical Review Letters}, 131\penalty0 (17):\penalty0 177301,
  2023.

\bibitem[Wang et~al.(2025)Wang, Cs{\'a}nyi, and Ortner]{wang2025many}
Yangshuai Wang, G{\'a}bor Cs{\'a}nyi, and Christoph Ortner.
\newblock Many-body coarse-grained molecular dynamics with the atomic cluster
  expansion.
\newblock \emph{arXiv preprint arXiv:2502.04661}, 2025.

\bibitem[Hern{\'a}ndez et~al.(2026)Hern{\'a}ndez, Win, O'Connor, Arratia, and
  Trask]{hernandez2026data}
Quercus Hern{\'a}ndez, Max Win, Thomas~C O'Connor, Paulo~E Arratia, and
  Nathaniel Trask.
\newblock Data-driven particle dynamics: {S}tructure-preserving coarse-graining
  for emergent behavior in nonequilibrium systems.
\newblock \emph{Proceedings of the National Academy of Sciences}, 123\penalty0
  (20):\penalty0 e2519631123, 2026.

\bibitem[Evans(1979)]{evans1979nature}
Robert Evans.
\newblock The nature of the liquid-vapour interface and other topics in the
  statistical mechanics of non-uniform, classical fluids.
\newblock \emph{Advances in Physics}, 28\penalty0 (2):\penalty0 143--200, 1979.

\bibitem[Hansen and McDonald(2013)]{hansen2013theory}
Jean-Pierre Hansen and Ian~R McDonald.
\newblock \emph{Theory of Simple Liquids: With Applications to Soft Matter}.
\newblock Academic Press, Oxford, 4th edition, 2013.

\bibitem[Rosenfeld(1989)]{rosenfeld1989free}
Yaakov Rosenfeld.
\newblock Free-energy model for the inhomogeneous hard-sphere fluid mixture and
  density-functional theory of freezing.
\newblock \emph{Physical Review Letters}, 63\penalty0 (9):\penalty0 980, 1989.

\bibitem[Roth(2010)]{roth2010fundamental}
Roland Roth.
\newblock Fundamental measure theory for hard-sphere mixtures: {A} review.
\newblock \emph{J. Phys.: Condens. Matter}, 22\penalty0 (6):\penalty0 063102,
  2010.

\bibitem[Rosenfeld(1993)]{rosenfeld1993free}
Yaakov Rosenfeld.
\newblock Free energy model for inhomogeneous fluid mixtures: {Y}ukawa-charged
  hard spheres, general interactions, and plasmas.
\newblock \emph{J. Chem. Phys.}, 98\penalty0 (10):\penalty0 8126--8148, 1993.

\bibitem[Gillespie et~al.(2002)Gillespie, Nonner, and
  Eisenberg]{gillespie2002coupling}
Dirk Gillespie, Wolfgang Nonner, and Robert~S Eisenberg.
\newblock Coupling {Poisson--Nernst--Planck} and density functional theory to
  calculate ion flux.
\newblock \emph{J. Phys.: Condens. Matter}, 14\penalty0 (46):\penalty0
  12129--12145, 2002.

\bibitem[H{\"a}rtel(2017)]{hartel2017structure}
Andreas H{\"a}rtel.
\newblock Structure of electric double layers in capacitive systems and to what
  extent (classical) density functional theory describes it.
\newblock \emph{J. Phys.: Condens. Matter}, 29\penalty0 (42):\penalty0 423002,
  2017.

\bibitem[Jiang et~al.(2018)Jiang, Ginzburg, and Wang]{jiang2018density}
Jian Jiang, Valeriy~V Ginzburg, and Zhen-Gang Wang.
\newblock Density functional theory for charged fluids.
\newblock \emph{Soft Matter}, 14\penalty0 (28):\penalty0 5878--5887, 2018.

\bibitem[Wang(2010)]{wang2010fluctuation}
Zhen-Gang Wang.
\newblock Fluctuation in electrolyte solutions: {T}he self energy.
\newblock \emph{Physical Review E}, 81\penalty0 (2):\penalty0 021501, 2010.

\bibitem[Nakamura et~al.(2011)Nakamura, Balsara, and
  Wang]{nakamura2011thermodynamics}
Issei Nakamura, Nitash~P Balsara, and Zhen-Gang Wang.
\newblock Thermodynamics of ion-containing polymer blends and block copolymers.
\newblock \emph{Physical Review Letters}, 107\penalty0 (19):\penalty0 198301,
  2011.

\bibitem[Nakamura and Wang(2012)]{nakamura2012salt}
Issei Nakamura and Zhen-Gang Wang.
\newblock Salt-doped block copolymers: {I}on distribution, domain spacing and
  effective $\chi$ parameter.
\newblock \emph{Soft Matter}, 8\penalty0 (36):\penalty0 9356--9367, 2012.

\bibitem[Sing et~al.(2014)Sing, Zwanikken, and Olvera de~la
  Cruz]{sing2014electrostatic}
Charles~E Sing, Jos~W Zwanikken, and Monica Olvera de~la Cruz.
\newblock Electrostatic control of block copolymer morphology.
\newblock \emph{Nature Materials}, 13\penalty0 (7):\penalty0 694--698, 2014.

\bibitem[Samm{\"u}ller et~al.(2023)Samm{\"u}ller, Hermann, de~Las~Heras, and
  Schmidt]{sammuller2023neural}
Florian Samm{\"u}ller, Sophie Hermann, Daniel de~Las~Heras, and Matthias
  Schmidt.
\newblock Neural functional theory for inhomogeneous fluids: {F}undamentals and
  applications.
\newblock \emph{Proceedings of the National Academy of Sciences}, 120\penalty0
  (50):\penalty0 e2312484120, 2023.

\bibitem[Bui and Cox(2025)]{bui2025learning}
Anna~T Bui and Stephen~J Cox.
\newblock Learning classical density functionals for ionic fluids.
\newblock \emph{Physical Review Letters}, 134\penalty0 (14):\penalty0 148001,
  2025.

\bibitem[Dijkman et~al.(2025)Dijkman, Dijkstra, van Roij, Welling, van~de
  Meent, and Ensing]{dijkman2025learning}
Jacobus Dijkman, Marjolein Dijkstra, Ren{\'e} van Roij, Max Welling, Jan-Willem
  van~de Meent, and Bernd Ensing.
\newblock Learning neural free-energy functionals with pair-correlation
  matching.
\newblock \emph{Physical Review Letters}, 134\penalty0 (5):\penalty0 056103,
  2025.

\bibitem[Bui and Cox(2026)]{bui2026unified}
Anna~T Bui and Stephen~J Cox.
\newblock A unified machine-learning framework for ab initio multiscale
  modeling of liquids.
\newblock \emph{Proceedings of the National Academy of Sciences}, 123\penalty0
  (30):\penalty0 e2610049123, 2026.

\bibitem[Batzner et~al.(2022)Batzner, Musaelian, Sun, Geiger, Mailoa,
  Kornbluth, Molinari, Smidt, and Kozinsky]{batzner20223}
Simon Batzner, Albert Musaelian, Lixin Sun, Mario Geiger, Jonathan~P Mailoa,
  Mordechai Kornbluth, Nicola Molinari, Tess~E Smidt, and Boris Kozinsky.
\newblock {E(3)}-equivariant graph neural networks for data-efficient and
  accurate interatomic potentials.
\newblock \emph{Nature Communications}, 13\penalty0 (1):\penalty0 2453, 2022.

\bibitem[Cheng(2026)]{cheng2026equivariant}
Bingqing Cheng.
\newblock Equivariant learning of a transferable three-dimensional classical
  density functional.
\newblock \emph{arXiv preprint arXiv:2608.13506}, 2026.

\bibitem[Lin and Oettel(2019)]{lin2019classical}
Shang-Chun Lin and Martin Oettel.
\newblock A classical density functional from machine learning and a
  convolutional neural network.
\newblock \emph{SciPost Physics}, 6\penalty0 (2):\penalty0 025, 2019.

\bibitem[Kelley et~al.(2024)Kelley, Quinton, Fazel, Karimitari, Sutton, and
  Sundararaman]{kelley2024bridging}
Michelle~M Kelley, Joshua Quinton, Kamron Fazel, Nima Karimitari, Christopher
  Sutton, and Ravishankar Sundararaman.
\newblock Bridging electronic and classical density-functional theory using
  universal machine-learned functional approximations.
\newblock \emph{J. Chem. Phys.}, 161\penalty0 (14):\penalty0 144101, 2024.

\bibitem[Glitsch et~al.(2025)Glitsch, Weimar, and Oettel]{glitsch2025neural}
Felix Glitsch, Jens Weimar, and Martin Oettel.
\newblock Neural density functional theory in higher dimensions with
  convolutional layers.
\newblock \emph{Physical Review E}, 111\penalty0 (5):\penalty0 055305, 2025.

\bibitem[Bjerrum(1926)]{bjerrum1926}
Niels Bjerrum.
\newblock {Untersuchungen {\"u}ber Ionenassoziation. I. Der Einfluss der
  Ionenassoziation auf die Aktivit{\"a}t der Ionen bei mittleren
  Assoziationsgraden}.
\newblock \emph{Kgl. Danske Vidensk. Selsk. Mat.-fys. Medd.}, 7\penalty0
  (9):\penalty0 1--48, 1926.

\bibitem[Fisher and Levin(1993)]{fisher1993criticality}
Michael~E. Fisher and Yan Levin.
\newblock Criticality in ionic fluids: {D}ebye-{H}{\"u}ckel theory, {B}jerrum,
  and beyond.
\newblock \emph{Physical Review Letters}, 71\penalty0 (23):\penalty0
  3826--3829, 1993.

\bibitem[Hermann and Schmidt(2021)]{hermann2021noether}
Sophie Hermann and Matthias Schmidt.
\newblock {N}oether's theorem in statistical mechanics.
\newblock \emph{Communications Physics}, 4\penalty0 (1):\penalty0 176, 2021.

\bibitem[Stillinger and Lovett(1968)]{stillinger1968general}
Frank~H. Stillinger, Jr. and Ronald Lovett.
\newblock General restriction on the distribution of ions in electrolytes.
\newblock \emph{J. Chem. Phys.}, 49\penalty0 (5):\penalty0 1991--1994, 1968.

\bibitem[Langreth and Perdew(1975)]{langreth1975exchange}
David~C. Langreth and John~P. Perdew.
\newblock The exchange-correlation energy of a metallic surface.
\newblock \emph{Solid State Communications}, 17\penalty0 (11):\penalty0
  1425--1429, 1975.

\bibitem[Gunnarsson and Lundqvist(1976)]{gunnarsson1976exchange}
O.~Gunnarsson and B.~I. Lundqvist.
\newblock Exchange and correlation in atoms, molecules, and solids by the
  spin-density-functional formalism.
\newblock \emph{Physical Review B}, 13\penalty0 (10):\penalty0 4274--4298,
  1976.

\bibitem[Samm{\"u}ller and Schmidt(2026)]{sammuller2026determining}
Florian Samm{\"u}ller and Matthias Schmidt.
\newblock Determining the chemical potential via universal density functional
  learning.
\newblock \emph{Physical Review Letters}, 136\penalty0 (6):\penalty0 068202,
  2026.

\bibitem[Chen et~al.(2026)Chen, Zhong, Zhang, and Li]{chen2026learning}
Mengyi Chen, Peichen Zhong, Zihan Zhang, and Qianxiao Li.
\newblock Learning ab initio phase-field models.
\newblock \emph{arXiv preprint arXiv:2610.01432}, 2026.

\bibitem[Lyu et~al.(2027)Lyu, Yu, and Schaeffer]{lyu2026mvnn}
Liyao Lyu, Xinyue Yu, and Hayden Schaeffer.
\newblock {MVNN}: A measure-valued neural network for learning {McKean--Vlasov}
  dynamics from particle data.
\newblock \emph{Journal of Computational Physics}, 569:\penalty0 115467, 2027.
\newblock ISSN 0021-9991.

\bibitem[Waisman and Lebowitz(1972)]{waisman1972mean}
Eduardo Waisman and Joel~L. Lebowitz.
\newblock Mean spherical model integral equation for charged hard spheres {I}.
  {M}ethod of solution.
\newblock \emph{J. Chem. Phys.}, 56\penalty0 (6):\penalty0 3086--3093, 1972.

\bibitem[Kremer and Grest(1990)]{KremerGrest1990}
Kurt Kremer and Gary~S. Grest.
\newblock Dynamics of entangled linear polymer melts: A molecular-dynamics
  simulation.
\newblock \emph{J. Chem. Phys.}, 92:\penalty0 5057--5086, 1990.

\bibitem[Hockney and Eastwood(2021)]{hockney2021computer}
Roger~W Hockney and James~W Eastwood.
\newblock \emph{Computer Simulation Using Particles}.
\newblock CRC Press, Boca Raton, FL, 2021.

\bibitem[Thompson et~al.(2022)Thompson, Aktulga, Berger, Bolintineanu, Brown,
  Crozier, In't~Veld, Kohlmeyer, Moore, Nguyen, et~al.]{thompson2022lammps}
Aidan~P Thompson, H~Metin Aktulga, Richard Berger, Dan~S Bolintineanu,
  W~Michael Brown, Paul~S Crozier, Pieter~J In't~Veld, Axel Kohlmeyer, Stan~G
  Moore, Trung~Dac Nguyen, et~al.
\newblock {LAMMPS}, a flexible simulation tool for particle-based materials
  modeling at the atomic, meso, and continuum scales.
\newblock \emph{Computer Physics Communications}, 271:\penalty0 108171, 2022.

\bibitem[Nos{\'e}(1984)]{nose1984unified}
Shuichi Nos{\'e}.
\newblock A unified formulation of the constant temperature molecular dynamics
  methods.
\newblock \emph{J. Chem. Phys.}, 81\penalty0 (1):\penalty0 511--519, 1984.

\bibitem[Hoover(1985)]{hoover1985canonical}
William~G Hoover.
\newblock Canonical dynamics: Equilibrium phase-space distributions.
\newblock \emph{Physical Review A}, 31\penalty0 (3):\penalty0 1695, 1985.

\bibitem[Perez et~al.(2018)Perez, Strub, de~Vries, Dumoulin, and
  Courville]{perez2018film}
Ethan Perez, Florian Strub, Harm de~Vries, Vincent Dumoulin, and Aaron
  Courville.
\newblock {FiLM}: Visual reasoning with a general conditioning layer.
\newblock In \emph{Proceedings of the AAAI Conference on Artificial
  Intelligence}, volume~32, pages 3942--3951, 2018.

\bibitem[Loshchilov and Hutter(2019)]{LoshchilovHutter2019}
Ilya Loshchilov and Frank Hutter.
\newblock Decoupled weight decay regularization.
\newblock In \emph{International Conference on Learning Representations}, 2019.

\bibitem[Dean(1996)]{dean1996langevin}
David~S Dean.
\newblock {L}angevin equation for the density of a system of interacting
  {L}angevin processes.
\newblock \emph{J. Phys. A: Math. Gen.}, 29\penalty0 (24):\penalty0 L613--L617,
  1996.

\bibitem[Avni et~al.(2022)Avni, Adar, Andelman, and
  Orland]{avni2022conductivity}
Yael Avni, Ram~M Adar, David Andelman, and Henri Orland.
\newblock Conductivity of concentrated electrolytes.
\newblock \emph{Physical Review Letters}, 128\penalty0 (9):\penalty0 098002,
  2022.

\bibitem[Lyu and Lei(2025)]{lyu2025generalization}
Liyao Lyu and Huan Lei.
\newblock On the generalization ability of coarse-grained molecular dynamics
  models for nonequilibrium processes.
\newblock \emph{Multiscale Model. Simul.}, 23\penalty0 (2):\penalty0 816--837,
  2025.

\end{thebibliography}

\begin{thebibliography}{23}
\providecommand{\natexlab}[1]{#1}
\providecommand{\url}[1]{\texttt{#1}}
\expandafter\ifx\csname urlstyle\endcsname\relax
  \providecommand{\doi}[1]{doi: #1}\else
  \providecommand{\doi}{doi: \begingroup \urlstyle{rm}\Url}\fi

\bibitem[Stillinger and Lovett(1968)]{SI-stillinger1968general}
Frank~H. Stillinger, Jr. and Ronald Lovett.
\newblock General restriction on the distribution of ions in electrolytes.
\newblock \emph{J. Chem. Phys.}, 49\penalty0 (5):\penalty0 1991--1994, 1968.

\bibitem[Evans(1979)]{SI-evans1979nature}
Robert Evans.
\newblock The nature of the liquid-vapour interface and other topics in the
  statistical mechanics of non-uniform, classical fluids.
\newblock \emph{Advances in Physics}, 28\penalty0 (2):\penalty0 143--200, 1979.

\bibitem[Langreth and Perdew(1975)]{SI-langreth1975exchange}
David~C. Langreth and John~P. Perdew.
\newblock The exchange-correlation energy of a metallic surface.
\newblock \emph{Solid State Communications}, 17\penalty0 (11):\penalty0
  1425--1429, 1975.

\bibitem[Gunnarsson and Lundqvist(1976)]{SI-gunnarsson1976exchange}
O.~Gunnarsson and B.~I. Lundqvist.
\newblock Exchange and correlation in atoms, molecules, and solids by the
  spin-density-functional formalism.
\newblock \emph{Physical Review B}, 13\penalty0 (10):\penalty0 4274--4298,
  1976.

\bibitem[Tsamopoulos and Wang(2024)]{SI-tsamopoulos2024ion}
Alexandros~J Tsamopoulos and Zhen-Gang Wang.
\newblock Ion conductivity in salt-doped polymers: {C}ombined effects of
  temperature and salt concentration.
\newblock \emph{ACS Macro Letters}, 13\penalty0 (3):\penalty0 322--327, 2024.

\bibitem[Kremer and Grest(1990)]{SI-KremerGrest1990}
Kurt Kremer and Gary~S. Grest.
\newblock Dynamics of entangled linear polymer melts: A molecular-dynamics
  simulation.
\newblock \emph{J. Chem. Phys.}, 92:\penalty0 5057--5086, 1990.

\bibitem[Shen and Hall(2020)]{SI-shen2020ion}
Kuan-Hsuan Shen and Lisa~M Hall.
\newblock Ion conductivity and correlations in model salt-doped polymers:
  {E}ffects of interaction strength and concentration.
\newblock \emph{Macromolecules}, 53\penalty0 (10):\penalty0 3655--3668, 2020.

\bibitem[Hockney and Eastwood(2021)]{SI-hockney2021computer}
Roger~W Hockney and James~W Eastwood.
\newblock \emph{Computer Simulation Using Particles}.
\newblock CRC Press, Boca Raton, FL, 2021.

\bibitem[Thompson et~al.(2022)Thompson, Aktulga, Berger, Bolintineanu, Brown,
  Crozier, In't~Veld, Kohlmeyer, Moore, Nguyen, et~al.]{SI-thompson2022lammps}
Aidan~P Thompson, H~Metin Aktulga, Richard Berger, Dan~S Bolintineanu,
  W~Michael Brown, Paul~S Crozier, Pieter~J In't~Veld, Axel Kohlmeyer, Stan~G
  Moore, Trung~Dac Nguyen, et~al.
\newblock {LAMMPS}, a flexible simulation tool for particle-based materials
  modeling at the atomic, meso, and continuum scales.
\newblock \emph{Computer Physics Communications}, 271:\penalty0 108171, 2022.

\bibitem[Nos{\'e}(1984)]{SI-nose1984unified}
Shuichi Nos{\'e}.
\newblock A unified formulation of the constant temperature molecular dynamics
  methods.
\newblock \emph{J. Chem. Phys.}, 81\penalty0 (1):\penalty0 511--519, 1984.

\bibitem[Hoover(1985)]{SI-hoover1985canonical}
William~G Hoover.
\newblock Canonical dynamics: Equilibrium phase-space distributions.
\newblock \emph{Physical Review A}, 31\penalty0 (3):\penalty0 1695, 1985.

\bibitem[Tschopp et~al.(2022)Tschopp, Samm{\"u}ller, Hermann, Schmidt, and
  Brader]{SI-tschopp2022force}
Salom{\'e}e~M Tschopp, Florian Samm{\"u}ller, Sophie Hermann, Matthias Schmidt,
  and Joseph~M Brader.
\newblock Force density functional theory in-and out-of-equilibrium.
\newblock \emph{Physical Review E}, 106\penalty0 (1):\penalty0 014115, 2022.

\bibitem[Hansen and McDonald(2013)]{SI-hansen2013theory}
Jean-Pierre Hansen and Ian~R McDonald.
\newblock \emph{Theory of Simple Liquids: With Applications to Soft Matter}.
\newblock Academic Press, Oxford, 4th edition, 2013.

\bibitem[Anderson(1965)]{SI-anderson1965iterative}
Donald~G Anderson.
\newblock Iterative procedures for nonlinear integral equations.
\newblock \emph{Journal of the ACM}, 12\penalty0 (4):\penalty0 547--560, 1965.

\bibitem[Walker and Ni(2011)]{SI-walker2011anderson}
Homer~F Walker and Peng Ni.
\newblock {A}nderson acceleration for fixed-point iterations.
\newblock \emph{SIAM Journal on Numerical Analysis}, 49\penalty0 (4):\penalty0
  1715--1735, 2011.

\bibitem[Kerker(1981)]{SI-kerker1981efficient}
GP~Kerker.
\newblock Efficient iteration scheme for self-consistent pseudopotential
  calculations.
\newblock \emph{Physical Review B}, 23\penalty0 (6):\penalty0 3082, 1981.

\bibitem[Samm{\"u}ller et~al.(2023)Samm{\"u}ller, Hermann, de~Las~Heras, and
  Schmidt]{SI-sammuller2023neural}
Florian Samm{\"u}ller, Sophie Hermann, Daniel de~Las~Heras, and Matthias
  Schmidt.
\newblock Neural functional theory for inhomogeneous fluids: {F}undamentals and
  applications.
\newblock \emph{Proceedings of the National Academy of Sciences}, 120\penalty0
  (50):\penalty0 e2312484120, 2023.

\bibitem[Bui and Cox(2025)]{SI-bui2025learning}
Anna~T Bui and Stephen~J Cox.
\newblock Learning classical density functionals for ionic fluids.
\newblock \emph{Physical Review Letters}, 134\penalty0 (14):\penalty0 148001,
  2025.

\bibitem[Dijkman et~al.(2025)Dijkman, Dijkstra, van Roij, Welling, van~de
  Meent, and Ensing]{SI-dijkman2025learning}
Jacobus Dijkman, Marjolein Dijkstra, Ren{\'e} van Roij, Max Welling, Jan-Willem
  van~de Meent, and Bernd Ensing.
\newblock Learning neural free-energy functionals with pair-correlation
  matching.
\newblock \emph{Physical Review Letters}, 134\penalty0 (5):\penalty0 056103,
  2025.

\bibitem[Cheng(2026)]{SI-cheng2026equivariant}
Bingqing Cheng.
\newblock Equivariant learning of a transferable three-dimensional classical
  density functional.
\newblock \emph{arXiv preprint arXiv:2608.13506}, 2026.

\bibitem[Waisman and Lebowitz(1972)]{SI-waisman1972mean}
Eduardo Waisman and Joel~L. Lebowitz.
\newblock Mean spherical model integral equation for charged hard spheres {I}.
  {M}ethod of solution.
\newblock \emph{J. Chem. Phys.}, 56\penalty0 (6):\penalty0 3086--3093, 1972.

\bibitem[Bjerrum(1926)]{SI-bjerrum1926}
Niels Bjerrum.
\newblock {Untersuchungen {\"u}ber Ionenassoziation. I. Der Einfluss der
  Ionenassoziation auf die Aktivit{\"a}t der Ionen bei mittleren
  Assoziationsgraden}.
\newblock \emph{Kgl. Danske Vidensk. Selsk. Mat.-fys. Medd.}, 7\penalty0
  (9):\penalty0 1--48, 1926.

\bibitem[Fisher and Levin(1993)]{SI-fisher1993criticality}
Michael~E. Fisher and Yan Levin.
\newblock Criticality in ionic fluids: {D}ebye-{H}{\"u}ckel theory, {B}jerrum,
  and beyond.
\newblock \emph{Physical Review Letters}, 71\penalty0 (23):\penalty0
  3826--3829, 1993.

\end{thebibliography}
\end{document}